\documentclass[authoryear]{elsarticle}

\usepackage{soul}

\usepackage[english]{babel}
\usepackage{fancyhdr} % um Header und Footer zu designen
\usepackage{url} % zum korrekten Einf\"ugen von URL mit dem \url Befehl, das Package hyperref erzeugt dazu noch Links
\usepackage{caption} % um Bilder neben Bildern oder Tabellen neben Tabellen zu erzeugen
\usepackage{subcaption} % um Bilder neben Bildern oder Tabellen neben Tabellen zu erzeugen
\usepackage{natbib} %erlaubt es, andere Zitatdarstellungen vorzunehmen als numerische
\usepackage{multirow} %erlaubt es, in einer Tabelle mehrerer Zeilen zu einer zusammenzufassen
\usepackage{float} % notwendig, um Abbildungen an der Stelle zu platzieren, an der sie im Quelltext stehen
\usepackage{amssymb}
\usepackage[intlimits]{amsmath} % beides sind Mathematikpakete
\usepackage{amsthm} % Um Theoreme, Beweise etc. zu verwalten
\usepackage[german]{nomencl}
\usepackage{booktabs} % f\"ur bessere Tabellenlinien
\usepackage{geometry}
\usepackage{enumitem}
\usepackage{multicol}

\makenomenclature
\usepackage[pdfborder={000},colorlinks=true,linkcolor=blue,citecolor=blue]{hyperref}
\usepackage{listings}
\usepackage[all]{xy}
\usepackage{color}

\makeatletter
\def\namedlabel#1#2{\begingroup
    #2%
    \def\@currentlabel{#2}%
    \phantomsection\label{#1}\endgroup
}
\makeatother

\definecolor{b}{rgb}{0,0,.8}	%%omega-blau
\definecolor{g}{rgb}{0,.6,0}	%%Tau-grün
\definecolor{n}{rgb}{0,0,0}	%%normal-schwarz
\definecolor{h}{rgb}{0.4,0.2,0.2}	%%hint
\definecolor{v}{rgb}{0.2,0.6,0}

\newtheorem{theorem}{Theorem}

\newcommand{\C}{{\mathbb C}}

\newcommand{\E}{{\mathbb E}}

\newcommand{\G}{{\mathbb G}}

\newcommand{\N}{{\mathbb N}}

\newcommand{\R}{{\mathbb R}}

\newcommand{\V}{{\mathbb V}}

\newcommand{\Z}{{\mathbb Z}}

\newcommand{\DD}{{\mathcal{D}}}

\newcommand{\II}{{\mathcal{I}}}
\newcommand{\JJ}{{\mathcal{J}}}

\newcommand{\OO}{{\mathcal{O}}}

\newcommand{\bss}{\boldsymbol s}

\newcommand{\bsv}{\boldsymbol v}
\newcommand{\bsw}{\boldsymbol w}

\newcommand{\bsA}{\boldsymbol A}

\newcommand{\bsL}{\boldsymbol L}

\newcommand{\bsW}{\boldsymbol W}
\newcommand{\bsX}{\boldsymbol X}
\newcommand{\bsY}{\boldsymbol Y}

\newcommand{\bsone}{\boldsymbol 1}

\newcommand{\bsnull}{\boldsymbol 0}

\newcommand{\bsalpha}{\boldsymbol \alpha}
\newcommand{\bsbeta}{\boldsymbol \beta}

\newcommand{\bseps}{\boldsymbol \varepsilon}

\newcommand{\bssigma}{\boldsymbol \sigma}
\newcommand{\bsSigma}{\boldsymbol \Sigma}

\newcommand{\bsGamma}{\boldsymbol \Gamma}

\newcommand{\eps}{{\varepsilon}}

\DeclareMathOperator*{\argmin}{arg\,min}

\DeclareMathOperator{\diag}{diag}
\DeclareMathOperator{\sign}{sign}

\DeclareMathOperator{\var}{\V ar}
\DeclareMathOperator{\cov}{\C ov}

\newcommand{\ov}\overline
\newcommand{\what}{\widehat}
\newcommand{\wtilde}{\widetilde}

\newcommand{\rig}\right
\newcommand{\lef}\left
\newcommand{\nf}\normalfont

\newcommand{\MAE}{\text{MAE}}

\begin{document}

\title{Iteratively reweighted adaptive lasso \\ for conditional heteroscedastic time series \\ with applications
to AR-ARCH type processes}
% with applications to fast subset selection for mulativariate AR-ARCH models}
\author{Florian Ziel}
\ead{ziel@europa-uni.de}
\address{Europa-Universit\"at Viadrina, Gro\ss e Scharrnstra\ss e 59, 15230 Frankfurt (Oder), Germany}
\journal{Computational Statistics and Data Analysis}

\begin{frontmatter}
\lhead{\nouppercase{\leftmark}}
\begin{abstract}
% Shrinkage algorithms are of great importance in almost every area of statistics, as well as in time series analysis due to the increasing impact of big data. In current literature of lasso type estimators for autoregressive time series the focus is still on models with homoscedastic residuals. An iteratively reweighted adaptive lasso algorithm for the estimation of time series models under conditional heteroscedasticity in a high-dimensional setting is presented. The asymptotic behaviour of the resulting estimator is analysed. It turns out that the proposed estimation procedure performs substantially better than its homoscedastic counterpart. A special case of this algorithm is suitable to estimate multivariate AR-ARCH type models in a very fast fashion. Several model extensions like periodic AR-ARCH, threshold AR-ARCH or ARMA-GARCH are discussed. Finally, different simulation results and applications to electricity market data and returns of metal prices are shown.

Shrinkage algorithms are of great importance in almost every area of statistics due to the increasing impact of big data. Especially time series analysis benefits from efficient and rapid estimation techniques such as the lasso. However, currently lasso type estimators for autoregressive time series models still focus on models with homoscedastic residuals. Therefore, an iteratively reweighted adaptive lasso algorithm for the estimation of time series models under conditional heteroscedasticity is presented in a high-dimensional setting. The asymptotic behaviour of the resulting estimator is analysed. It is found that the proposed estimation procedure performs substantially better than its homoscedastic counterpart. A special case of the algorithm is suitable to compute the estimated multivariate AR-ARCH type models efficiently. Extensions to the model like periodic AR-ARCH, threshold AR-ARCH or ARMA-GARCH are discussed. Finally, different simulation results and applications to electricity market data and 
returns of metal prices are shown.

\end{abstract}
\end{frontmatter}

\section{Introduction} \label{Introduction}

High-dimensional shrinkage and parameter selection techniques are of increasing importance in statistics in the past years. 
% \textcolor{red}{TH: }
In recent years, high-dimensional shrinkage and parameter selection techniques have been of increasing importance.
In many statistical areas, lasso (least absolute shrinkage and selection operator) estimation methods, as introduced by \cite{tibshirani1996regression},
% \textcolor{red}{TH: ,} 
are very popular. 
In time series analysis the influence of lasso type estimators is growing, especially as the asymptotic properties of stationary time series are
usually very similar for stationary time series to the standard regression case,
see e.g. \cite{wang2007regression}, \cite{nardi2011autoregressive} and \cite{yoon2013penalized}.
% The shrinkage property of the lasso make it attractive for 
% subset selection in autoregressive models. 
% \textcolor{red}{TH: 
Hence, given the lasso's shrinkage properties, it is attractive for subset selection in autoregressive models.
 In big data settings, it provides an efficient estimation technique, see \cite{hsu2008subset}, \cite{ren2010subset}, and \cite{ren2013two} for more details.

%zou2006adaptive (adaptive lasso)

Unfortunately, almost the entire literature about $\ell_1$-penalised least square estimation, like the lasso, deals with
 homoscedastic models. The case of heteroscedasticity and conditional heteroscedasticity is rarely has rarely been %\textcolor{red}{TH: has rarely been} 
covered so far. Recently, \cite{medeiros2012estimating} showed that the adaptive lasso estimator 
is consistent and asymptotically normal under very weak assumptions. 
They proved that the consistency and the asymptotic normality hold if the residuals %follow 
% \textcolor{red}{TH: are of/ adhere to/ are described by} 
are described by a weak white noise process.
This includes the case of conditional heteroscedastic ARCH and GARCH-type residuals.
Nevertheless, their classical lasso approach does not make use of the structure of the conditional heteroscedasticity within 
the residuals. Without going into detail,
it is clear that the estimators might be improved if the structure of the conditional
heteroscedasticity in the data is used.
% Further \textcolor{red}{TH: Furthermore}
Furthermore, \cite{yoon2013penalized} analysed the lasso estimator in an autoregressive regression model.
Additionally, they formulated the lasso problem in a time series setting with ARCH errors.
However, they did not provide a solution to the estimation problem and left this for future research.

Recently, \cite{wagener2012bridge} and \cite{wagener2013adaptive} analysed the
properties of weighted lasso-type estimators in a classical heteroscedastic regression setting. 
They showed that their estimators are consistent and asymptotically normal. In addition, their estimators perform significantly better
than their homoscedastic counterpart. 
Their results, conditioned on the covariates, can be used to construct a reweighted estimator that
also works in time series settings.

% In this paper 
We derive an iteratively reweighted adaptive lasso algorithm that
addresses the above mentioned problems. It enables the estimation of high-dimensional sparse time series models
under conditional heteroscedasticity. % \textcolor{red}{TH: \st{possible}}. 
% The assumed regression structure of the model that is satisfied by the majority of important time series processes
% allows for fast estimation methods. \textcolor{red}{TH: We assume a regression structure which is satisfied by the majority of the important time series processes and which admits fast estimation methods.} 
We assume a regression structure which is satisfied by the majority of the important time series processes and which admits fast estimation methods.
The computational complexity of the algorithm 
is essentially the same as the coordinate descent algorithm of \cite{friedman2007pathwise}. 
This very 
fast estimation method for convex penalised models, such as the given $\ell_1$ situation, % \textcolor{red}{TH: ,}
% \textcolor{red}{TH: \st{is suitable for the application} can be applied} 
can be applied to the iteratively reweighted adaptive lasso algorithm.

% We apply and suggest this very 
% fast estimation method for convex penalised models, such as the given $\ell_1$ situation to our iteratively reweighted 
% lasso algorithm.

The algorithm is based on the results of \cite{wagener2013adaptive}, as their results 
% \textcolor{red}{TH: \st{are suitable to be generalised for} can be generalised to}
can be generalised to models 
with conditional heteroscedasticity. 
The sign consistency and asymptotic normality for the proposed estimator is adduced.
% \textcolor{red}{TH: Furthermore,} 
Furthermore, a general high-dimensional setting, where in which the underlying process might have
an infinite amount of parameters, is considered. Note that all the time series results hold in a classical regression setting as well. 

However, we restrict ourself to $\ell_1$-penalised regressions as they are popular in time series settings 
(see e.g. \cite{wang2007regression}, \cite{nardi2011autoregressive} and \cite{yoon2013penalized}).
In general, other $\ell_q$-penalty could also be considered, e.g. the $\ell_2$ penalty.
The $\ell_2$ penalty, which gives the ridge regression, is suitable for shrinkage as well, but
does not allow for sparsity.
% But \textcolor{red}{TH: However,}
However in $\ell_q$-penalised regression, the case $q=1$ is the greatest %\textcolor{red}{TH: \st{largest}} 
case of %\textcolor{red}{TH: greatest} 
practical intereststill allowing %\textcolor{red}{TH: \st{that allows} still allowing} 
for sparsity. %, so some parameters are estimated
%to be exactly zero. 
This sparsity property can be used in applications to select the required tuning parameter based 
on information criteria that are popular in time series analysis. 
% \textcolor{red}{TH: Die ben\"otigten Tuningparameter k\"onnen mit sparsity und Informationskriterien bestimmt werden, die popul\"are Zeitreihenanalysen sind? Dieser Satz ist komisch und ich wei\ss{} 
% nicht wirklich, was du damit sagen willst.} 
% Obviously, extExtensionsRecently 
% \cite{gefang2014bayesian} applied them sucessfully to multivariate AR processes.

%\cite{gefang2014bayesian}. %TODO

% In the second section the general problem is stated. \textcolor{red}{TH: The general problem ist stated in section 2.} 
The general problem ist stated in section 2.
% The third section motivates and provides the estimation algorithm. \textcolor{red}{TH: In section 3, we motivate and provide the estimation algorithm.} 
In section 3, we motivate and provide the estimation algorithm.
Subsequently, the asymptotics are discussed in in section 4.% the next one \textcolor{red}{TH: in section 4}.
In Section 5, an application to multivariate AR-ARCH type processes is considered. 
This includes several extensions such as periodic AR-ARCH, AR-ARCH with structural breaks, threshold AR-ARCH and ARMA-GARCH models. 
The section 6 shows simulation which underline the results given above. It provides evidence 
that incorporating the heteroscedasticity in a high-dimensional setting is more important than in low dimensional problems.
Finally, we consider the proposed algorithm as a model for the electricity market and metal prices returns data.
% Finally, the proposed algorithm is considered to model electricity market and metal prices returns data. 
% \textcolor{red}{TH: Dieser Satz besagt ``Zuletzt bedenken wir ob der vorgeschlagene Algorithmus Strom und Metall Preis Daten modelliert.'' Ich denke du meinst: Finally, we consider the proposed algorithm as a model for the electricity market and metal prices returns data.} 
A two-dimensional AR-ARCH type model is used in both applications to the hourly data,
in the first one to electricity price and load data and in the second one to gold and silver price %prices \textcolor{red}{TH: price}
returns.

% to the German/Austrian day-ahead electricity spot price of the European Power Exchange (EPEX), 
% such as the electricity load for Germany and 
% The paper ends with some conclusions and final remarks.

% 

\section{The considered time series model}
The considered model is basically similar to the one used by \cite{yoon2013penalized} or \cite{medeiros2012estimating}.
Let $(Y_t)_{t\in \Z}$ be the considered causal univariate time series. 
We assume that 
it follows the
linear equation
\begin{equation}
 Y_{t} 
 = \bsX_{\infty,t} \bsbeta^0_{\infty}  + \eps_t
 \label{eq_main_model} ,
\end{equation}
where %$(\bsX_t)_{t\in \Z}$ with 
$\bsX_{\infty,t} = (X_{1,t}, X_{2,t}, \ldots)$ is a possibly infinite 
vector of covariates of weakly stationary processes $(X_{i,t})_{t\in \Z}$, $(\eps_t)_{t\in \Z}$ is an error process,
and the parameter vector is $\bsbeta^0_{\infty} = (\beta^0_1, \beta^0_2, \ldots)'$ with $\sum_{i=1}^\infty |\beta^0_i|< \infty$. 
% $\bsX_t = (X_{1,t}, \ldots, X_{p_n,t})$ is a vector of covariates of weakly stationary processes $(X_{1,t})_{t\in \Z},\ldots, 
% (X_{p_n,t})_{t\in \Z}$ and $\bsbeta$. 
The covariates can also contain lagged versions of $Y_t$, which allows flexible modelling of autoregressive processes.

A simple example of a process that helps for understanding this paper is an invertable seasonal MA(1) process.
In particular, the AR($\infty$) representation of a seasonal MA(1) with seasonality $2$ is useful.
It is given by $Y_t = \eps_t-\theta \eps_{t-2} = \theta Y_{t-2} + \theta^{2}Y_{t-4} + \theta^{3}Y_{t-6} + \ldots + \eps_t$,
choosing
$\bsX_{\infty,t} = (Y_{t-1}, Y_{t-2}, \ldots)$ with
$\bsbeta_\infty^0=(0, \theta, 0, \theta^2, 0, \theta^3, 0, \ldots)'$.
The error process $(\eps_t)_{t\in \Z}$ is assumed to follow a zero mean process
with $\epsilon_t$ being uncorrelated to the covariates $\bsX_{\infty,t}$. Hence we require
 $\E(\eps_t) = 0$ and $\cov(\eps_t, X_{i,t})=0$ for all $i\in \N$. 
 %$\cov(\eps_t, \eps_{t-k}) = 0$ for $k>0$ 
 Moreover, we assume that $\eps_t$ is a weak white noise process, such that
\begin{equation}
 \eps_t = \sigma_t Z_t \text{ where } \sigma_t = g( \bsalpha^0_\infty ; \bsL_{\infty,t}) \text{ and } 
(Z_{t})_{t\in \Z} \text{ is i.i.d. with } \E(Z_{t})=0 \text{ and } \var(Z_t)=1. 
 \label{eq_main_model_cond_var} 
\end{equation}
 Here, $g$ is a positive function,  
 $\bsL_{\infty,t} = (L_{1,t}, L_{2,t}, \ldots)$ is a possibly infinite 
vector of covariates of weakly stationary processes $(L_{i,t})_{t\in \Z}$, 
and $\bsalpha^0_\infty = (\alpha^0_1, \alpha^0_2, \ldots)'$ is a parameter vector.
% and $\bsL_t = (L_{1,t}, \ldots, L_{l,t})$. 
Similarly to the covariates $\bsX_{\infty,t}$ in \eqref{eq_main_model}, $\bsL_{\infty,t}$ 
can also include lags of $\sigma_t$ or $\eps_t$. This allows for a huge class of 
 popular conditional variance models, like ARCH or GARCH type models. Choosing
 \begin{equation*}
   g( \bsalpha^0_\infty ; \bsL_{\infty,t}) = g( (\alpha_0,\alpha_1, \ldots ) ; (\eps_{t-1}, \sigma_{t-1}, 0, \ldots))  = 
 \sqrt{ \alpha_0 + \alpha_1 \eps_{t-1}^2 + \alpha_2 \sigma_{t-1}^2 }
 \end{equation*}
 leads to the very popular GARCH(1,1) process. Note that the introduced setting is more general than the
 conditional heteroscedastic problem stated by \cite{yoon2013penalized}, who mentioned only ARCH errors.
 %TODO maybe say sthmth about simple heteroscedasticity
 
 For the following we assume that the time points $1$ to $n$
 are observable for $Y_t$. %Moreover we will assume that $p_n$ variables
%  and $\bsX_t$, 
 Thus, we denote by
\begin{align*}
 \bsY_n = \left(
   \begin{array}{c} Y_1 \\ \vdots \\ Y_n \end{array} \right),
  \bsX_n = \left(
   \begin{array}{ccc}
     X_{1,1} & \cdots & X_{1,p_n} \\
     \vdots  &\ddots & \vdots \\
     X_{n,1} &\cdots & X_{n,p_n}
   \end{array}
\right), \bsbeta_n =\left(
   \begin{array}{c} \beta_1 \\ \vdots \\ \beta_{p_n} \end{array} \right) \text{, and }
   \bseps_n = \bsY_n - \bsX_n \bsbeta_n 
\end{align*} %\left(
%   \begin{array}{c} \eps_1 \\ \vdots \\ \eps_n \end{array} \right)
the response vector $\bsY_n$, the $ n \times p_n $ matrix of the covariates $\bsX_n$, the parameter vector $\bsbeta_n$ and the corresponding errors
$\bseps_n$. 
Furthermore let $X_1, \ldots, X_{n}$ be the rows of $\bsX_n$. 
% $\bsbeta_n = (\beta_1, \ldots, \beta_{p_n})'$ the parameter vector, 
% and $\bseps_n = (\eps_{n,1}, \ldots,\eps_{n,n} )' = \bsY_n - \bsX_n \bsbeta_n$ the errors. %Additionally denote $X_1, \ldots X_{n}$ the $n$ columns of $\bsX_n$.
 
Since we deal with a high-dimensional setting
we are interested in situations where the number of possible parameters $p_n$ increases with sample size $n$.
Therefore, denote $\bsbeta^0_{n} = (\beta_1^0, \ldots, \beta_{p_n}^0)'$ the restriction of $\bsbeta_\infty^0$ to its first $p_n$ coordinates.
%  The vector \beta_infty^0 is therefore limited by beta=(...)' to its first p_n parameters.
Due to $\sum_{i=1}^\infty |\beta^0_i|< \infty$ it follows for 
$\bseps^0_n = (\eps_{n,1}^0, \ldots,\eps^0_{n,n} )' = \bsY_n - \bsX_n \bsbeta^0_n$ that 
there is a positive decreasing sequence $(\zeta_n)_n$ with $\zeta_n \to 0$ such that
$\lim_{n\to\infty} P( \max_{1\leq t\leq n}|\eps_{n,t}^0-\eps_t|< \zeta_n ) \to 1$ holds. 
Thus, for a sufficiently large $n$ we can approximate $Y_t$ by $\bsX_{n,t} \bsbeta_n^0$ 
arbitrarily well.

However, under the
assumption of sparsity, meaning that only some of the regressors attribute significantly to the model, 
we can conclude that only $q_n$ of the $p_n$ parameters are non-zero.
% 
% the sparsity assumption, we have that out of these $p_n$ parameters only $q_n$ 
% with $0 \leq q_n \leq p_n$ are non-zero.
% However, in a lasso-type framework we assume that out of these $p_n$ parameters only
% $q_n$ with $0 \leq q_n \leq p_n$ are non-zero. 
Hence, there are $p_n-q_n$ parameters that are exactly zero.
 Without loss of generality we assume that $\bsX_n$ and $\bsbeta^0_n$ are 
 arranged so that the first $q_n$ components of $\bsbeta^0_n$ are non-zero, whereas the following are zero. 
Obviously we have $\bsbeta^0_{n} = (\beta^0_1, \ldots, \beta^0_{q_n}, 0, \ldots, 0)' = ( \bsbeta^0_n(1)' , \bsnull')'$.
This arrangement of the non-zero components is only used to simplify the notation, it is especially not required
by the estimation procedure.
Additionally we introduce the naive partitioning of $\bsX_n$ and $\bsbeta_n$, in such a manner that
$\bsbeta_{n} = ( \bsbeta_n(1)' , \bsbeta_n(2)' )'$, $\bsX_n = (\bsX_n(1), \bsX_n(2))$ 
and $\bsX_{n,t} = (\bsX_{n,t}(1)', \bsX_{n,t}(2)')' $ holds.

Subsequently, we focus on the estimation of $\bsbeta_n^0$, for which we will utilize a lasso-based approach for $\bsbeta_n$.
Henceforth, we achieve  never a
direct estimate for $\bsbeta^0_\infty$, but we can
approximate it by $(\bsbeta_{n}', \bsnull')'$.

\section{Estimation algorithm}

The proposed algorithm is based on the classical iteratively reweighted least squares procedure.
An example for an application of it to time series analysis can be found e.g. in \cite{mak1997estimation}. 
However, similar approaches are not popular in time series modelling, as there are usually better alternatives
if the number of parameters is small. 
In that case, we can simply perform an estimation of the joint likelihood function of \eqref{eq_main_model}, see e.g. \cite{bardet2009asymptotic}.
But when facing a high-dimensional problem, it is almost impossible to maximise the non-linear loss function 
with many parameters. In contrast, our algorithm can be based on the coordinate descent lasso estimation technique 
as suggested by \cite{friedman2007pathwise} which provides a feasible and fast estimation technique. Other techniques, 
like the LARS algorithm introduced by \cite{efron2004least} which provides the full lasso solution path, can be used as well.

%And even if we eventually find a local optimum by some optimisation method.
%\cite{mak1997estimation} ling2007self -selfweighted 

For motivating the proposed algorithm, we divide equation \eqref{eq_main_model} by its volatility, resp.
conditional standard deviation $\sigma_t$.
Thus, we obtain %\textcolor{red}{TH: obtain} 
 \begin{equation}
 \wtilde{Y}_{t} 
 =  \wtilde{\bsX}_{\infty,t} \bsbeta_\infty+ Z_t
 \label{eq_stand_main_model} ,
\end{equation}
where $ \wtilde{Y}_{t} =  \frac{1}{\sigma_t} {Y}_{t} $ and $ \wtilde{\bsX}_{\infty,t} =  \frac{1}{\sigma_t} {\bsX}_{\infty,t}$. Here, the 
noise $Z_t$ is homoscedastic with variance $1$.
% \textcolor{red}{TH: \st{So} Hence,} 
Hence, if the volatility $\sigma_t$ of the process $Y_t$ is known, we can simply apply common lasso time series techniques
under homoscedasticity. Unfortunately, this is never the case in practice. 
The basic idea is now to replace $\sigma_t$ by a suitable estimator $\what{\sigma}_t$, which allows 
us to  perform a lasso estimate on a homoscedastic time series as in equation \eqref{eq_stand_main_model}.
% However the used estimator $\what{\sigma}_t$ will have a big impact on the chosen model.
% Furthermore it will be hard to provide a good estimator for the volatility in a high dimensional setting
% of equation \eqref{eq_main_model}. 
% without knowing $\eps_t$, as many volatility models depend 
% However if the volatility model is a ARCH resp. GARCH type model that depends only on 

For estimating ARMA-GARCH processes, practitioners sometimes use a multi-step estimator. 
This estimation technique involves computing ARMA parameters in a homoscedastic setting first and then use the resulting estimated
% First, the ARMA parameters are computed in an homoscedastic setting,
% then the resulting estimated 
residuals are used to estimate the GARCH part in a second step, see e.g. 
\cite{mak1997estimation} or \cite{ling2007self}. We will apply a similar step-wise estimation technique here. 

In general, we have no a priori information about $\sigma_t$, hence we should
assume homoscedasticity in a first estimation step. % based on $n$ given observations. 
We start with the estimation of the regression parameters 
$\bsbeta_n^0$, resp. $\bsbeta^0_\infty$, and obtain the residuals $\what{\eps}_{n,1}, \ldots, \what{\eps}_{n,n}$. 
We use the residuals to estimate the conditional variance parameters $\bsalpha^0_\infty$ and thus 
$(\sigma_1, \ldots, \sigma_n)$ by $(\what{\sigma}_{n,1}, \ldots, \what{\sigma}_{n,n})$ afterwards.
Afterwards, we reweight model \eqref{eq_main_model} by $\what{\sigma}_t^{-1}$ to get a homoscedastic 
model version which we utilise in order to reestimate $\bsbeta_n^0$ again.
% we utiliize in order to reestimate \beta_n^0.
We can use this new estimate of 
$\bsbeta^0_n$ to repeat this procedure. Thus, we will end up in an iterative algorithm
that hopefully converges in some sense to $\bsbeta_n^0$, resp. $\bsbeta^0_\infty$, with increasing sample size $n$.

%As we mentioned, w
We use an adaptive weighted lasso estimator
to estimate $\bsbeta^0_n$ within each iteration step.
%The adaptive weighted lasso estimator 
It is given by
$$\bsbeta_{n,\text{lasso}}(\lambda_n, \bsv_n, \bsw_n) = \argmin_{\bsbeta} 
\sum_{t= 1}^n  w_{n,t}^2\bigg(Y_{t} - \sum_{i=1}^{p_n} X_{t,i} \beta_i \bigg)^2 
+ \lambda_n  \sum_{j=1}^{p_n} v_{n,j} |\beta_j|
$$
or in vector notation 
$$ \bsbeta_{n,\text{lasso}}(\lambda_n, \bsv_n, \bsw_n) = \argmin_{\bsbeta} 
%\sum_{t= 1}^n  w_{n,t}^2\left(Y_{t} - \bsX_{n,t} \bsbeta \right)^2 
(\bsY_n - \bsX_n \bsbeta )' \bsW_n^2 (\bsY_n - \bsX_n \bsbeta )
+ \lambda_n  \bsv_{n}' |\bsbeta|, 
$$
where $\bsW_n = \diag(\bsw_n) $, $\bsw_n = (w_{n,1},\ldots, w_{n,n})$ are the heteroscedasticity weights, $\bsv_n = (v_{n,1},\ldots, v_{n,p_n})$ 
are the penalty weights and $\lambda_n$ is a penalty tuning parameter.
As described above, in the iteratively reweighted adaptive lasso algorithm we  
 have the special choice $\bsw_n = (w_{n,1},\ldots, w_{n,n}) = (\what{\sigma}_{n,1}^{-1},\ldots, \what{\sigma}_{n,n}^{-1} )$ 
for the heteroscedasticity weights within each iteration step.  
We require $\bsw_n = \bsone$ for the homoscedatic initial step.

Like \cite{zou2006adaptive} we consider, for the tuning parameter $\bsv_n$, 
the choice $\bsv_n = {\bsbeta}_{n,\text{init}}^{-\tau}$ for some $\tau \geq 0$
and some initial parameter estimate $\bsbeta_{n,\text{init}}$.
With $\tau = 0$ we obtain $\bsv_n = \bsone$ which is the usual lasso estimator. Obviously, there is no initial estimator
required in this case. 
However, we consider the case of $\tau=0$ and the adaptive lasso approach for our practical application, as they resulted in different perfomances. 
% However, it is worth considering the adaptive lasso as well,
% as it showed a different performance in application.

The selection of the tuning parameters $\lambda_n$ and $\tau$ such as the choice of the 
initial estimate ${\bsbeta}_{n,\text{init}}$ is crucial for the application and might demand some computational cost.
We discuss this issue in more detail at the end of the next section.

Subsequently, we denote 
$\what{\bsalpha}_n = \what{\bsalpha}_n( \bsbeta_n; \bsX_n, \bsY_n )$ as a known plug-in estimator for $\bsalpha_n^0$, which is 
the projection of $\bsalpha_\infty^0$ to its first $l_n$ coordinates. 
We denote $g_n$ as restriction of $g$ that corresponds to $\bsalpha_n^0$.
Thus, $g_n$ is defined such that $\bsalpha_\infty^0$ is restricted to $\bsalpha_n^0$
and $\bsL^0_n$ is a restriction of $\bsL_\infty = (\bsL_{\infty,t})_{t\in \Z}$ to its first $m_n(l_n)$ coordinates.
% \textcolor{red}{TH: Den Satz verstehe ich nicht. Und ganz wichtig ``so that'' gibt es nicht im Englischen}. 
Similarly, let $\what{\bsL}_n = \what{\bsL}_n( \bsbeta_n; \bsX_n, \bsY_n )$ be an estimator for 
$(\bsL^0_{n,1}, \ldots, \bsL^0_{n,n})'$. 
%We will further assume that $l_n$ resp. $m_n(l_n)$ 

For example, if $\eps_t$ follows a GARCH(1,1) process 
we receive $\sigma_t = g_n(\bsalpha_n^0, \bsL_{n,t}^0 )$ for all $n\in \N$, where
$\bsalpha_n^0 = (\alpha_0, \alpha_1, \alpha_2) $ with $l_n = 3$
and $\bsL_{n,t}^0 = (\eps_{t-1}, \sigma_{t-1})$ with $m_n(l_n)=2$ for all $n\in\N$.
%so we directly have . 
This is similarly feasible for every variance model with a finite amount of parameters.
However, if $\sigma_t$ follows an infinite parameterised 
process, e.g. through an ARCH($\infty$) process, $l_n$ and $m_n(l_n)$ should tend to infinity as $n\to \infty$.
%So function $g_n(\a,)$ grows 

%  the restriction of $\bsL_{\infty}$
%  to its first $m_n(l_n)$ coordinates.
% 
%  where $\bsL_{n} 
% = (\bsL_{n,1}, \ldots, \bsL_{n,n})$. 
%  and

 The estimation scheme of the described iteratively reweighted adaptive lasso algorithm is given by: 

 \framebox{
\parbox{.93\textwidth}{
\begin{enumerate}
   \item initialise $\lambda_n\geq 0$, %$\bstheta^{[0]} = (\bsbeta^{[0]}, \bsalpha^{[0]})$  
   $\bsv_n(\tau) = (v_{n,1}(\tau),\ldots, v_{n,p_n}(\tau)) = \bsbeta_{n,\text{init}}^{-\tau}$ with $\tau \geq 0$ and 
   $\bsw_n^{[0]} = \bsone$, $k=1$
%     \item estimate $\phi^{[init]}$ by OLS and define $\bsv = (v_{1},\ldots, v_{m})$ with $v_j= \frac{1}{\beta_{j}^{
%     [init]}}^\tau$ [CITE lassots]
%    , $k=0$ so $\theta^{[0]}$ admits a homoscedastic solution.
   \item estimate by weighted lasso:
% \begin{equation*}
$\bsbeta_n^{[k]} = \bsbeta_n^{[k]}(\bsw_n^{[k-1]}) =  
   %\text{IC}^{-1}\left( \min_{\lambda, \tau} \text{IC}\left(
   \bsbeta_{n,\text{lasso}}(\lambda_n, \bsv_n(\tau), \bsw_n^{[k-1]}) 
   $%\right) \right)  
% \end{equation*}
   \item estimate the conditional variance model:
%   \begin{equation*}
$ \bsalpha_n^{[k]} = \what{\bsalpha}_n( \bsbeta_n^{[k]}; \bsX_n, \bsY_n  )$ and $\bsL_n^{[k]} 
 = \what{\bsL}_n( \bsbeta_n^{[k]}; \bsX_n, \bsY_n )$
 %   \end{equation*}
    
   \item compute new weights
   $\bsw^{[k]}_n = (\bsw_{n,1}^{[k]},\ldots,\bsw_{n,n}^{[k]}) $   with
   $\bsw_{n,t}^{[k]} = g_n(\bsalpha_n^{[k]}, \bsL_{n,t}^{[k]})^{-1} $
  \item if the stopping criterion is not met, $k=k+1$ and back to 2. otherwise, return 
 estimate $\bsbeta_n^{[k]}$  and volatilities $\what{\bssigma}_{n,t}^{[k]} 
 = g_n(\bsalpha_n^{[k]}, \bsL^{[k]}_{n,t})$ %for $t\in \{1,\ldots, n\}$ 
   \end{enumerate}
\vspace{-3mm} 
   }
   } 
 
\vspace{2mm} 

We can summarise that we have to specify the tuning parameter $\lambda_n$, the initial estimator $\bsbeta_{n,\text{init}}$ with an inital value of $\tau$,
and the initial heteroscedasticity weights $\bsw_n$.
% and
% the information criteria selection method represented by $\text{IC}$.
To reduce the computation time it can be convenient in practice to choose $\tau=0$ (lasso) 
or $\tau =1$ (almost non-negative garotte).
% without
% searching for the optimal $\tau$ value within the adaptive lasso optimisation. 

%Note that this $\text{IC}$ might be a total different approach 

The stopping criterion in step 5 has to be chosen as well, such that the algorithm eventually stops. 
A plausible stopping criterion should measure the convergence of $\bsw_n^{[k]}$, resp. $\bssigma_n^{[k]}$. 
We suggest  to stop the algorithm if $\|\bssigma_n^{[k]} - \bssigma_n^{[k-1]}\|< \epsilon$ for a selected vector norm $\|\cdot\|$ and some 
small $\epsilon > 0$.
Nevertheless, in our simulation study, we realised that the difference in the later steps are marginal,
so that stopping at  $k=2$ or $k=3$ seems to be reasonable for practice. 
This will be underlined by the asymptotics of the algorithm as analysed below; it
can be shown that, under certain conditions, $k=2$ is sufficient to get an optimal estimator if $n$ is large.

\section{Asymptotics of the algorithm}

For the general convergence analysis it is clear that the asymptotic of the estimator $\bsbeta_{n}^{[k]}$ will 
strongly depend on the (cond.) heteroscedasticity
models \eqref{eq_main_model_cond_var} (esp. the formula for $g$)
such as on the linked estimators $\what{\bsalpha}_n$ and $\what{\bsL}_n$.
Despite that strong dependence we are able to prove sign consistency as introduced by \cite{zhao2006model} and asymptotic 
normality of the non-vanishing components of $\bsbeta_{n}^{[k]}$ in a time series framework.
% Though, we consider the sign consistency as introduced by \cite{zhao2006model} and asymptotic normality of 
% the non-vanishing components of $\bsbeta_{n}^{[k]}$. 

If we assume that the number of parameters $p_n$ does not depend on the sample size $n$, 
then we could make use of the results from \cite{wagener2012bridge}
to obtain asymptotic properties, 
as they prove sign consistency and asymptotic normality under some conditions for the weighted adaptive lasso estimator.

% for getting asymptotic properties.
% They prove sign consistency and asymptotic normality for the weighted lasso and weighted adaptive lasso 
% estimator under some conditions. 
% However they noted that as in classical lasso theory, the weighted lasso does not 
% enjoy the optimal rate of convergence. 
% In contrast they showed that the adaptive lasso can
% perform optimal in terms of convergence rate for parameter selection and estimation.
% However, in praxis it might be worth considering the weighted lasso instead of the weighted adaptive lasso
% as it is computationally easier to deal with. 

The case where the number of parameters $p_n$ increases with $n$ is analysed euivalently in a regression framework by
\cite{wagener2013adaptive}, but only for the adaptive lasso case with $\tau=1$.
They basically achieve the same asymptotic behaviour as for the fixed $p_n$ case, 
but it is clear that the conditions are more complicated 
compared to those of \cite{wagener2012bridge}.

In the following we will introduce several assumptions, which allow us to generalise the results of \cite{wagener2013adaptive}.%Wagener and Dette
% The following assumptions will generalise the results of \cite{wagener2013adaptive}. 
One crucial point is the assumption that the process $Y_t$ can be 
parameterised by infinitely many parameters, 
%As discussed above, we estimate $\bsbeta^0_n$ by $\bsbeta_n$. 
so that the error term
$\bseps^0_n = \bsY_n - \bsX_n \bsbeta^0_n$, based on the restriction $\bsbeta^0_n$ of the true parameter vector $\bsbeta_{\infty}^0$,
is not identical to the true error restriction $\bseps_{\infty, n}^0$. 
In contrast to $\bseps_{\infty, n}^0$, the term $\bseps^0_n$ is in general correlated. This has to be taken into account 
for the proof concerning the asymptotic behaviour.

For the asymptotic properties we introduce a few more notations.
Let $\wtilde{\bsX}^{[k]}_n = {\bsW_n^{[k-1]}}\bsX_n  $ and $\wtilde{\bsY}_n^{[k]} = {\bsW^{[k-1]}_n} \bsY_n  $,
where $\bsW_n^{[k]} = \diag(\bsw_n^{[k]})$. Let $\bsSigma_n^0$ denote the true volatility matrix and 
$\bsSigma_n^{[k]}= {\bsW_n^{[k]}}^{-1}$ its estimate in the $k$-th iteration.
Additionally, we introduce $\wtilde{\bsGamma}_n^{[k]}  = \frac{1}{n} {(\wtilde{\bsX}_n^{[k]}) }' \wtilde{ \bsX}_n^{[k]}$ as the scaled Gramian,
where $\bsGamma_n  = \wtilde{\bsGamma}_n^{[1]} = \frac{1}{n} \bsX_n' \bsX_n$ is the unscaled Gramian. 
Furthermore, let $\bsW^0_n $ and $\wtilde{\bsGamma}^0_n $ denote the weight matrix and the
Gramian that correspond to 
the true matrix $\bsSigma_n^0$. 
% scaled Gramian 
The submatrices to $\bsbeta_n^0(1)$ are denoted by $\wtilde{\bsGamma}_{n}^{[k]}(1)$, $\bsGamma_{n}(1)$, and
$\wtilde{\bsGamma}_{n}^0(1)$.
% $\wtilde{\bsGamma}_{p,q}^{[k]}$ and $\wtilde{\bsGamma}_{q,q}^{[k]}$ 
% the submatrices that corresponds to the $\bsbeta_n^0$.

Similarly to \cite{wagener2013adaptive}, we require the following additional assumptions, which we extended to carry out our proof:
% Similarly to \cite{wagener2013adaptive}, we require the following additional assumptions:
\begin{itemize}
 \item[\namedlabel{asump_stationarity}{(a)}] The process $(Y_t, Z_t, X_{1,t}, \ldots, X_{m,t}, \sigma_t)_{t\in \Z}$ %L_{1,t},\ldots, L_{m,t})_{t\in \Z}$ 
 is weakly stationary with zero mean
 for all $m\in \N$.
% \item[(b)] $\E(\eps_{t} ) = 0$ for all $t\in $ %\{1,\ldots, n\}$.
\item[\namedlabel{asump_normalise}{(b)}] 
The covariates are standardised so that 
$\E ( X_{i,t}^2 ) = 1$ for all $t\in \Z$ and $i\in \N$.

\item[\namedlabel{asump_Xnt_vartheta}{(c)}] 
For the sequence of covariates $(\bsX_{n,t})_{n\in\N}$ of a fixed $t$ there is a
positive sequence $(\vartheta_n)_{n\in \N}$ such that
$$\max_{1\leq t \leq n} \| \bsX_{n,t}(1)\|_2 = \OO_P(\vartheta_n\sqrt{q_n}). $$

\item[\namedlabel{asump_bn}{(d)}] 
For the minimum of the absolute non-zero parameters $b_n = \min \{|\bsbeta^0_n(1)|\} $
%\footnote{The absolute value $|\cdot|$ is considered element-wise.}  
and the initial estimator $\bsbeta_{n,\text{init}}$ there exists a constant $b>0$ so that 
\begin{equation*}
\lim_{n\to\infty} P \left( b \min\{ |\bsbeta_{n,\text{init}}(1)|^{\tau} \} < b_n \right) = 0 .
\end{equation*}

\item[\namedlabel{asump_rn}{(e)}] 
There exists a positive sequence $(r_n)_{n\in \N}$ with $r_n \to \infty$ such that
\begin{equation*}
\lim_{n\to\infty} P( \max \{ |\bsbeta_{n,\text{init}}(2)|^{\tau}\} < r_n^{-1} ) = 0. 
\end{equation*}

\item[\namedlabel{asump_eigenval}{(f)}] 
There are constants $0<\lambda_{0,\min}<\lambda_{0,\max}$ and $0< \lambda_{1,\min}$ such that the eigenvalues satisfy
\begin{equation*}
P( \lambda_{0,\min} < \lambda_{\min}(\bsGamma_n(1)) \leq \lambda_{\max}(\bsGamma_n(1)) < \lambda_{0,\max} ) \to 1 ,
\end{equation*}
\begin{equation*}
P\left( \lambda_{1,\min}  < \lambda_{\min}(\wtilde{\bsGamma}^0_n(1)) \leq \lambda_{\max}(\wtilde{\bsGamma}^0_n(1)) \right) \to 1  ,
\end{equation*}
for $n\to \infty$.

\item[\namedlabel{asump_variance_bounds}{(g)}] 
There is a positive constant $\sigma_{\min}$ %and $\sigma_{\max}$ 
such that 
\begin{equation*}
0<\sigma_{\min} 
< g_n(\what{\bsalpha}_n(\bsbeta_n, \bsX_n, \bsY_n) , \what{\bsL}_{n,t}(\bsbeta_n; \bsX_n, \bsY_n)) %< 0
\end{equation*}
for all $n>N$ with $N\in \N$, $t\in \{1, \ldots, n\}$ and $\bsbeta_n$ in an open neighbourhood of $\bsbeta^0_n$.
% [[TODO think about $g_n$ and $g$  and sigmaMAX]]

\item[\namedlabel{asump_variance_mom}{(h)}]
% The functions $\bsbeta_n \mapsto g_n( \what{\bsalpha}_n(\bsbeta_n, \bsX_n, \bsY_n) 
% , \what{\bsL}_{n,t}(\bsbeta_n; \bsX_n, \bsY_n) )$ are twice 
% differentiable in a neighbourhood of $\bsbeta^0_n$ all $n\in \N$ and $t\in \{1, \ldots, n\}$.
% The corresponding partial derivatives are uniformly bounded on its first $q_n$ coordinates. 
The volatilities have afinite fourth moment, so $\E(\sigma^4_t) = \E(g(\bsalpha_{\infty}^0, \bsL_{\infty,t})^4) < \infty$ for all $t$.

\item[\namedlabel{asump_variance_estimators}{(i)}] For all $n\in \N$ 
the estimator $\what{\bsalpha}_n$ and $\what{\bsL}_n$ are consistent for $\bsalpha^0_n$
and $\bsL^0_{n,1}, \ldots, \bsL^0_{n,n}$,
additionally 
$$ |g(\bsalpha_{\infty}^0, \bsL_{\infty,t})^{-2} - g_n(\what{\bsalpha}_n(\bsbeta^0_n; \bsX_n, \bsY_n) ,
\what{\bsL}_{n,t}(\bsbeta^0_n; \bsX_n, \bsY_n))^{-2} | = \OO_P(\frac{h_n}{\sqrt{n}})$$
for some $(h_n)_{n\in\N}$ with  $h_n n^{-\frac{1}{2}} \to 0$ as $n\to \infty$.

\item[\namedlabel{asump_convergence}{(j)}] 
It holds for $\lambda_n$, $\vartheta_n$, $p_n$, $q_n$, $b_n$, $r_n$, and $h_n$ that

\begin{multicols}{2}
\begin{enumerate}[label=(\roman*), ref=(\roman*)]
 \item $\frac{\log(n)^{\bsone\{d=1\}} \log(q_n)^{\frac{1}{d}}}{\sqrt{n}b_n} \to 0$  %[[ OLD: $\frac{\lambda_n \sqrt{q_n}}{\sqrt{n b_n}} \to 0$ as $n\to \infty$ ]]
 \label{asump_convergence_item1}
 \item $\frac{h_n}{\sqrt{n} b_n} \to 0$ %$\frac{ \sqrt{n} |\log(n)|^{\ind\{d=1\}}}{\lambda_n r_n |\log(p_n-q_n)|^{d}} \to 0$ as $n\to \infty$
\item $ \frac{\lambda_n \sqrt{q_n} } {n b_n^{1.5}}  \to 0$
 \item $\frac{ \sqrt{n} \log(n)^{\bsone\{d=1\}} \log(p_n - q_n)^{\frac{1}{d}}  }{\lambda_n r_n } \to 0$ %as $n\to \infty$
 \label{asump_convergence_item4}
 \item $ \frac{h_n \sqrt{n}}{\lambda_n r_n} \to 0$ %as $n\to \infty$
 
\item $\frac{\lambda_n \sqrt{q_n}}{\sqrt{b_n}} \to 0$
 \item $ \frac{ \vartheta_n \sqrt{q_n} }{\sqrt{n}} \to 0$ %as $n\to \infty$
 \item $ \frac{ h_n \sqrt{q_n} }{\sqrt{n}} \to 0$ %as $n\to \infty$
%   \item[(vi+vii)] $ (\vartheta_n+h_n)\frac{ \sqrt{q_n} }{\sqrt{n}} \to 0$ %as $n\to \infty$
 \end{enumerate}
\end{multicols}
as $n\to \infty$.

\item[\namedlabel{asump_tail}{(k)}] 
There are positive constants $C_1$, $C_2$ and $d$ with $1\leq d \leq 2$ such that 
\begin{equation*}
P(|\eps_t|>x)\leq C_1 \exp({-C_2 x^d}).
\end{equation*}

\item[\namedlabel{asump_tailalt}{(k')}] 
It holds for  $\lambda_n$, $p_n$, $q_n$ and $r_n$ that 
$\frac{\sqrt{n}\sqrt{p_n - q_n}}{\lambda_n r_n} \to 0$
% \begin{multicols}{2}
% \begin{itemize}
% %  \item[(i)] $\frac{1}{\sqrt{n} b_n} \to 0$
%  \item[(ii)] $\frac{\sqrt{n}\sqrt{p_n - q_n}}{\lambda_n r_n} \to 0$
%  \end{itemize}
% \end{multicols}
as $n\to \infty$.

% \item[\namedlabel{asump_l}{(l)}] $$ \frac{1}{n} \max_{1\leq i \leq n }   \sum_{j=1}^{q_n} X_{i,j}^2  \to 0$$
%TODO CHECK
%  \item[\namedlabel{asump_correlation}{(l)}]
%  For the correlations $\rho(k) = \cor( \eps_{t}, \eps_{t-k} )$ we have $\rho(k) \to 0$ as $k\to \infty$.
%  Additionally it holds either $\E|Z_t|^{2+\varrho}< \infty$ for some $\varrho>0$
%  or $\sum_{k=1}^\infty \rho(2^k) < \infty$.
\end{itemize}

% As mentioned above, most of these assumptions are adjusted from the setting of \cite{wagener2013adaptive}.
Assumption \ref{asump_stationarity} is standard in a time series setting. 
\ref{asump_normalise} is the scaling that is required in a lasso framework.
\ref{asump_bn} and \ref{asump_rn} are usual assumptions in an adaptive lasso setting (see e.g. \cite{zou2006adaptive} or 
\cite{huang2008adaptive}). 
\ref{asump_eigenval} gives bounds for the weighted and unweighted Gramian.
\ref{asump_variance_bounds}, \ref{asump_variance_mom} and \ref{asump_variance_estimators} 
postulate properties required for the heteroscedasticity in the 
model. 
\ref{asump_convergence} states some convergence properties that make
restrictions to the grow behaviour within the model, especially the number of parameters $p_n$ and the number of relevant parameters $q_n$. 
\ref{asump_tail} makes a statement about the tails of the errors.

% \ref{asump_correlation} is required to get the $\rho$-mixing central limit theorem to work,
% it is a very weak assumption.
% However
% \cite{wagener2013adaptive} showed that an homoscedastic version can satisfy it.

Using the assumptions above we can prove sign consistency and asymptotic normality. %The first one is concerning the sign consistency of the corresponding estimator.
\begin{theorem} \label{thm_asymtotic}
Under conditions \ref{asump_stationarity} to \ref{asump_convergence}, where either \ref{asump_tail} or \ref{asump_tailalt} holds, 
it holds for all $k\geq 1$ that 
\begin{equation*}
\lim_{n\to \infty} P \left( \sign(\bsbeta_{n}^{[k]}) = \sign(\bsbeta_{0}) \right)  = 1.
\end{equation*}
% \end{theorem}
% 
% \begin{theorem}
Moreover it holds for $\xi_n\in \R^{q_n}$ with $\|\xi_n\|_2=1$ that 
%Under the conditions above for $\bsgamma\in \R^p$ with $\bsgamma^\top \bsgamma =1$ we have
\begin{equation*}
\sqrt{n}  s_n(k)^{-1} \xi_n' \left( {\bsbeta_{n}^{[k]}(1)} - \bsbeta_{n}^0(1) \right)  \to 
N(0, 1)
\end{equation*}
%\sqrt{n} \bsa^\top \left(\bsbeta_{OLS}^{(p)} - \bsbeta_0^{(p)} \right) + o_q(1) $$
 in distribution, 
 where $s^2_n(1) = \xi_n'  ( \bsGamma_n(1) )^{-1} \xi_n $
and $s^2_n(k) = \xi_n'  ( \wtilde{\bsGamma}_n^0(1) )^{-1} \xi_n $ for $k\geq 2$.%\bsgamma$.
\end{theorem}
The proof is given in the appendix. Note that the variance $s^2_n(k)$ for $k\geq 2$ is substantially smaller than $s^2_n(1)$.
Hence the estimator $\bsbeta_{n}^{[k]}$ has minimal asymptotic variance for all $k\geq 2$.

Due to the general formulation of the theorem assumption  \ref{asump_convergence} contains several assumptions on problem characterizing
sequences. The convergence rate $h_n n^{-\frac{1}{2}}$ of the volatility model is relevant as well. If
we have that $h_n = \OO_P(1)$ (e.g. the variance model is asymptotic normal) 
then the three conditions involving $h_n$ are automatically satisfied by the other conditions. This reduces %the simplifies
the relevant conditions in \ref{asump_convergence} a lot.

There is one condition in assumption \ref{asump_convergence} involving $\vartheta_n$ that is given through assumption \ref{asump_Xnt_vartheta}.
As it holds that
$\max_{1\leq t \leq n} \| \bsX_{n,t}(1)\|_2 = \OO_P(\vartheta_n\sqrt{q_n})$
it characterises the structure of regressors. Obviously it holds that $\vartheta_n = \OO_P(1)$ if $\bsbeta^0_{\infty}$ contains only 
a finite amount of non-zero parameters, so $q_n\to c$ for some $c\in \N$ as $n\to \infty$.
However, there are many other situations where $\vartheta_n = \OO_P(1)$ holds. For example, if
we have that $\bsX_{\infty,t}(1)$ is stationary. In the example above, where $Y_t$ follows
a seasonal moving average process the process, $\bsX_{\infty,t}(1) = (Y_{t-2}, Y_{t-4}, Y_{t-6}, \ldots)$ is stationary.

% [[TODO req. $\sqrt{q_n}\vartheta_n /\sqrt{n} \to 0$  and 
% $\max_{1\leq t \leq n}\| \bsX_{n,t}(1) \|_2 = \OO(\sqrt{q_n} \vartheta_n)$  +  discussion e.g. const. no problem. 
% if stationary (or periodic stat. no problem. e.g SMA2, if different something dependend on tail kicks in]] 

Furthermore, there is the option of \ref{asump_tail} or \ref{asump_tailalt} in the theorem.
\ref{asump_tail} restricts the residuals to have an exponential decay in the tail, like the normal or the Laplace distribution.  
However, this can be replaced by the stronger condition \ref{asump_tailalt} in the theorem.
In this situation, polynomially decaying tails in the residuals are possible. Here 
a specification of the constant $d$ in \ref{asump_convergence} is not required, as \ref{asump_tailalt} implies directly 
\ref{asump_convergence} \ref{asump_convergence_item4}, 
which means that \ref{asump_convergence} \ref{asump_convergence_item1} is not used in this case. 
More details are given in the proof.

%, as long as they have finite variance.
As discussed in \cite{wagener2013adaptive} the assumption \ref{asump_tail} or \ref{asump_tailalt} has an impact on the maximal possible growth
of the amount of parameters $p_n$ in the estimation. 
There are situations where under assumption \ref{asump_tail} $p_n$ can grow with every polynomial order, even slow
exponential growth is possible. 
In contrast, given assumption \ref{asump_tailalt} this is impossible. 
Here \cite{wagener2013adaptive} argued that
sign consistency is possible for rates that increase 
slightly faster than linearly, such as $p_n \sim n\log(n)$, but not for polynomial rates like $p_n \sim n^{1+\delta}$
for some $\delta>0$. \cite{wagener2013adaptive} do not discuss this case 
for the asymptotic normality.
%TODO from here
In this situation, we can get an optimal rate of $n^{1- \delta}$ for the 
number of relevant parameters $q_n$ (having $b_n \sim 1$, $r_n\sim n^{\frac{1}{2}}$, $h_n\sim 1$ and $\vartheta_n \sim 1$),
when we have a polynomial growth for $p_n$. %But even slow exponential growth in $p_n$ is possible.

% The maximal possible growth in $p_n$ is slow exponential growth. However, this comes at the 
% cost that $q_n $ has rate $ n^{2\delta}$, so the amount of relevant parameters in the model is almost fixed for small $\delta$.

The quite general formulation in \ref{asump_variance_estimators} 
can be replaced by a more precise assumption when a 
variance model is specified. For example, if we have a finite dimensional conditional variance model
where $\bsalpha_n$ is asymptotic normal, i.e. converges with rate of $n^{-\frac{1}{2}}$, 
and  $\bsbeta_n \mapsto g_n( \what{\bsalpha}_n(\bsbeta_n, \bsX_n, \bsY_n),
\what{\bsL}_{n,t}(\bsbeta_n; \bsX_n, \bsY_n) )$ is twice continously differentiable with uniformly
bounded derivatives, then \ref{asump_variance_estimators} can be satisfied by $h_n  = c$ under some regularity
conditions on $\bsalpha_n$ and $\bsL_n$
or its estimated counterparts $\what{\bsalpha}_n$ and $\what{\bsL}_n$.
If in contrast $l_n$ is increasing we will usually tend to get worse rates for $h_n$. 

% Thus, if the number of parameters in the
% volatility model increases, the rate must be slower than linearly to satisfy the assumptions of the theorem

% TODO EXPLAIN HOW. 
%For practical purpose a linear should be sufficient for most of the applications in an high-dimensional setting

% In assumption \ref{asump_convergence}, the last convergence $\frac{q_n^{3}}{n} \to 0$ is relaxed
% in comparison to the analogous part in \cite{wagener2013adaptive}. They
% required $\frac{q_n^{5}}{n} \to 0$ to achieve asymptotic normality, in their settings the maximal rate 
% for the number of relevant parameters $q_n$ in the model is consequently only $n^{1/5 - \delta}$.

% So we
% allow the that the amount of relevant parameters can grow with a rate of $n^{1/3}$ instead of $n^{1/5}$.

In empirical applications, practitioners often just want to apply a lasso type algorithm without
caring much about the chosen size of $n$ and $p_n$. They tend to stick all available 
$n$ and $p_n$ into their model as long as it is computational feasible. 
However, usually it is feasible to validate the convergence assumptions in \ref{asump_convergence} at least partially.
Therefore, we have to estimate the model for several sample sizes $n$ and a specified growth rate 
for %the selected number of possible covariates 
$p_n$ and $\lambda_n$. 
As we can observe the estimated values for $q_n$ of the model 
we can get clear indications for the asymptotic convergence properties. This also helps
to find the optimal tuning parameter $\lambda_n$. 
The tail assumption \ref{asump_tail} can be checked using log-density plots and related tests. 
The moment restriction \ref{asump_variance_mom} to the volatilities can be validated using tail-index estimation techniques, like the Hill estimator.

%Unfortunately a comprehensive study might be time consuming.

%, as well as the best information criterion.

% 
% The assumptions are essential. (a) to (h) are conditions regarding the 
% as used by \cite{medeiros2012estimating}, including minor adjustments. The assumptions (i)-(k) make some statements about
% the volatlity structure of the model. Here \ref{asump_sigmaeps} is very important, as it controls the
% quality of the approximation of $\bssigma$ by $\bssigma^{[k]}$. In praxis this assumption always has to be checked
% given a used conditional variance model and estimator. However as $\bsbeta^{[k]}$ is consistent 
% it should not be to difficult to provide an estimator that can estimate $\bssigma$ consistently.
% However it might be more difficult to proof the required rate of convergence.
% Furthermore note that for the sign consistency itself condition 
% \ref{asump_sigmaeps} can be relaxed such that we require only $\sigma_{\epsilon}^{[k],*} = O( 1 )$.
 
Note that in the algorithm $\lambda_n$ is assumed to be the same in every iteration. It is clear that if we have two different sequences 
$(\lambda_n)_{n\in\N}$ and $(\wtilde{\lambda}_n)_{n\in\N}$ that satisfy the assumptions of the theorem, we can use them both in
the algorithm. For example we can use $(\lambda_n)_{n\in\N}$ for the first iteration and $(\wtilde{\lambda}_n)_{n\in\N}$ 
for the subsequent iterations. This might help in practice to achieve better finite sample results.
 
 For finding the optimal tuning parameters we suggest to use common time series methods that
are based on information criteria.
\cite{zou2007degrees}, \cite{wang2007regression}, \cite{zhang2010regularization} and \cite{nardi2011autoregressive} 
analyse information criteria in the lasso and adaptive lasso time series framework. %; CITE apllied them as well.
% Subsequently, let $\text{IC}(\bsbeta_n)$ 
% be an information criterion that corresponds to a model fitted by $\bsbeta_n$ and takes the minimal value in the optimum.
% We will assume that this information criterion is almost surely unique in its minimum.
Possible options for this information criteria are the 
Akaike information criterion (AIC), Bayes information criterion (BIC) or a cross-validation based criterion. 
Here, it is worth mentioning that \cite{kim2012consistent} discusses the generalised information criterion (GIC) in 
a classical homoscedastic lasso framework where the amount of parameters $p_n$ depends on $n$.
They establish that under some regularity conditions the GIC can be chosen so that a consistent model selection is possible.

For the initial estimate ${\bsbeta}_{n,\text{init}}$ that is required 
for the penalty weights there are different options available. The simplest is the OLS estimator,
which is available if $p_n< n$. Another alternatives are the lasso ($\tau=0$), elastic net or ridge regression estimator,
see e.g. \cite{zou2005regularization}. 
Remember that we require an initial estimate ${\bsbeta}_{n,\text{init}}$ only for the adaptive lasso case if $\tau>0$.

Note that \cite{wagener2013adaptive} described a setting with two initial estimators. One for 
the adaptive lasso weights as we do, 
and another one for the weight matrix $\bsW_n$.
% and a second one for the variance resp. the weight matrix $\bsW_n$. 
The first estimator 
corresponds to our ${\bsbeta}_{n,\text{init}}$, whereas the second inital estimator is not required, 
as we can initialise the volatility weight matrix $\bsW_n$ by the homoscedastic setting. A similar result was achieved by 
\cite{wagener2013adaptive} who showed that the homoscedastic estimator can be used as initial estimator in their setting.

\section{Applications to AR-ARCH type models}

In the introduction we mentioned that one of the largest fields of application might be the estimation of high-dimensional AR-ARCH type processes.
Therefore, we discuss a standard multivariate AR-ARCH model in detail. Afterwards, we briefly deal with several extensions, 
the periodic AR-ARCH model, change point AR-ARCH models, threshold AR-ARCH models, interaction models and ARMA-GARCH models.

Let $\bsY_t = (Y_{1,t}, \ldots, Y_{d,t})'$ be a $d$-dimensional multivariate process and $\DD = \{1, \ldots, d\}$.

\subsection{AR-ARCH model}

The multivariate AR model is given by
\begin{equation}
 Y_{i,t} = \phi_{i,0} + \sum_{j \in \DD} \sum_{k\in I_{i,j}} \phi_{i,j,k} Y_{j,t-k}+  \eps_{i,t}
 \label{eq_main_ar_model}
\end{equation}
for $i \in \DD$, where $\phi_{i,j,k}$ are non-zero autoregressive coefficients, $I_{i,j}$ 
are the index sets of the corresponding relevant lags and $\eps_{i,t}$ is 
the error term. The error processes $(\eps_{i,t})_{t\in \Z}$ 
follow the same conditional variance structure as in \eqref{eq_main_model_cond_var}, so
%$$\eps_{i,t} = \sigma_{i,t} Z_{i,t} \text{ where } (Z_{i,t})_{t\in \Z} \text{ is i.i.d. with } \E(Z_{i,t})=0 \text{ and } 
%\var(Z_{i,t}) = 1$$
% \begin{equation}
 $\eps_{i,t} = \sigma_{i,t} Z_{i,t}$ where  $\sigma_{i,t} = g_i( \bsalpha_i ; \bsL_i)$ and  
$(Z_{i,t})_{t\in \Z}$  is i.i.d. 
% \label{eq_main_model_cond_var} ,
% \end{equation}
with $\E(Z_{i,t})=0$ and $\var(Z_{i,t})=1$. 

% Note that if we choose $I_{i,j}=\{1,\ldots, p\}$ for $i,j\in \DD$ we
% get a common multivariate AR($p$) model for $\bsY_t$ 
% with $\bseps_t = \left(\eps_{1,t}, \ldots, \eps_{d,t} \right)$ as error term. 
% The multivariate representation of
% \eqref{eq_main_model} is given by
% \begin{equation}
%  \bsY_{t} = \Phi_{0} + \sum_{k\in I} \Phi_{k} \bsY_{t-k}+ \bseps_{t}
%  \label{eq_main_model_multivariate}
% \end{equation}
% with intercept vector $\Phi_{0} = (\phi_{1,0}, \ldots, \phi_{d,0})$, index set  $I= \bigcup_{i,j\in \DD} I_{i,j}$ and coefficient matrices
% $$\Phi_k= (\varphi_{i,j,k})_{i\in \DD,j\in \DD} \text{ where }\varphi_{i,j,k} = \begin{cases}
%           \phi_{l,h,k} & , l,h\in I_{i,j} \\
%           0 & , \ow
%          \end{cases} .$$
%          
% The multivariate representation for $\bseps_t$ is given by 
% $$\bseps_t = \Sigma_t \bsZ_t \text{ where } (\bsZ_{t})_{t\in \Z} \text{ is i.i.d. with } \E(\bsZ_{t})=\bsnull 
% \text{ and } \diag(\var(\bsZ_{t})) = \bsone$$
% where $\bsZ_t = (Z_{1,t},\ldots,Z_{d,t})$. 

Now, we define the representation \eqref{eq_main_ar_model} that matches the general representation \eqref{eq_main_model}
by 
\begin{equation*}
Y_{i,t} =  \bsX_{i,t} \bsbeta_i + \eps_{i,t}
\end{equation*}
for $i \in \DD$ where the parameter vector
$\bsbeta_i = (\phi_{i,0}, (\phi_{i,1,k})_{k \in I_{i,1}}, \ldots, (\phi_{i,d,k})_{k \in I_{i,d}} )$
and the corresponding regressor matrix $\bsX_{i,t}= (\bsone, (X_{i,1,t-k})_{k\in I_{i,1}},\ldots, (X_{i,d,t-k})_{k\in I_{i,d}} )$.
Note that this 
definition of $\bsbeta_i$ is only well defined if all $I_{i,j}$ for $j\in \DD$ are finite, if
one index set is infinite we have to consider another enumeration, but everything holds in the same way.

Furthermore, we assume that $\bseps_t = \left(\eps_{1,t}, \ldots, \eps_{d,t} \right)'$ follows an ARCH type model. 
In detail we consider a multivariate power-ARCH process which generalises the common multivariate ARCH process slightly.
Recently, \cite{francq2013optimal} discussed the estimation of such power-ARCH($\infty$) processes and showed applications to finance.
It is given by
 \begin{equation}
\sigma_{i,t}^{\delta_i} = \alpha_{i,0} + \sum_{j \in \DD} \sum_{k \in J_{i,j}} \alpha_{i,j,k} |\eps_{j,t-k}|^{\delta_i} ,
  \label{eq_main_arch_model}  
 \end{equation}
with $J_{i,j}$ as index set and $\delta_i$ as power of the corresponding $\sigma_t$. 
The parameters satisfy the positivity restriction, so $\alpha_{i,0}>0$ and $\alpha_{i,j,k} \geq 0$. 
Moreover we require that the $\delta_i$'s absolute moment $\E|Z_t|^{\delta_i}$ exists.
Obviously, we have 
\begin{equation*}
g_{i}(\bsalpha_i, \bsL_i) = 
%\sqrt[{\delta_i}]{ 
\bigg(\alpha_{i,0} + \sum_{j \in \DD} \sum_{k \in J_{i,j}} \alpha_{i,j,k} |\eps_{j,t-k}|^{\delta_i} \bigg)^{1/\delta_i}
\end{equation*}
where $\bsalpha_i= (\alpha_{i,0}, (\alpha_{i,1,k})_{k\in J_{i,1}}, \ldots, (\alpha_{i,d,k})_{k\in J_{i,d}} )$
and $\bsL_i = ( (\eps_{1,t-k})_{k\in J_{i,1}}, \ldots, (\eps_{d,t-k})_{k\in J_{i,d}} )$.
Similarly as for $\bsbeta_i$, $\bsalpha_i$ is only well defined if all $J_{i,j}$ for $j\in \DD$ are finite. Otherwise 
we have to consider another enumeration.
% If
% one is infinite we can consider another enumeration. Everything holds in the same way.}
%(Frank an zoikan analysed the power transformed detailled in CITE royB). 
The case $\delta_i=2$ leads to the
well known ARCH process which turns into a multivariate ARCH($p$) if $J_{i,j}=\{1,\ldots,p\}$.
         
For estimating the ARCH part parameters we will make use of a recursion that holds for the residuals.
%Using REF and REF we get
This is given by
\begin{equation}
|\eps_{i,t}|^{\delta_i} = \wtilde{\alpha}_{i,0} +
\sum_{j \in \DD} \sum_{k \in J_{i,j}} \wtilde{\alpha}_{i,j,k} |\eps_{i,t-k}|^{\delta_i} + u_{i,t}
\label{eq_absarch} 
\end{equation}
where $\wtilde{\alpha}_{i,0} = \gamma_i \alpha_{i,0}$, $\wtilde{\alpha}_{i,j,k} = \gamma_i \alpha_{i,j,k}$ 
and $u_{i,t} = \sigma_{i,t}(|Z_{i,t}| - \gamma_i)$ with 
$\gamma_i=\gamma_i(\delta_i) = \E|Z_{i,t}|^{\delta_i}$. Here, $u_{i,t}$ is a weak white noise process with $\E(u_{i,t}) = 0$.
The fitted values $\wtilde{\sigma}^i_t$ 
of equation \eqref{eq_absarch} are proportional to the $\sigma^i_t$ up to the constant $\gamma_i$.
 As  $\gamma_i$ is the $\delta_i$'s absolute moment 
  of $Z_{i,t}$, it holds that $\gamma_i = 2$, if $\delta_i = 2$. 
  %which might help to characterize the distribution of $\eps^i_t$. 
  If $\delta_i=1$ and $\eps_{i,t}$ follows a normal distribution $\gamma_i$ it is
  $\sqrt{2\pi^{-1}} \approx 0.798$. 
  If $\eps_{i,t}$ exhibits e.g. a standardised t-distribution we will observe larger first absolute moments $\gamma_i$.
%   , whereas e.g. the standardised t-distributions have a larger first absolute moment.

Clearly, the true index sets $I_{i,j}$ and $J_{i,j}$ are unknown in practice.
Thus we fix some index sets $\II_{i,j}(n)$ and $\JJ_{i,j}(n)$ for the estimation
that can depend on the underlying sample size $n$. 
If the true index sets $I_{i,j}$ and $J_{i,j}$ are finite, then
the choices $\II_{i,j}(n) = \{1, \ldots, \max(I_{i,j}) \}$ and $\JJ_{i,j}(n) = \{1, \ldots, \max(J_{i,j}) \}$
are obvious.
If $I_{i,j}$ and $J_{i,j}$ are infinite, 
$\II_{i,j}(n)$ and $\JJ_{i,j}(n)$ should be chosen so that they
are monotonically increasing in the sense that $\II_{i,j}(n-1) \subseteq \II_{i,j}(n)$ and $\JJ_{i,j}(n-1) \subseteq \JJ_{i,j}(n)$
with $\bigcup_{n\in \N} \II_{i,j}(n) = \N$ and $\bigcup_{n\in \N} \JJ_{i,j}(n) = \N$.
The size of $\II_{i,j}(n)$ and $\JJ_{i,j}(n)$ is directly related to the 
size of the estimated parameters $p_{i,n}$ for $\bsbeta_{i,n}$ and $l_{i,n}$ for $\bsalpha_{i,n}$.
It holds that $p_{i,n} = 1+ \sum_{j\in \DD} \II_{i,j}(n)$ and 
$l_{i,n} = 1+ \sum_{j\in \DD} \JJ_{i,j}(n)$. Here, $\bsbeta_{i,n}$ and $\bsalpha_{i,n}$ are the
restrictions of $\bsbeta_i$ and $\bsalpha_i$ to their first $p_{i,n}$ and $l_{i,n}$ coordinates.

For the estimation of $\bsbeta_i$ and $\bsbeta_{i,n}$ we can apply the iteratively reweighted 
adaptive lasso algorithm as described in the previous section. However, we have to specify 
an estimation method for the variance part. 
In particular we require the estimators $\what{\bsalpha}_i$ and $\what{\bsL}_i$, or more precisely their restrictions
$\what{\bsalpha}_{i,n}$ and $\what{\bsL}_{i,n}$ to its $l_{i,n}$ and $m_{i,n}(l_{i,n})$ coordinates.
For $\what{\bsL}_{i,n}(\bsbeta_{i,n}; \bsX_{i,n}, \bsY_{i,n})$ we have the estimator
\begin{equation*}
\what{\bsL}_{i,n,t} = \what{\bsL}_{i,n,t}(\bsbeta_{i,n}; \bsX_{i,n,t}, \bsY_{i,t}) = 
|Y_{i,t} -  \bsX_{i,n, t} \bsbeta_{i,n} |^{\delta_i}
\end{equation*}
which provides an estimate for $|\eps_{i, t}|^{\delta_i}$ and $|\eps_{i,n, t}|^{\delta_i}$.
For the estimation of $\what{\bsalpha}_{i,n}$ we suggest to minimise the problem
\begin{equation}
\| \what{\bsL}_{i,n,t} - \bsA_{i,t} \bsalpha_i \|_2,
\label{eq_minimise_alpha} 
\end{equation}
where $\bsA_{i,t} = (1, (\what{\bsL}_{1,n,t-k})_{k\in J_{i,1}}, \ldots, (\what{\bsL}_{d,n,t-k})_{k\in J_{i,d}}  )$, which corresponds to
the plug-in version of equation \eqref{eq_absarch}. 
For the estimation of \eqref{eq_minimise_alpha} a common non-negative least squares (NNLS) estimation technique can be considered.
If the variance equation is high-dimensional approaches like the positive lasso are suitable as well. Hence high-dimensional
lasso type algorithms with positivity constraint can be applied for the parameter estimation. 
But as the residuals in \eqref{eq_absarch} only follow a weak white noise process,
there are more advanced results for the asymptotic of this procedure required
% for the asymptotic of this procedure slightly more advanced results are required.
For the non-restricted adaptive lasso \cite{medeiros2012estimating} show
sign consistency and asymptotic normality under certain conditions for such a situation
with a weakly stationary error process.

However, the simple NNLS estimation procedure can act as a shrinkage procedure as well, 
as some parameters can be estimated to be 0. 
This well known sparsity effect of NNLS settings was recently analysed by \cite{meinshausen2013sign} and \cite{slawski2013non}.
\cite{slawski2013non} provided evidence that the NNLS approach is potentially 
superior to the positive lasso. 
We use the NNLS algorithm for the computational applications as described by \cite{lawson1974solving}.

\subsection{Periodic AR-ARCH model}

Another class of models where we can apply the proposed estimation technique 
is the class of periodic AR-ARCH models. Here, we assume a model as described above, but all parameters
are allowed to vary periodically over time. This is very suitable for modelling seasonal effects in high-dimensional data.

Thus, the model for the conditional mean equation is given by 
\begin{equation}
 Y_{i,t} = \phi_{i,0}(t) + \sum_{j \in \DD} \sum_{k\in I_{i,j}} \phi_{i,j,k}(t) Y_{j,t-k}+  \eps_{i,t}
 \label{eq_main_ar_model_timevar}
\end{equation}
 and for the conditional variance equation
\begin{equation}
\sigma_{i,t}^{\delta_i} = \alpha_{i,0}(t) + \sum_{j \in \DD} \sum_{k \in J_{i,j}} \alpha_{i,j,k}(t) |\eps_{j,t-k}|^{\delta_i} .
 \label{eq_main_arch_model_timevar}
\end{equation}
As mentioned, the time dependent parameters vary periodically over time.
Assuming a periodicity of $S$ we have 
$\phi_{i,0}(t) = \sum_{l} B_{i,0,l}(t) \phi_{i,0,l}$, $\phi_{i,j,k}(t) = \sum_{l} B_{i,j,k,l}(t) \phi_{i,j,k,l}$,
$\alpha_{i,0}(t) = \sum_{l} B_{i,0,l}(t) \alpha_{i,0,l}$, and $\alpha_{i,j,k}(t) = \sum_{l} B_{i,j,k,l}(t) \alpha_{i,j,k,l}$,
where $B_{i,0,l}$ and $B_{i,j,k,l}$ are $S$-periodic basis functions. 

Note that the processes is in general not weakly stationary anymore. However,
they are periodically weakly stationary (also known as weakly cyclostationary). 
So if $S\in \N$ then the subsequences $(\bsY_{St+s})_{t\in \Z}$ follow a weakly stationary 
process. For more details see e.g. \cite{aknouche2012asymptotic}.

As choice for the periodic basis functions, periodic indicator functions are suitable
if $S$ is small, the parameter space will be blown up by a factor of $S$.
If $S$ is large, a Fourier approximation, periodic B-splines or periodic wavelets might be a good choice as basis
to keep the parameter space reasonable.

As mentioned, the process $\bsY_t$ is not stationary in general, so the asymptotic theory given above can not be applied.
Nevertheless, a similar theorem is likely to hold true for periodic stationary processes. 
In order to proof this statement one would have to focus on the
% In the proof we have to work one a 
level of the mentioned weakly stationary subsequences,
similarly as in \cite{ziel2015quasi}.
%   $Y_{i,t} = \phi_{i,i,0}(t) + \sum_{j = 1}^d \sum_{k\in I_{i,j}} \phi_{i,j,k}(t) Y_{j,t-k}+ \eps_{i,t}$
%   with $\phi_{i,j,k}(t) = \sum_l \phi_{i,j,k,l} B_l(t)$ where $B_l(t)$ basis functions
The estimation procedure can be then performed as in the AR-ARCH model part.

\subsection{AR-ARCH with structural breaks}

Another field of possible applications is the one of change point models, i.e.
models where we have at least one structural break.
Here, the basic model is a time-varying AR-ARCH model as defined in
equations \eqref{eq_main_ar_model_timevar} and \eqref{eq_main_arch_model_timevar} for the periodic AR-ARCH model. 
The basis functions are defined so that they can capture structural breaks instead of periodic effects.
The resulting model is of the same structure as the change point model used by \cite{chan2013group}.
If we have a priori information about the change point
we can take this into account. 
If we have no information, some clever segmentation of the time should be considered.
One option is to allow a change in every parameter (especially $\phi_{i,0}$) and at
every time point. This can be handled by choosing $n$ basis functions for each parameter
so that they build a triangular matrix. The resulting model is a special 
case of the so called fused lasso (see e.g. \cite{tibshirani2005sparsity}) and suitable for change point analysis. 
This particular mentioned approach of modelling change points is analysed in \cite{levy2008catching} and \cite{harchaoui2010multiple}.
However, this increases the parameter space enormously, in every case we receive $p_n> n$.
%Thus, this is only suitable if $n$ and $p$ are not too large. 

A general problem of the change point model is that the theorem above cannot be applied
due to the structural breaks. Even though the proposed algorithm might be a powerful tool to solve the problem,
we have to use it carefully. Any inference after estimating the model
should be backed up by some Monte-Carlo studies.

\subsection{Threshold AR-ARCH model}

Threshold AR-ARCH models are popular when the mean or variance reversion properties change 
dependent on the past of the process. Threshold AR models are popular as they are simple but powerful examples
for regime switching models. Threshold ARCH processes have many applications in finance, because they are suitable
to capture the so called leverage effect.

The general model is given by
\begin{equation*}
 Y_{i,t} = \phi_{i,0} + \sum_{j \in \DD} \sum_{k\in I_{i,j}} \sum_l   \phi_{i,j,k,l} \bsone{\{ Y_{j, t-k}> a_{k,l} \}} Y_{j,t-k}+ \eps_{i,t}
\end{equation*}
with thresholds $a_{k,l}$ %for the conditional mean equation 
and
\begin{equation*}
\sigma_{i,t}^{\delta_i} = \alpha_{i,0} + 
  \sum_{j \in \DD} \sum_{k\in I_{i,j}} \sum_l \alpha_{i,j,k,l} \bsone{\{ \eps_{j, t-k}> b_{k,l} \}} |\eps_{j,t-k}|^{\delta_i} 
  + \eps_{i,t}
  \end{equation*}
  with thresholds $b_{k,l}$. The option of one threshold at $b_{1,k}=0$ in the conditional variance model
  is very popular. This leads to the well known TARCH model, introduced by \cite{rabemananjara1993threshold}. 
  \cite{ziel2015efficient} applied the proposed algorithm to a similar multivariate AR-TARCH type model
%   successfully 
  to electricity market data.
Here, we can use the algorithm proposed above, because all 
covariate processes and $\bsY_t$ can be weakly stationary. % (if the true parameters satisfy some not mentioned conditions on the model).
  The mentioned zero-threshold option is often suitable in practice as 
it only doubles the volatility parameter space.

\subsection{AR-ARCH model with quadratic interactions}

Interaction models are very popular in classical regression settings, especially in medicine.
This type of model was e.g. analysed by \cite{choi2010variable} or \cite{bien2013lasso}, but not in a time series context.
In general we can apply the theorem for these models as well, as the 
interactions are in general weakly stationary processes, if they
have still a finite second moment.
The full quadratic interaction model is given by
\begin{equation*}
  Y_{i,t} = \phi_{i,0} + \sum_{j \in \DD} \sum_{k\in I_{i,j}} 
  \phi_{i,j,k} Y_{j,t-k}  + \sum_{j \in \DD} \sum_{l\in \DD} \sum_{k\in I_{i,j}} \sum_{m\in I_{i,j}}
  \phi_{i,j,k,l,m} Y_{j,t-k} Y_{l,t-m} + \eps_{i,t} .
\end{equation*}
% For the general interaction model the parameter restrictions to $\phi_{i,j,k}$ and $\phi_{i,j,k,l,m}$ for receiving a stationary 
% process are quite complicated.
  A problem that arises is the size of the parameter space which is $p_n(p_n+1)/2$, where
  the standard AR-ARCH model has $p_n$ parameters. 
 
\subsection{ARMA-GARCH model}

The last extension considers a very popular class of models. 
We know that every ARMA($p$, $q$) model can be rewritten as an AR($\infty$). Similarly a univariate GARCH($p$, $q$)
can be expressed as an ARCH($\infty$). Hence, it is clear that 
every ARMA-GARCH model can be written as an AR($\infty$)-ARCH($\infty$). This
AR($\infty$)-ARCH($\infty$) can be well approximated by an AR($\wtilde{p}$)-ARCH($\wtilde{q}$)
for large $\wtilde{p}$ and $\wtilde{q}$.
However, this gives an approximation and will likely include more parameters than the original ARMA-GARCH model.

Recently, \cite{chen2011subset} proposed a method of how to estimate ARMA processes in a lasso framework,
using this kind of approximation.
The idea is simple: 
Given the ARMA model
\begin{equation*}
 Y_{i,t} = \phi_{i,0} + \sum_{j \in \DD} \sum_{k\in I_{i,j}} \phi_{i,j,k} Y_{j,t-k}
 + \sum_{k\in K_{i,j}} \theta_{i,j,k} \eps_{j,t-k} +  \eps_{i,t} 
 \label{eq_main_arma_model}
\end{equation*}
we consider first an AR($\wtilde{p}$)-model with large $\wtilde{p}$ that can approximate
the true ARMA model sufficiently well. 
The residuals of this fitted model are used
for constructing the regressor matrix that contains the lagged autoregressive part and moving average part.
We repeat the lasso estimation with this regressor matrix.
So this procedure leads automatically to a two step approach. Clearly, we can iterate this more often to receive 
better stability, similarly to the algorithm we presented. % in this paper.
\cite{chen2011subset} showed that under certain conditions this estimation principle based on the adaptive lasso 
can lead to consistent estimates.

The same principle can be applied to the GARCH model as well. So we first estimate
a high dimensional ARCH model and take the estimated conditional variances for 
constructing the response matrix required for the GARCH model. This method opens a lot of possibilities
for applications in financial frameworks. In multivariate settings, we have to specify a special GARCH model.
In fact we can use every GARCH model that we can express in regression form, so even the BEKK-GARCH
is possible.

% either a
% non-negative least square or a positive lasso approach.
% 
% %They we might cover 
% 
% 
% Now we define the representation \eqref{eq_main_ar_model} that matches \eqref{eq_main_model}
% by 
% $Y_{i,t} = \bsbeta_i^\top \bsX_{i,t} + \eps_{i,t}$
% for $i \in \DD$ where 
% $\bsbeta = (\phi_{i,0}, \phi_{i,1,\min(I_{i,1})}, \ldots, \phi_{i,1,\max(I_{i,1})}, \phi_{i,2,\min(I_{i,2})} 
% , \ldots, \phi_{i,\max(I_{i,s}),d} )$
% and 
% $\bsX_{i,t}= (\bsone, X_{i,1,t-\min(I_{i,1})},\ldots, X_{i,1,t-\max(I_{i,1})},
% X_{i,2,t-\min(I_{i,2})} , \ldots, X_{i,\max(I_{i,d}),d})$ the corresponding regressor matrix.
% 
% Using the equation \eqref{eq_absarch} 
% we can define $\bsL_i( \bsbeta; \bsX, Y ) = (\ldots, \bsL_{i,t-1}, \bsL_{i,t})$ 
% with $ \bsL_{i,t} = (L_{i,k,t})_{k\in J_{i,j}}$  by
% $$ L_{i,k,t} = B^k|Y_{i,t} - \bsbeta_i^\top \bsX_{i,t}|^\delta $$ 
% for $k\in J_{i,j}$ and $j\in \DD$.
% where $B$ is the Backshift operator, so $B(x_t)=x_{t-1}$.
% % Hence we can use 
% % $\what{\bsL}( \bsbeta; \bsX, Y )$
% 
% 
%  $$\bsalpha^{[k]} = \what{\bsalpha}( \bsbeta^{[k]}; \bsL^{[k]} ) \text{ with } \bsL^{[k]} 
%  = \what{\bsL}( \bsbeta^{[k]}; \bsX, Y )$$
% 
% where additionally

% Nevertheless for some conditional variance models that depend on the past residuals $\eps_{t-1}, \eps_{t-2}, \ldots$
% like an ARCH process this is easily possible.
\section{Simulation study}

\begin{figure}[hbt!]
\centering
\begin{subfigure}[b]{0.49\textwidth}
 \includegraphics[width=1\textwidth, height=.75\textwidth]{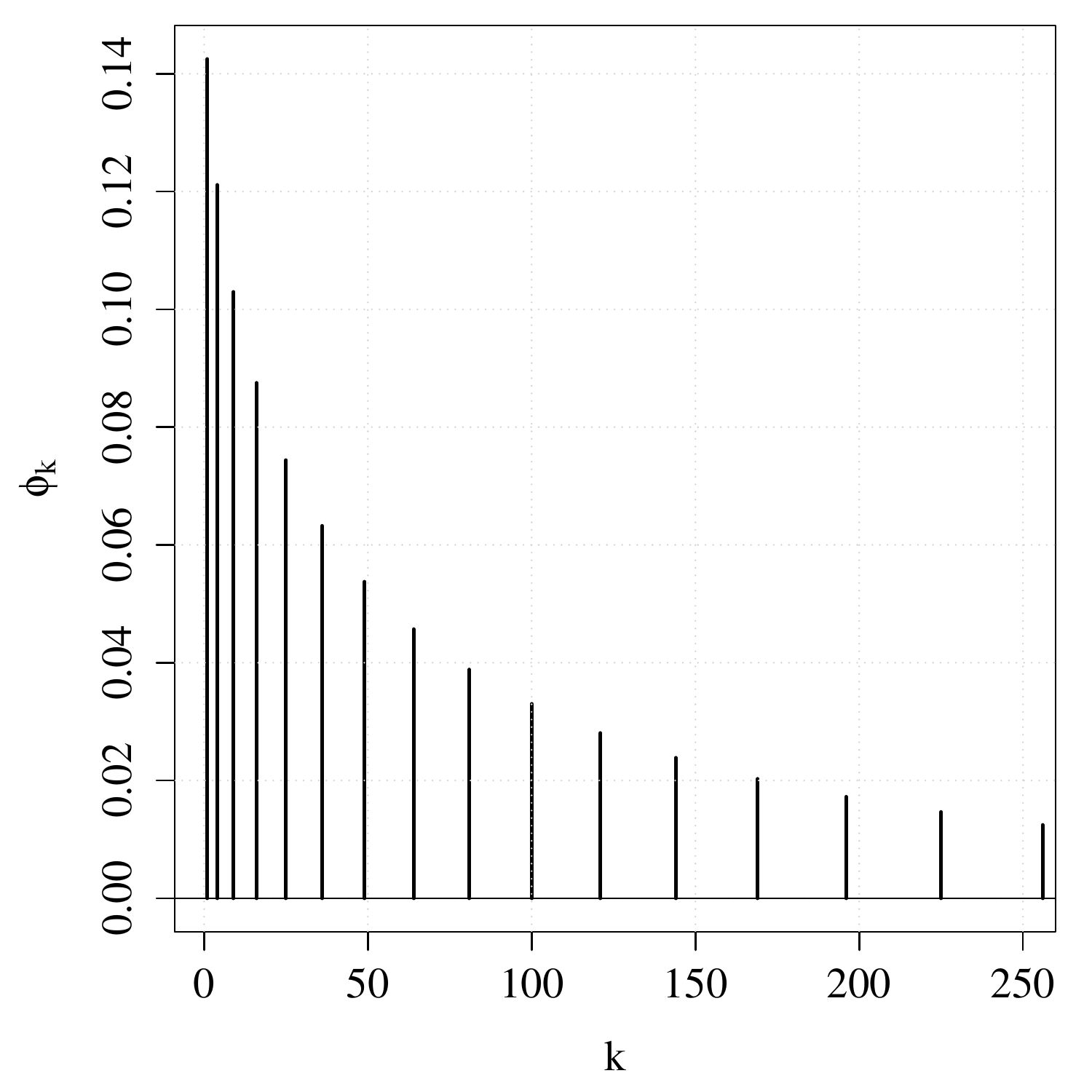}
   \caption{First considered coefficients with corresponding lag.}
  \label{fig_sim_sub1}
\end{subfigure}
\begin{subfigure}[b]{0.49\textwidth}
 \includegraphics[width=1\textwidth, height=.75\textwidth]{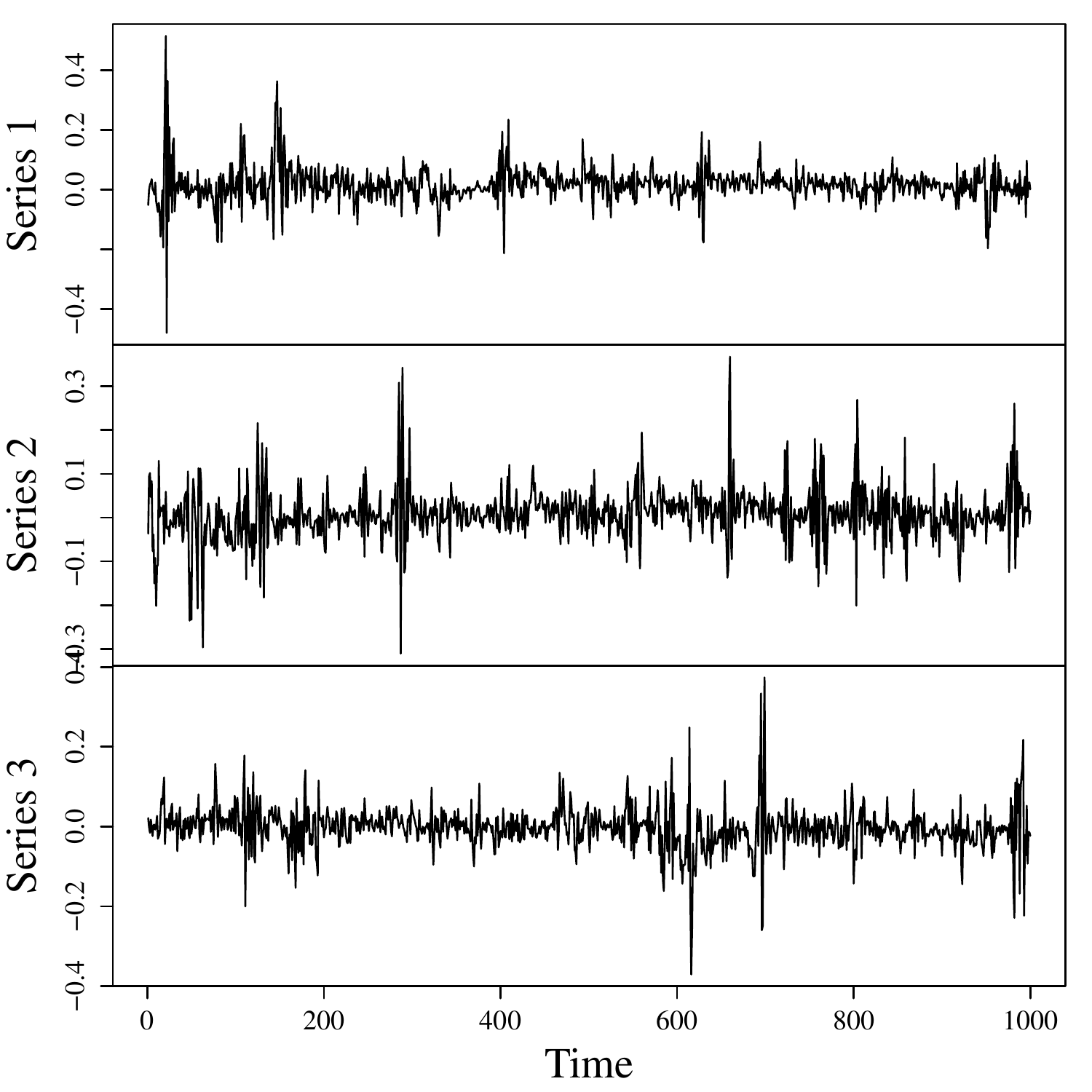}
   \caption{Sample of three simulated processes.}
  \label{fig_sim_sub2}
\end{subfigure}
 \caption{Considered parameters in \ref{fig_sim_sub1} and simulated sample paths of 3 time series in \ref{fig_sim_sub2} of model \eqref{eq_sim}.}
 \label{fig_sim}
\end{figure}

In this section we perform Monte-Carlo simulations to learn about the finite sample properties of the model algorithm.
Of course the results of the simulation will very much depend on the true model.
For illustration purposes we restrict ourselves to a univariate settings where both $p_n$ and $q_n$ are 
increasing with a rate of $\sqrt{n}$. For all simulations we consider a one-dimensional AR-ARCH-type process
\begin{equation}
Y_t = \sum_{k\in I_{1,1}} \phi_k Y_{t-k} + \eps_t 
\label{eq_sim}
\end{equation}
where $\eps_t = \sigma_t Z_t$ with $Z_t \stackrel{\text{iid}}{\sim} N(0,1)$ and
$$\sigma_t = \alpha_0 + \alpha_1 |\eps_{t-1}| + \alpha_2 |\eps_{t-2}|$$
with $\alpha_0 = 0.01$ and $\alpha_1 = \alpha_2 = 0.49$.
 The true subset $I_{1,1}$ of relevant lags of model \eqref{eq_sim} 
 is given by 
 $I_{1,1} = \{n^2 | n\in \N\} = \{1, 4, 9, 16, 25, \ldots\} $.
For parameters $\phi_k$ with $k\in I_{1,1}$ we define $\phi_k = 0.95   (\phi^{-1}-1) \phi^{\sqrt{k}}   $ 
with $\phi= 0.85$.
As $ (\phi^{-1}-1) \sum_{k\in I_{1,1} } \phi^{\sqrt{k}} = (\phi^{-1}-1) \sum_{k\in \N } \phi^{k} = 1 $ we have $\sum_{k\in I_{1,1}} \phi_k = 0.95$.
So the considered process has a clear autoregressive structure and is stationary.
In Figure \ref{fig_sim} the considered coefficient structure and some simulated sample paths are visualised.
In the sample paths we observe the clear conditional heteroscedasticity.
%  is generated by sampling $M$ times uniformly without replacement from $ \II_{1,1}$  
For the estimation the proposed superset $\II_{1,1}$ will be important as well. 
We consider the set $\II_{1,1} =  \{1, 2, \ldots, \lfloor 5\sqrt{n} \rfloor\}$, so we have that $p_n \sim \sqrt{n}$.
  
  Subsequently we want evaluate the estimation procedure on the full tuning parameter path. 
Therefore we estimate \eqref{eq_sim} for all $\lambda$ values on a given exponential grid 
$\Lambda = \{ 2^g | g\in \G \}$ where $\G$ is a equidistant grid from $-4$ to $-18$ of length $100$.
Additionally, we want to illustrate the impact of different information criteria.
The information criteria that we consider are
the Akaike information criterion (AIC), the Hannan-Quinn criterion (HQC) and the
 Bayesian information criterion (BIC). These are all special cases of the generalised information
 criterion (see e.g. \cite{kim2012consistent}) that is given by $\text{GIC}(\kappa_n) = \log( \what{\sigma}_t^2 ) + \kappa_n K /n$,
where $K$ represents the number of parameters in the model. We get the AIC, HQC and BIC by choosing either
$\kappa_n=2$, or $\kappa_n=2\log(\log(n))$ or $\kappa_n=\log(n)$, respectively.
The volatility model is estimated by the methods explained in the section above. The model order is assumed to be known.
In all adaptive lasso estimation procedures we choose only the lasso itself, so $\tau = 0$.
We simulated for $n \in \{300,600, 1200\}$ with a Monte Carlo sample size $N=1000$.

% Furthermore we choose the following setting for this simulation study:
%    \begin{itemize}
%   \item  $\eps_t = \sigma_t Z_t$ with $\sigma_t = \alpha_0 + \sum_{i=1}^{10} \alpha_i |\eps_t|$ 
%   with $\alpha_0 = 0.01$, $\alpha_i = 0.09$ for $i\in \{1, \ldots, 10\}$ and \\
% (i)    $Z_t \sim N(0,1)$ resp. (ii) $Z_t \sim t_3(0,1)$ which represents the standardised t-distribution with $3$ degrees of freedom 
% %  \end{itemize}
%   \item proposed superset $\II_{1,1} = \{1, \ldots, 100\}$
%   \item true subset $I_{1,1}$ is generated by sampling $M$ times uniformly without replacement from $ \II_{1,1}$
%   \begin{enumerate}
%    \item $M= 20$ with $\phi_i = 0.049$ for $i\in I_{1,1}$
%    \item $M= 25$ with $\phi_i = 0.039$ for $i\in I_{1,1}$
%   \end{enumerate}
% \item simulated for $n \in \{1000,2000,4000\}$ with a Monte Carlo sample size $N=200$
% %\item simulation for information criteria: AIC, HQC, BIC
%  \end{itemize}

\begin{figure}[hbt!]
\centering
\begin{subfigure}[b]{0.49\textwidth}
 \includegraphics[width=1\textwidth,height=.75\textwidth]{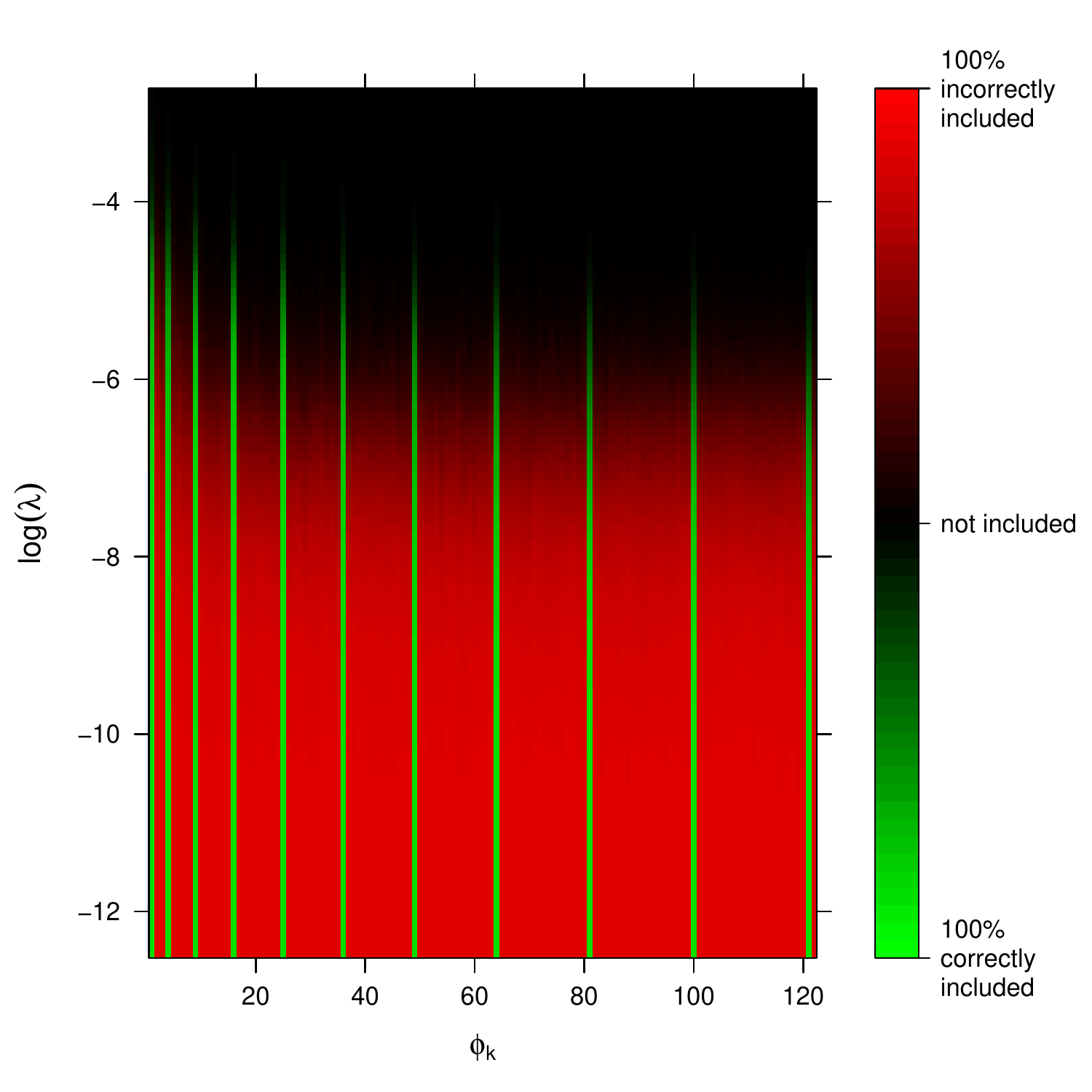}
   \caption{$k=1$}
  \label{fig_tval_sub1}
\end{subfigure}
\begin{subfigure}[b]{0.49\textwidth}
 \includegraphics[width=1\textwidth, height=.75\textwidth]{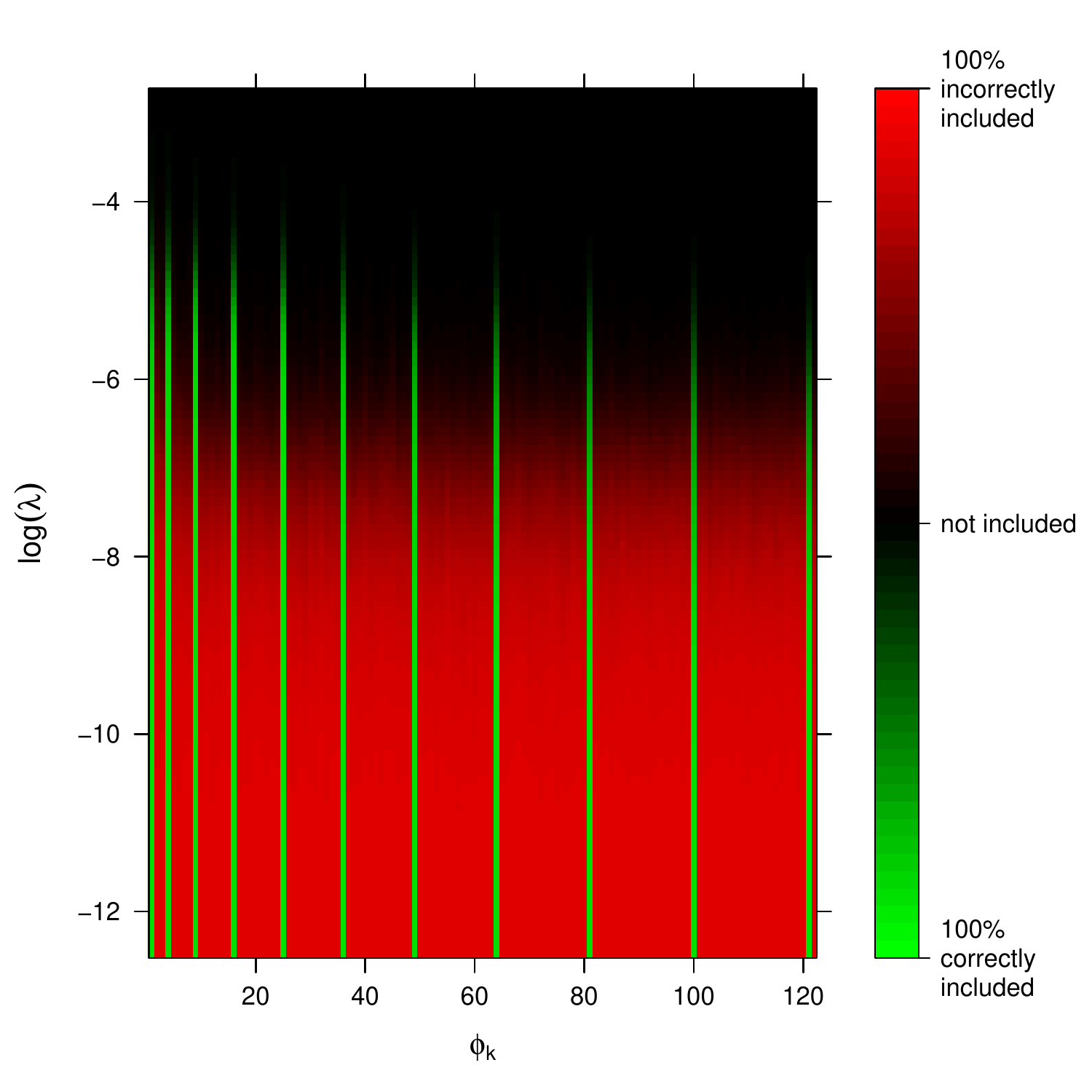}
   \caption{$k=2$}
  \label{fig_tval_sub2}
\end{subfigure}
 \caption{Proportion of irrelevant included parameters (black to red) and relevant included parameters (black to green) for $n=600$
 and $\lambda \in \Lambda$.}
 \label{fig_props}
\end{figure}

%  To get more realistic results, we take models where the relevant lags in $I_{1,1}$ are not fixed for all simulation, but 
%  uniformly sampled without replacement from $\{1, \ldots, 100\}$.
 After simulating the process, we estimate by the proposed iteratively reweighted lasso algorithm.
 The first simulation result is given in Figure \ref{fig_props}.
There we see the proportions of both the irrelevant and relevant included parameters
of all estimated parameters for the homoscedastic case ($k=1$) and for the heteroscedastic with one additional replication ($k=2$)
given a situation with $n=600$ observations 
and the exponential grid tuning parameter grid $\Lambda$.
Obviously, we observe that for both models the probability to include a parameter increases with decreasing $\lambda$.
We see that parameters $\phi_k$ with $k\in I_{1,1}$ and small $k$ are easier to detect than
those with larger $k$. This is clear as $\phi_k$ with $k\in I_{1,1}$ is decreasing in $k$.
Further, we can observe that for both cases ($k=1$ and $k=2$) the algorithm
seems to distinguish well between relevant parameters and irrelevant parameters. In this situation a reasonable 
choice of the tuning parameter could be $\log(\lambda) = -6$. There we see that proportion of relevant parameters (green colored) included 
is clearly closer to 100\% than the proportion of irrelevant included parameters (dark red to black).
It seems that the heteroscedastic algorithm can distinguish better than its homoscedastic counterpart.

\begin{figure}[hbt!]
\centering
 \includegraphics[width=.75\textwidth, height=.45\textwidth]{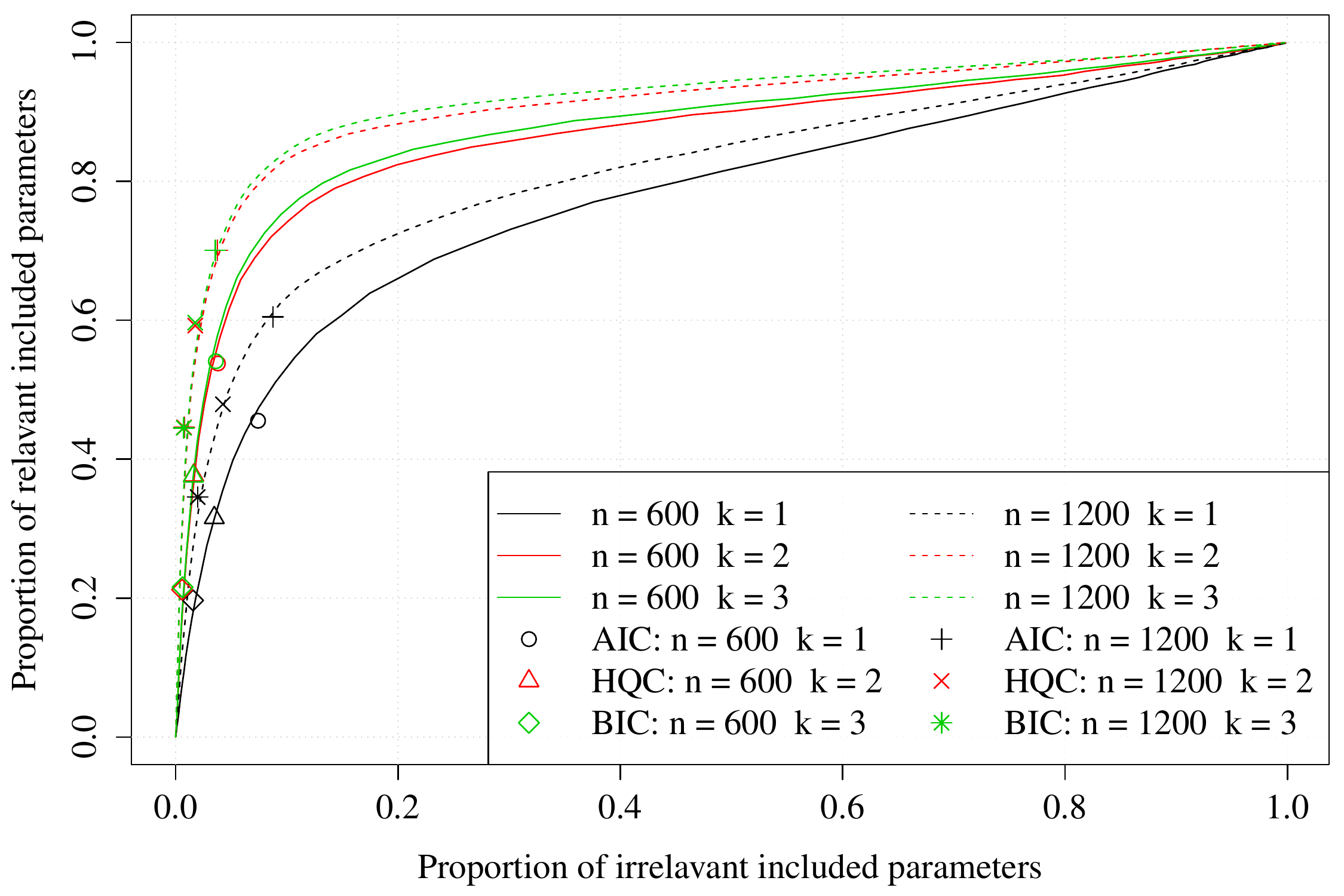}
 \caption{Mean proportion of all irrelevant included parameters against 
 mean proportion of all relevant included parameters for the first three iterations and $n\in \{600,1200\}$ on the 
 full $\lambda$ grid and for the considered information criteria.}
 \label{fig_propic}
\end{figure}

To emphasis this fact we created a new plot where we
visualise the computed mean proportion of all irrelevant included parameters against the mean proportion of all relevant included parameters. 
The mentioned plot is given in Figure \ref{fig_propic}.
We additionally added the corresponding values for the considered information criteria.
To understand the impact of the sample size and the number of iterations we plot the cases for $n=600$ and $n=1200$ and the first three iterations of 
the algorithm.
The bottom left corner corresponds to very large $\lambda$ values where no parameter at all is included in the model.
The top right corner covers the ordinary least square estimate with $\lambda=0$.

Roughly speaking we are aiming for estimators that are as close as possible to the upper left corner. 
It is particularly important to mention that for increasing $n$ we should get close to the upper left corner.
This seems to be satisfied for the relevant tuning parameter path. We see that
with the heteroscedastic cases with $k=2$ and $k=3$ have better selection properties than 
the homoscedastic case. The improvement from the case $k=2$ to $k=3$ is very small, but it is still there.
The same holds for the considered information criteria. Note that even though it is well known
that the AIC is inconsistent in parameter selection in a finite sample setting it seems to perform quite well.

% the different information criteria show the expected behaviour. So 
% the conservative BIC includes the least amount of parameters. Hence, the PIIP is the smallest for BIC in comparison to HQC and AIC,
% but the PRIP is smaller as well. 
% The PRIP is always larger for the proposed algorithm that takes the conditional heteroscedasticity into account as the homoscedastic analog.
% This seems to be independent of the sample size, the chosen information criteria and residual distribution.
% For the PIIP this situation in not that clear, but for larger $n$ we have smaller PIIP for the proposed model
% which matches our expectations. Moreover, the results seems to be very robust with regards to the tails of the errors.
% So the results for the t-distribution with 3 degrees of freedom are still satisfying in comparison to their
% normally distributed analog.

Nevertheless, it is not clear how the algorithm performs in an out-of-sample forecasting study. 
Therefore, we conduct another simulation study where we focus on the out-of-sample forecasting error.
We compute the $1$-step ahead mean absolute forecast error ($\MAE$)
% and the mean of $j$-step ahead mean absolute forecast error until step $h$ ($\MMAE_h$).
which is defined by $\MAE = \frac{1}{N} \sum_{i=1}^N |\what{Y}_{n+1} - Y_{n+1}| $
% and 
% $\MMAE_h = \frac{1}{h} \sum_{i=1}^h \MAE_h $,
where $\what{Y}_{n+1}$ denotes the forecast of $Y_{n+1}$. % We consider a forecast horizons of $h\in \{1, \ldots, 10\}$.
Additionally, we calculate the forecasting error for the corresponding oracle model. For the oracle we assume
that the underlying lag structure of the autoregressive model is known. 

\begin{figure}[hbt!]
\centering
\begin{subfigure}[b]{0.49\textwidth}
 \includegraphics[width=1\textwidth, height=.75\textwidth]{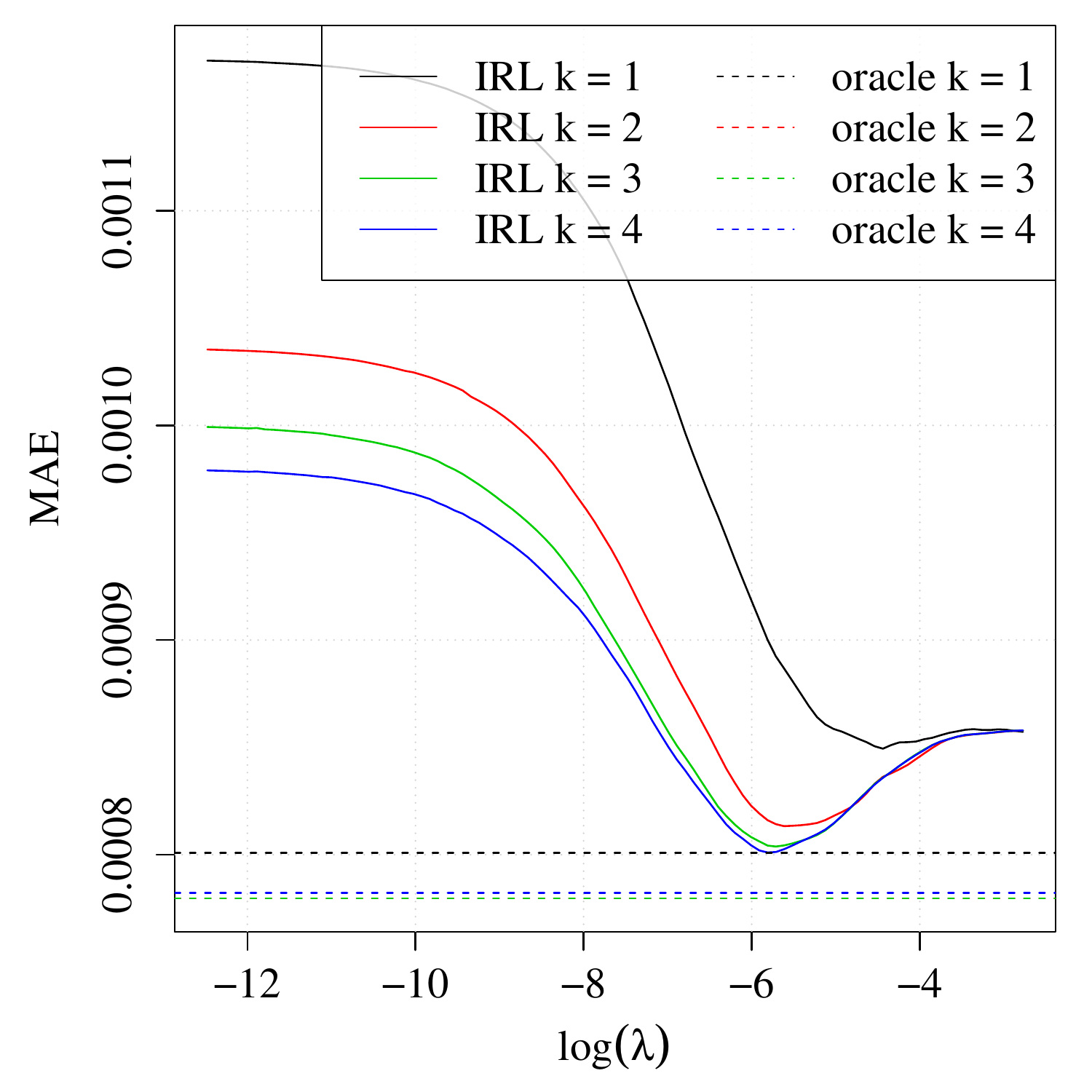}
   \caption{$n=300$}
  \label{fig_mae_sub1}
\end{subfigure}
\begin{subfigure}[b]{0.49\textwidth}
 \includegraphics[width=1\textwidth, height=.75\textwidth]{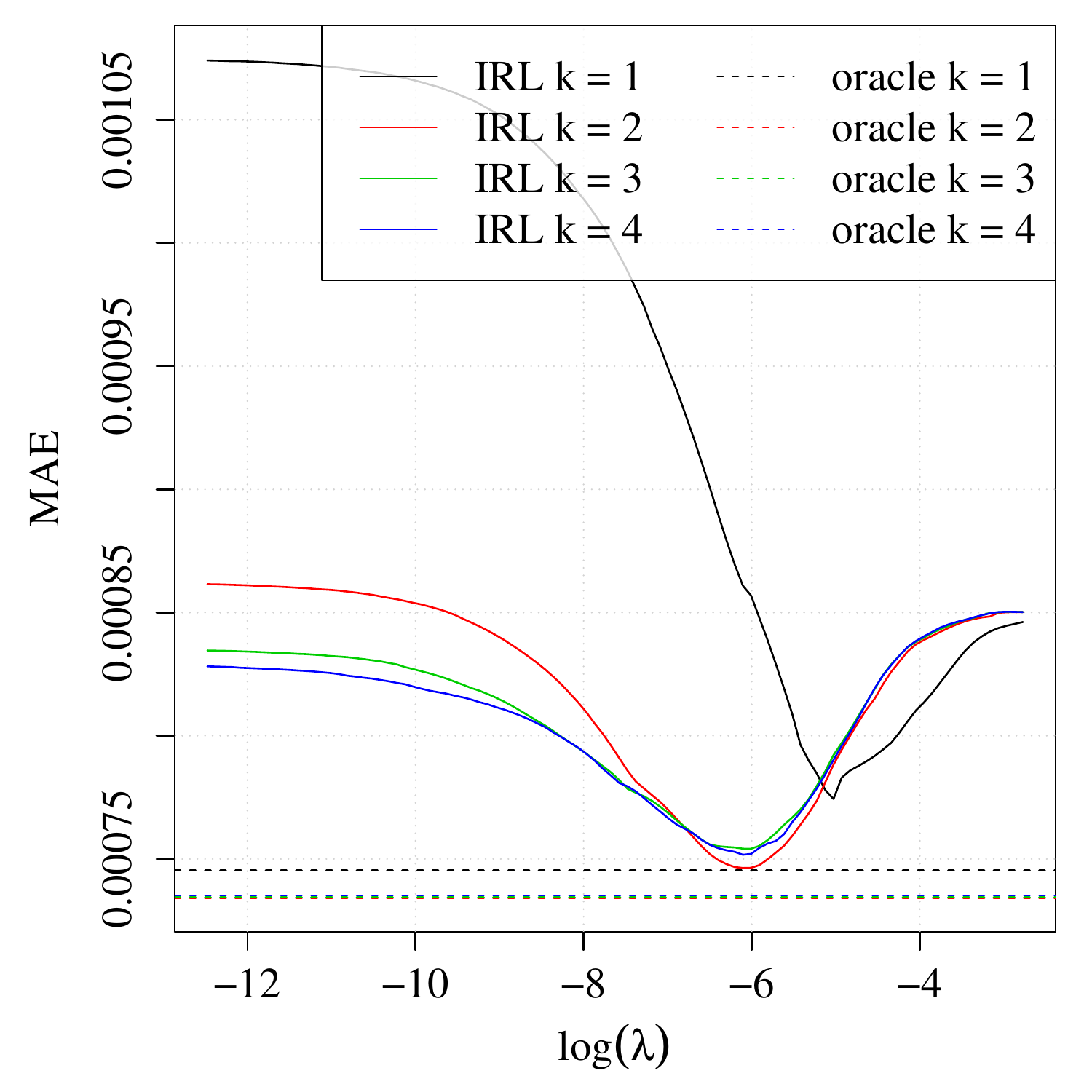}
   \caption{$n=600$}
  \label{fig_mae_sub2}
\end{subfigure}
 \caption{$\MAE$ for $n=300$ (\ref{fig_mae_sub1}) and 
$n=600$  (\ref{fig_mae_sub2}) of the iteratively reweighted lasso (IRL) method for several iterations $k\in \{1,2,3,4\}$, such as their
 oracle estimators for the AR-ARCH model.}
 \label{fig_mae}
\end{figure}

The simulation results for $n=300$ and 
$n=600$ are given in Figure  \ref{fig_mae}.
% Here, we observe additionally to the estimated $\MAE_h$ and $\MMAE_h$ their estimated 2-$\sigma$ ranges, namely
% $\MAE_h \pm 2 \sigma( \MAE_h )$ resp. $\MMAE_h \pm 2 \sigma( \MMAE_h )$. We observe basically observe the same relationships
% for $\MAE_h$ and $\MMAE_h$, but the $\MMAE_h$ 
% is better to interpret due to smaller confidence bands. 
We see that
the homoscedastic algorithm performs significantly worse than the heteroscedastic one with $k=1$, except for large $\lambda$ values in 
the $n=600$ situation. Interestingly for $n=300$ and $k=1$ the $\MAE$ hardly goes below the value of
the case with very large $\lambda$ where $\what{\phi}_k=0$ for all $k\in \II_{1,1}$. In contrast 
for $k>1$ an improvement in the forecasting performance to the case with very large $\lambda$ where $\what{\phi}_k=0$ is possible.
The same fact can be observed, within the oracle procedures, but the 
improvement is not that obvious.
From an applications perspective, this is extremely significant. It indicates that we can benefit more from taking the heteroscedasticity into account
in settings with unknown model structure than in a setting where the underlying structure is known.
However, we usually do not know the true underlying model as the oracle does, especially in high-dimensional settings. 
It shows that the proposed 
estimation algorithm can lead to crucial improvements in a high-dimensional setting.
This is also observed by \cite{ziel2015efficient} in applications of the proposed estimation algorithm to electricity market data.

As a robustness check we replicate the simulation study with a different volatility model.
We assume a TARCH process for the residuals. 
TARCH models are popular in financial applications as they are able to capture leverage effects.
% In detail we consider 
The considered TARCH process for the simulation study is parameterised through
$$\sigma_t = \alpha_0 + \alpha_1 |\eps_{t-1}| + \alpha^{-}_1 \bsone\{\eps_{t-1}<0\} |\eps_{t-1}| + \alpha_2 |\eps_{t-2}| + 
\alpha^{-}_2 \bsone\{\eps_{t-2}<0\} |\eps_{t-2}| $$
where the leverage effect is modelled by the two parameters $\alpha^{-}_1$ and $\alpha^{-}_2$ 
which give an additional impact on negative past residuals to the volatility. The selected parameter setting 
is $\alpha_1 = \alpha_2 = 0.245 $ and  $\alpha^{-}_1 = \alpha^{-}_2 = 0.49 $.

\begin{figure}[hbt!]
\centering
\begin{subfigure}[b]{0.49\textwidth}
 \includegraphics[width=1\textwidth, height=.75\textwidth]{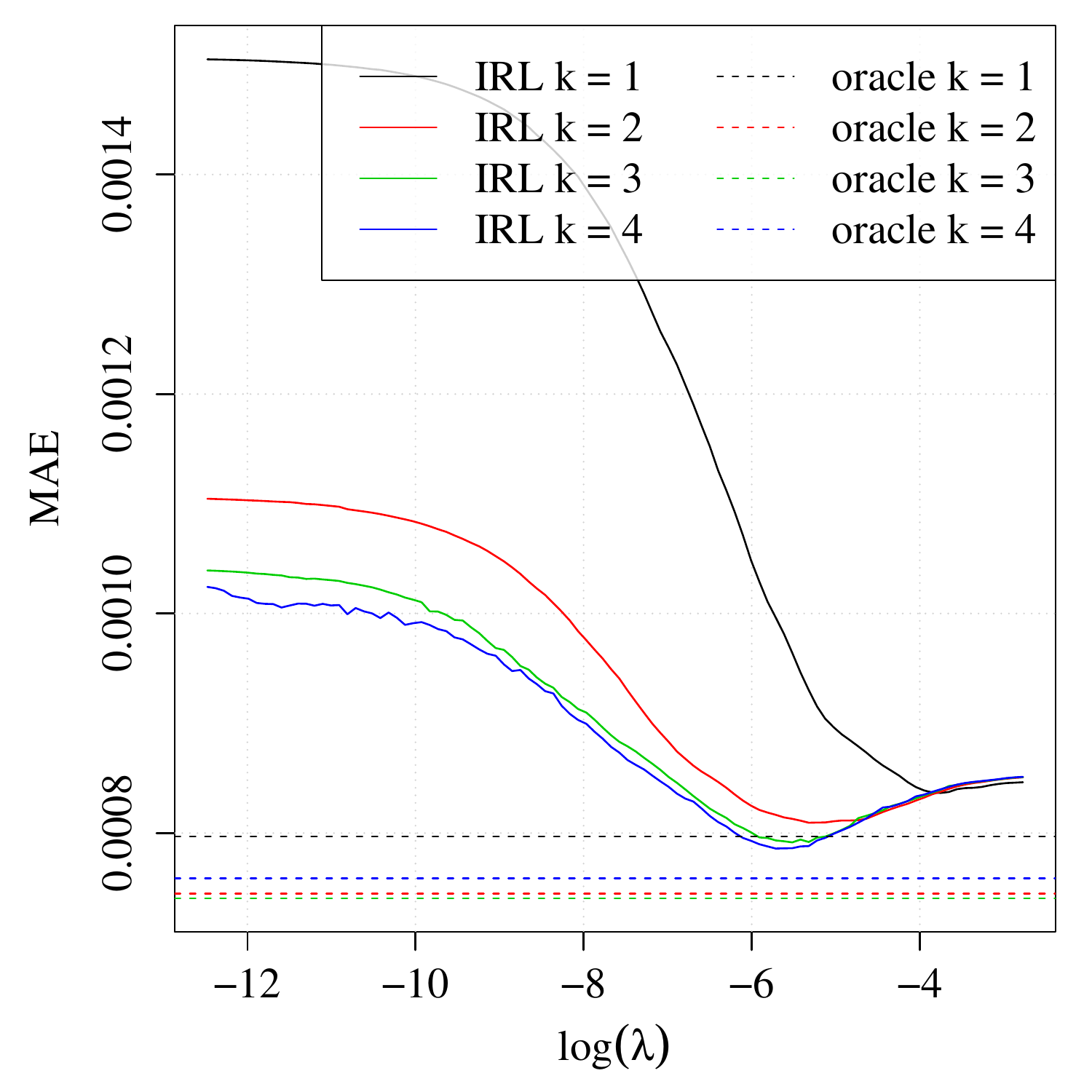}
   \caption{$n=300$}
  \label{fig_tarch_mae_sub1}
\end{subfigure}
\begin{subfigure}[b]{0.49\textwidth}
 \includegraphics[width=1\textwidth, height=.75\textwidth]{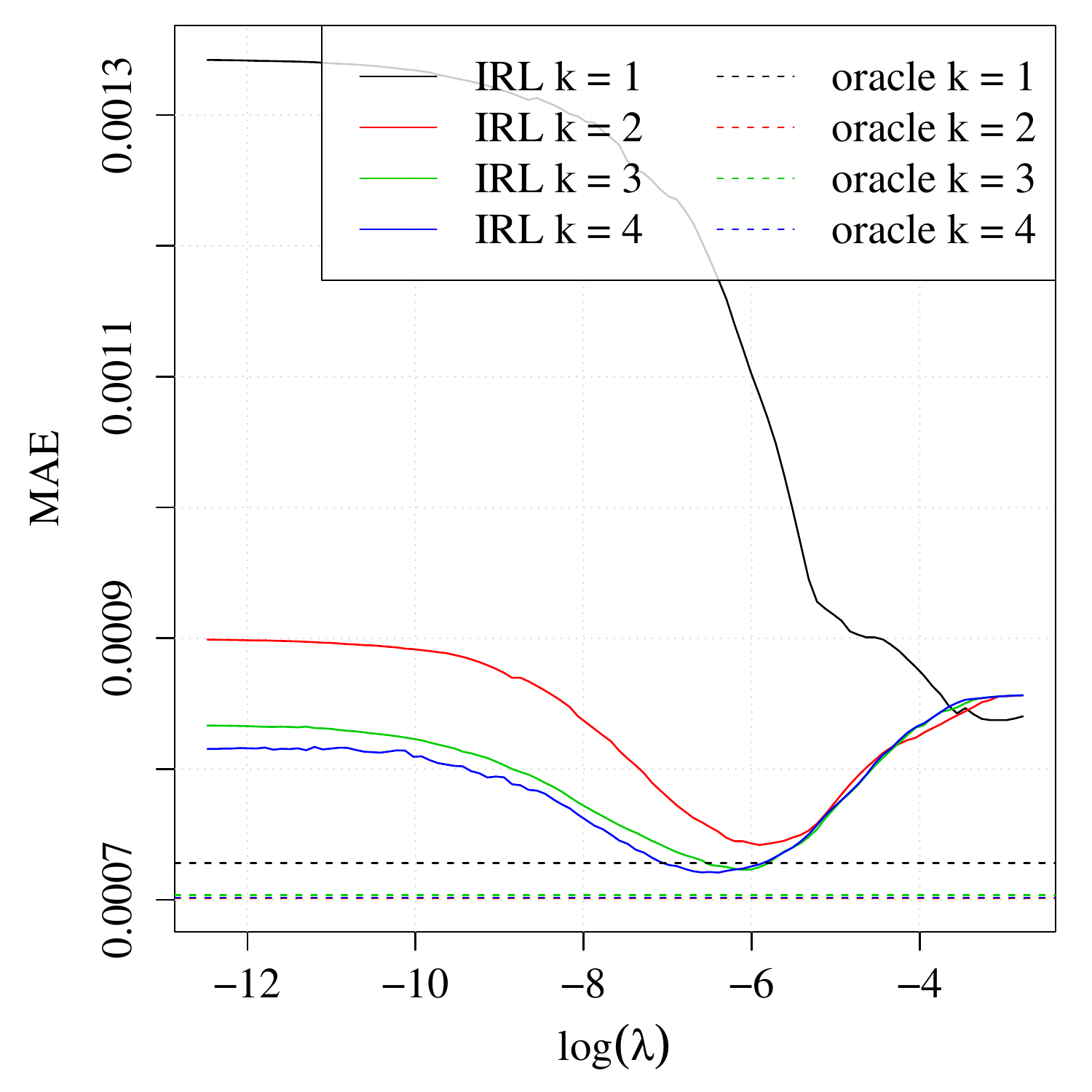}
   \caption{$n=600$}
  \label{fig_tarch_mae_sub2}
\end{subfigure}
 \caption{$\MAE$ for $n=300$ (\ref{fig_tarch_mae_sub1}) and 
$n=600$  (\ref{fig_tarch_mae_sub2}) of the iteratively reweighted lasso (IRL) method for several iterations $k\in \{1,2,3,4\}$, such as their
 oracle estimators for the AR-TARCH model.}
 \label{fig_tarch_mae}
\end{figure}

We compute the $1$-step ahead mean absolute forecast error ($\MAE$)
for $n=300$ and $n=600$. 
The simulation results with the corresponding oracles are given in Figure \ref{fig_tarch_mae}.
There we observe similar behaviour as for the AR-ARCH model in Figure \ref{fig_mae}. 
As there is a clear improvement in the MAE it shows that the iteratively reweighted lasso algorithm 
can work well for data with asymmetric volatility.

% Another issue that was not discussed so far is the convergence of the algorithm. 
% As mentioned above a plausible 
% stopping criteria for the algorithm ist 
% $\|\bssigma_n^{[k]} - \bssigma_n^{[k-1]}\|< \epsilon$ for a chosen vector norm $\|\cdot\|$ and $\epsilon$ clos to zero.
% We suggest the 
% TODO show first convergence properties, mentioning problem of finite sample non-convergence.

\section{Applications to electricity market data and metal prices returns}

In this section we briefly show two applications of the proposed model to real data.
For both applications a two-dimensional AR-ARCH model 
to the process $(\bsY_t)_{t\in \Z} = (Y_{1,t}, Y_{2,t})_{t\in \Z}$ is considered.

In the first application we use the hourly day-ahead electricity spot price for Germany/Austria 
at the European Power exchange (EPEX) as one process $(Y_{1,t})_{t\in \Z}$ and the hourly electricity load of Germany 
as $(Y_{2,t})_{t\in \Z}$. The considered time range is from 28.09.2010 to 17.04.2014. 
%, so we have $n=31128$ observations.
%2010-09-28 01:00:00 GMT" "2014-04-17
For the second example we take the hourly intra-day returns of gold and silver prices in U.S. dollar 
(from London Bullion Market Association), denoted as XAU/USD and XAG/USD.
Here $(Y_{1,t})_{t\in \Z}$ represents the gold and $(Y_{2,t})_{t\in \Z}$ the silver price returns. 
The data covers 12 years of observations from 01.01.2002 to 31.12.2013.

Note that electricity prices are known to have a strong correlation structure. In contrast, 
we expect either no or a very weak autoregressive dependency structure for the commodity returns.

For both applications we suppose that $\bsY_t$ follows an AR-ARCH model as given in \eqref{eq_main_ar_model}
and \eqref{eq_main_arch_model}. 
As the electricity data has usually a long memory we propose for the autoregressive parameters 
the lags $\II_{i,j}=\{1,\ldots, 700\}$ for $i,j\in \{1,2\}$ and similarly for 
the ARCH part parameters we take $\JJ_{i,j} = \{1, \ldots, 700\}$. 
This covers a memory of more than 4 weeks.
For the metal prices we take $\II_{i,j}=\{1,\ldots, 200\}$ for the conditional mean and $\JJ_{i,j} = \{1, \ldots, 200\}$ for 
the volatility part. The index sets are sufficiently large to capture possible weekly dependencies.
We consider the conservative BIC as information criterion and for the adaption parameter $\tau$ we take the lasso case with $\tau=0$. 
Then we apply the iteratively reweighted algorithm and stop after $R_{\max}=3$ iterations.
Hence we solve $dR_{\max} = 2\times3$ lasso problems in for each application. %The computation time for this full iterative problem is about 20 minutes on a standard $3GHz$ computer.

\begin{figure}[hbt!]
\centering
\begin{subfigure}[b]{0.49\textwidth}
 \includegraphics[width=1\textwidth, height=.62\textwidth]{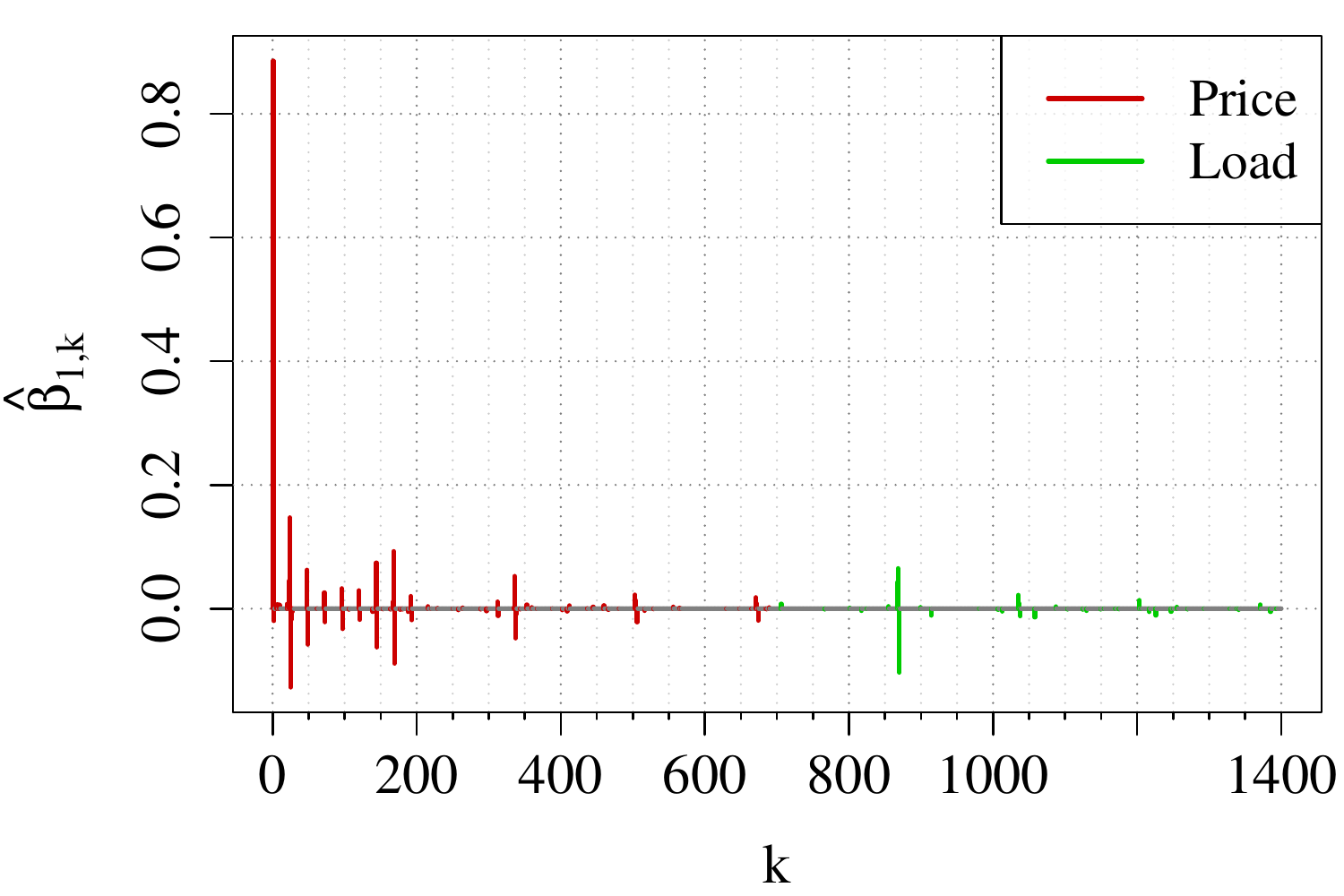}
   \caption{Estimated coefficients for the electricity price}
  \label{fig_example_sub1}
\end{subfigure}
\begin{subfigure}[b]{0.49\textwidth}
 \includegraphics[width=1\textwidth, height=.62\textwidth]{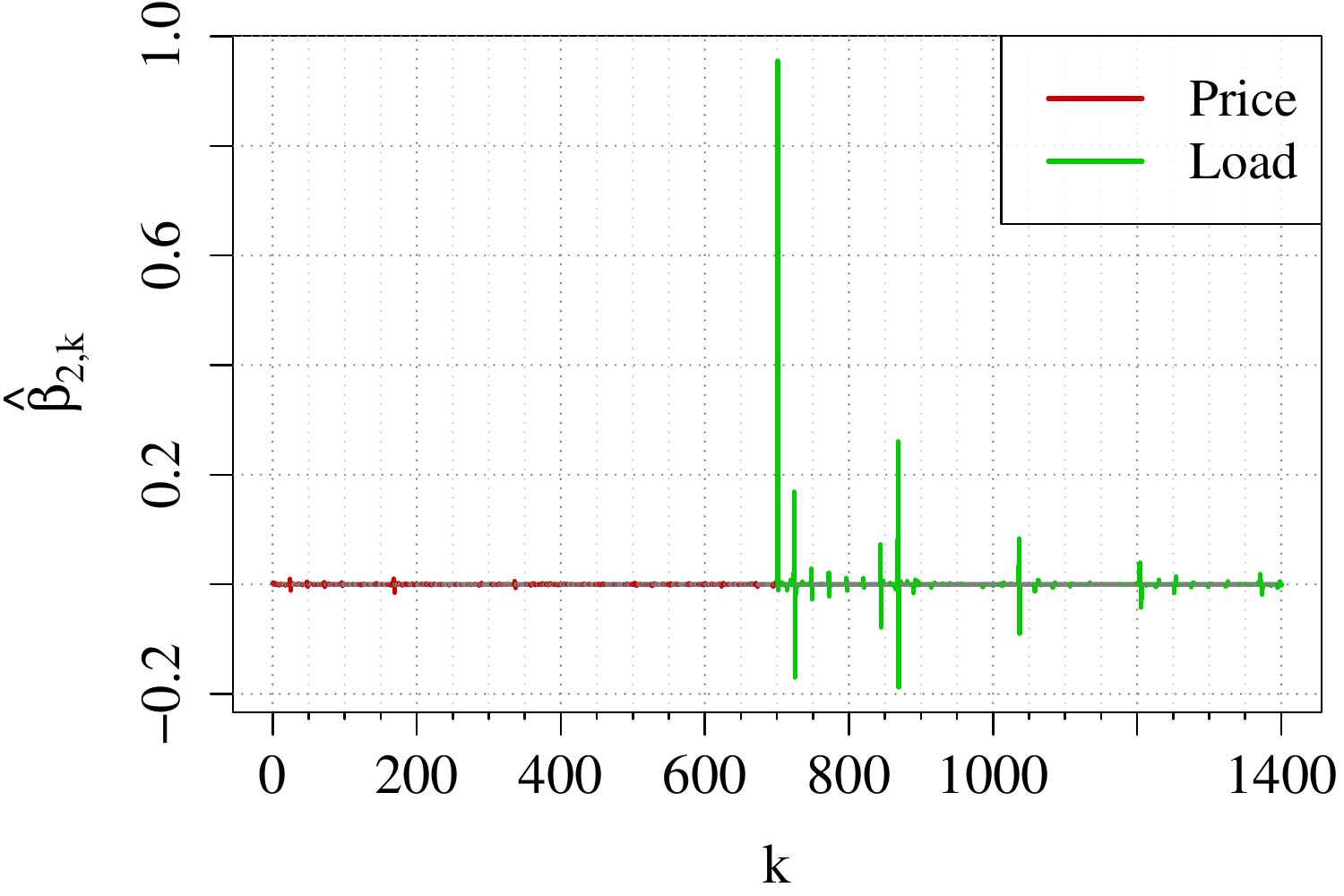}
   \caption{Estimated coefficients for the electricity load}
  \label{fig_example_sub2}
\end{subfigure}
\begin{subfigure}[b]{0.49\textwidth}
 \includegraphics[width=1\textwidth, height=.62\textwidth]{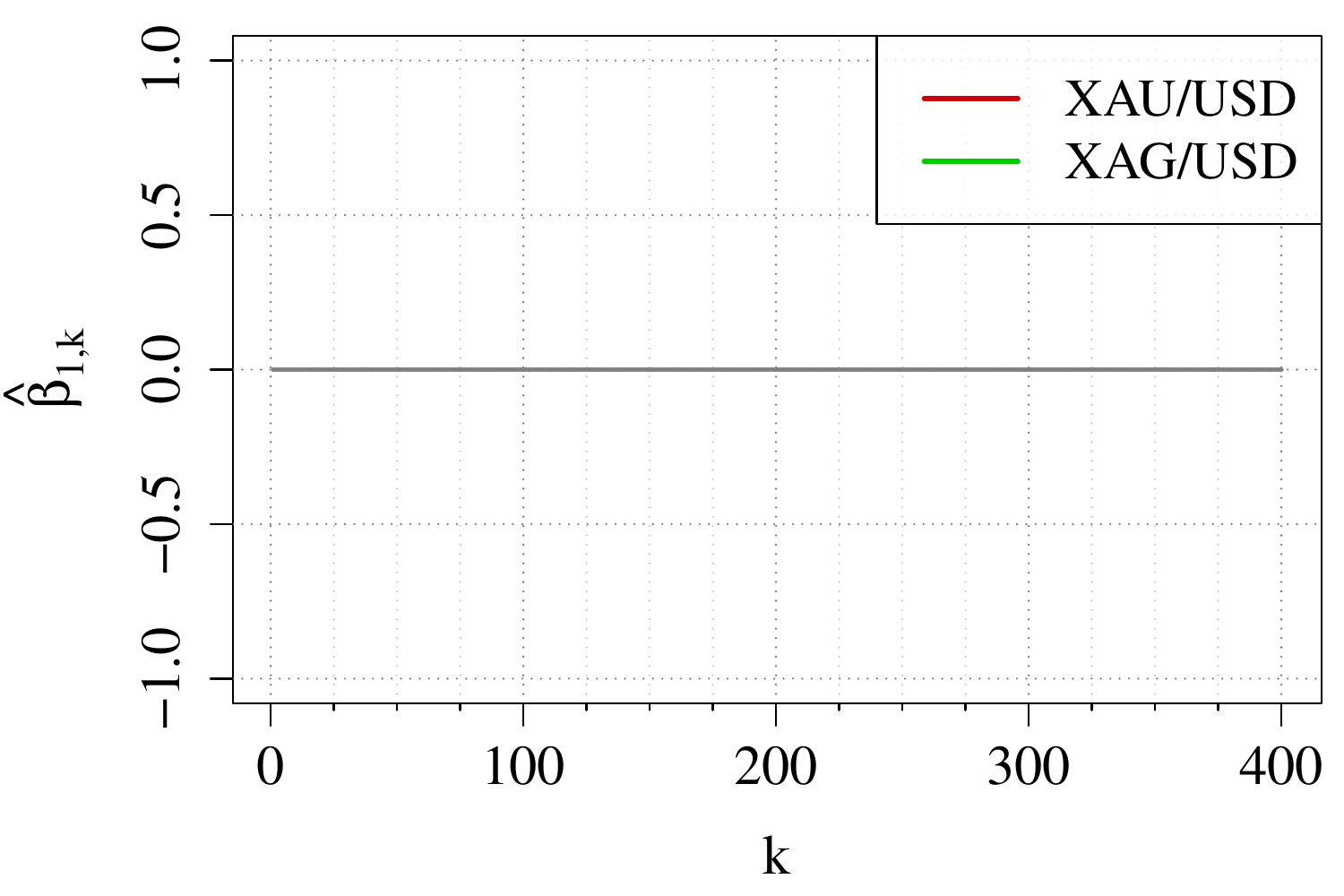}
   \caption{Estimated coefficients for the gold price returns}
  \label{fig_example_sub3}
\end{subfigure}
\begin{subfigure}[b]{0.49\textwidth}
 \includegraphics[width=1\textwidth, height=.62\textwidth]{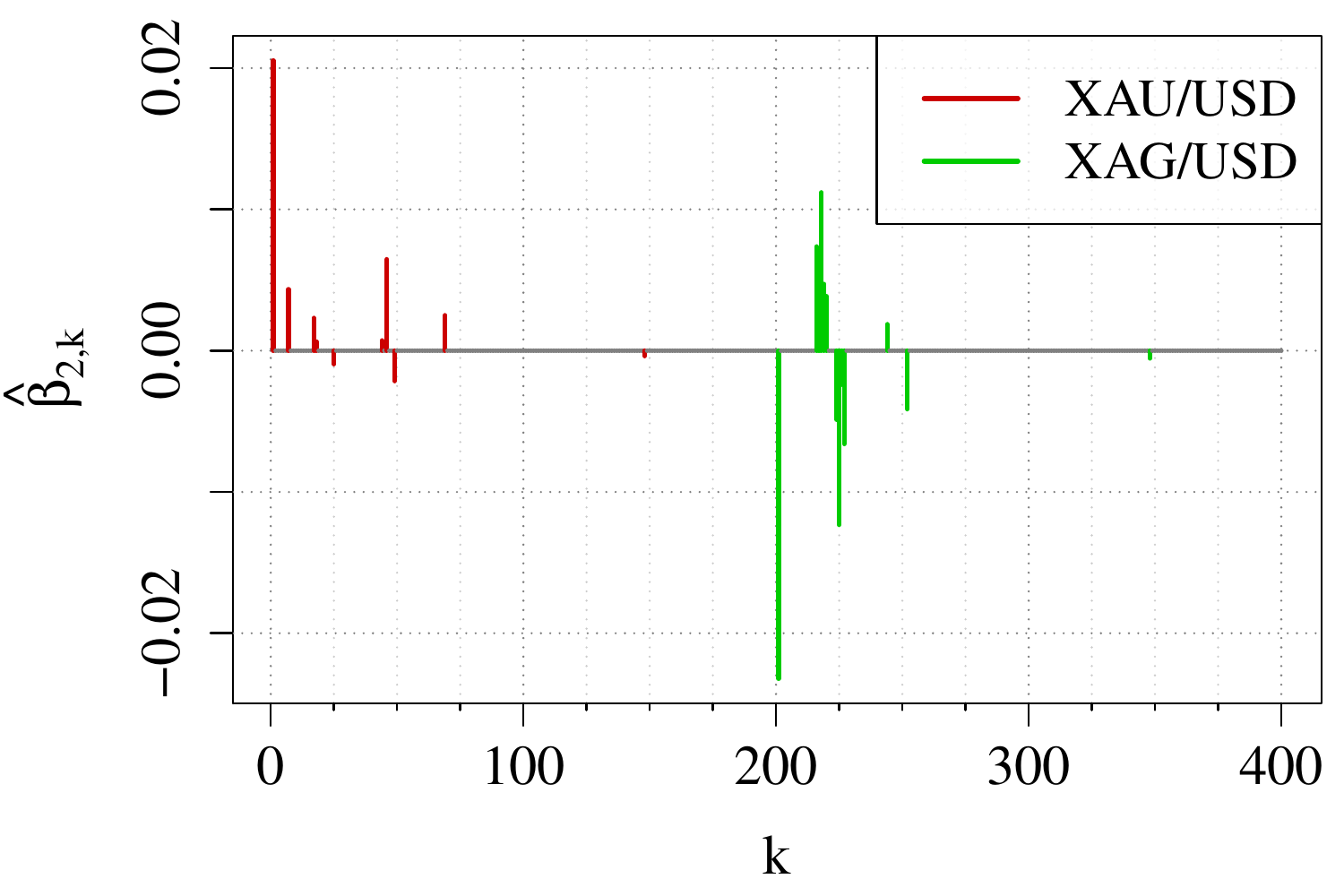}
   \caption{Estimated coefficients for the silver price returns}
  \label{fig_example_sub4}
\end{subfigure}

\caption{Estimated parameters $\what{\bsbeta}_{i,n}$ %for $i\in \{1,2\}$ 
for the electricity market model in \ref{fig_example_sub1} and \ref{fig_example_sub2} and for the metal prices in
\ref{fig_example_sub3} and \ref{fig_example_sub4}.
 }
 \label{fig_electricity_tvals}
\end{figure}

The estimated $\what{\bsbeta}_{i,n}$ for $i\in \{1,2\}$ and both applications are given in Figure \ref{fig_electricity_tvals}.
%TODO PARAMETER ELECTR: 133, LOAD: 416
%TODO PARAMETER GOLD:0, SILVER 23 
Here we see that in general most of the parameters are not included in the model.
For the electricity price model there are 133 parameter included and for the load model 416.
This matches a proportion of included parameters of $9.5\%$ and $29.7\%$.
We see that the complex autocorrelation structure that is driven by daily and weekly seasonal effects is well captured.

For the metal prices we observe a different situation. 
HHere the gold price returns have no significant parameter at all.
However, the silver time series exhibits a weak dependency structure. 
Most distinct is the first lag pattern with a positive coefficient for 
the gold returns and a negative one for the silver returns. Furthermore, we have 
two small silver coefficient clusters, a positive one around a lag of 16 hours and a negative one 
around a lag of 24.

% Using a estimated model we can perform asymptotic inference using theorem \ref{thm_asymtotic} and 
% forecast the time series. A $120$-step ahead forecast is given in Figure \ref{fig_electricity_forc}.
%  \begin{figure}[hbt!]
% \centering
%  \includegraphics[width=1\textwidth, height=.3\textheight]{price_load_forecast.pdf}
%  \caption{Observed sample with forecasts and their prediction region given with dotted and dashed lines.}
%  \label{fig_electricity_forc}
% \end{figure}
% There we can see how the conditional mean and variance structure influences the forecasts. 
% The complex autocorrelation structure that 
% is driven by multiple seasonal effects is well captured.
 
% % latex table generated in R 3.0.2 by xtable 1.7-1 package
% % Wed May 28 16:11:24 2014
% \begin{table}[ht]
% \centering
% \begin{tabular}{r|rr|rr|rr}
%   \hline
%  & prop. inrel. &  prop. rel.  & prop. inrel. &  prop. rel. & prop. inrel. &  prop. rel.\\ 
%   \hline
% lasso 		& 0.018 & 0.274  & 0.050 & 0.486  &   0.125 0.650  	\\
% lasso-homo 	& 0.022 & 0.286 &  0.046 & 0.462  &   0.129 0.652 \\
%    \hline
% \end{tabular}
% \end{table}

\section{Summary and Conclusion} \label{Summary}

An iterative algorithm to solve adaptive lasso time series problems with conditionally heteroscedastic residuals is described. 
We showed the sign consistency and asymptotic normality in a rather general time series setting.
The asymptotic theory shows that a significant estimation improvement is possible if the conditional heteroscedasticity 
is considered. We discussed %\textcolor{red}{TH: discussed} 
the application to AR-ARCH type models and showed % \textcolor{red}{TH: showed} 
applications 
to intra-day electricity market and commodity data.

The simulation studies underline the asymptotic results.
Additionally, we showed % \textcolor{red}{TH: showed} 
that considering the heteroscedasticity in high-dimensional settings with unknown parameter specification
is more important than in cases where the true underlying model is known, as it 
can substantially improve %\textcolor{red}{TH: \st{the}} 
forecasting performance.
This observation will likely have a strong impact %\textcolor{red}{TH: \st{to} on} 
on high-dimensional time series modelling,
as almost every time series exhibits conditional heteroscedasticity, especially in economics and finance. 

The asymptotic theory shows that only two iterations are required for receiving optimal asymptotic behaviour.
Thus, the algorithm is suitable for applications, as the computational effort is only doubled in comparison to 
standard homoscedastic situations.

For future research it might be important to analyse the mentioned model extensions more carefully. Another
very important issue is to identify the optimal penalty parameter $\lambda_n$ in high-dimensional time series settings.
A different direction of further research might concern the robustness of the algorithm. The 
% \textcolor{red}{TH: performed} 
performed simulation study %\textcolor{red}{TH: \st{carried out}} showed 
carried out that the algorithm works %\textcolor{red}{TH: works} 
well in a finite sample setting. % and even with quite heavy tailed residuals.
However, in a heavy tailed situation , it might be worth considering the LAD-lasso (see e.g. \cite{wang2007robust}), 
which minimises the sum of the absolute residuals, instead of their squares (as in lasso type algorithms). 
Another direction that seems to be a promising extension concerns the $\ell_q$ penalty itself. 
% \textcolor{red}{TH: Den Satz kann ich nicht entziffern.} 
An extension to elastic net estimators, which combine $\ell_1$ and $\ell_2$ penalties, could also improve estimation power. Recently 
\cite{gefang2014bayesian} applied the elastic net method successfully to homoscedastic multivariate AR processes.

% DISCUSS:
% 
% TUNING PARAMETER / SELECTION of INITIAL
% 
% HIGHER MOMENTS(!?)
% 
% TAIL PROPERTY (!?)
% 
% AUTOMATIC learner for PERIODIC 
% 
% THEORY for structural breaks...
% 
% for more robust LAD-lasso, see \cite{wang2007robust}.

\section{Appendix}

\begin{proof}[Proof: Theorem \ref{thm_asymtotic}]
 We show the sign consistency first and then the asymptotic normality. As mentioned, the proof extents mainly methods
 from \cite{wagener2013adaptive}.
 Denote $e_{n,j}$ the $j$'th unit vector in $\R^{q_n}$, 
  $a =_s b$ holds if $\sign(a) = \sign(b)$ and %$\|\cdot\|_F$ the Frobenius norm, 
  $\|\cdot\|_{\psi_d}$ Orlicz norm with ${\psi_d}(x) = \exp(x^d) -1$. %We consider only the case where we assume that \ref{asump_tail} is true.
%  If \ref{asump_tailalt} is true, the same follows by arguments illustrated in \cite{wagener2013adaptive}.
 In proof we will introduce at some points several constants $c_k$ that are positive.
 
 %We assume that 
 Let $k>1$ and assume that the theorem holds for $k-1$.
%  so that 
%  $\| \bsbeta_n^{[k-1]} - \bsbeta_n^0 \|_2 = \OO_P(\frac{a_n}{\sqrt{n}})$ for some $(a_n)_{n \in \N}$ such that 
%  $\| \bsbeta_n^{[k-1]} - \bsbeta_n^0 \|_2 \to 0$.  
 Following the Karush-Kuhn-Tucker conditions we have that
\begin{equation*}
 (\bsY_n - \bsX_n \bsbeta )' (\bsW_n^{[k-1]})^2 (\bsY_n - \bsX_n \bsbeta )
+ \lambda_n  \bsv_{n}' |\bsbeta|
\end{equation*}
is minimised by $\bsbeta = (\bsbeta(1)', \bsnull')' \in \R^{p_n}$ if and only if
$$X_j(1)' (\bsW_n^{[k-1]})^2 (\bsY_n - \bsX_n \bsbeta ) = \frac{\lambda_n}{2} v_j \sign(\beta_j) \text{ if } \beta_j\neq 0 \ \ \ \text{ and}$$
 %$|X_j'W_n^2 (\bsY_n - \bsX_n \bsbeta )|< \frac{\lambda_n}{2} v_j$ 
$$|X_j(1)' (\bsW^{[k-1]}_n)^2 (\bsY_n - \bsX_n \bsbeta )|< \frac{\lambda_n}{2} v_j \text{ if } \beta_j = 0$$
holds.
Thus, we have the estimator $\bsbeta^{[k]}_n=(\bsbeta^{[k]}_n(1)',\bsnull')' \in \R^{p_n}$ where
\begin{align}
\bsbeta^{[k]}_n(1) = \bsbeta_n^0(1) + \frac{1}{n} (\wtilde{\bsGamma}_n^{[k]}(1) )^{-1} \bsX_n(1) (\bsW_n^{[k-1]})^2 %\bsSigma_n^0 
\bseps_n^0 - 
\frac{\lambda_n}{2n} (\wtilde{\bsGamma}_n^{[k]}(1))^{-1} \bss^0_n(1)
\label{eq_proof_beta_exp} 
\end{align}
where $\bss^0_n(1) = (v_1, \ldots, v_{q_n})' \sign(\bsbeta_n^0(1))$.

Now we define the expressions
\begin{align*}
\eta_{1,j} &= e_{n,j}' (\wtilde{\bsGamma}_n^{[k]}(1))^{-1} \bsX_n(1)' (\bsW_n^{[k-1]})^2  \bseps_n^0\\ 
\eta_{2,j} &= e_{n,j}' (\wtilde{\bsGamma}_n^{[k]}(1))^{-1} \bss^0_n(1)\\
\eta_{3,j} &= X_j(1)' (\bsW_n^{[k-1]})^2 
(I_n - n^{-1} \bsX_n(1) (\wtilde{\bsGamma}_n^{[k]}(1))^{-1} \bsX_n(1)'  (\bsW_n^{[k-1]})^2 )  \bseps_n^0 \\
\eta_{4,j} &= \lambda_n (2n)^{-1} X_j(1)' (\bsW_n^{[k-1]})^2 
\bsX_n(1) (\wtilde{\bsGamma}_n^{[k]}(1))^{-1}   \bss^0_n(1) . 
\end{align*}

As in \cite{wagener2013adaptive} we can use the argument of \cite{huang2008adaptive} that the KKT conditions are satisfied if 
\begin{align}
| \eta_{3,j} - \eta_{4,j}| < \frac{\lambda_n}{2} v_j 
\label{eq_proof_KKT34} 
\end{align}
holds for all $j>q_n$.

Hence we receive with \eqref{eq_proof_beta_exp} 
 and \eqref{eq_proof_KKT34} that
$$P\left(\bsbeta_n^{[k]} \neq_s \bsbeta_n^0\right) \leq P(A_1) + P(A_2) +P(A_3) +P(A_4), \ \ \ \text{ with}$$
\begin{align*}
A_1 &= \left\{ \frac{1}{n}| \eta_{1,j}| \geq \frac{1}{2}|\beta^0_j| \text{ for some } j\leq q_n \right\}, \ \ \ 
A_2 = \left\{ \frac{\lambda_n}{n}| \eta_{2,j}| \geq |\beta^0_j| \text{ for some } j\leq q_n \right\}, \\
A_3 &= \left\{ | \eta_{3,j}| \geq \frac{\lambda_n}{4} v_j \text{ for some } j> q_n \right\} \ \text{ and }  
A_4 = \left\{ | \eta_{4,j}| \geq \frac{\lambda_n}{4} v_j \text{ for some } j> q_n \right\} .
\end{align*}
So we only need to show that $P(A_j)\to 0$ as $n\to \infty$.

Regarding $P(A_1)$ we have with definition of $b_n$ (see \ref{asump_bn}) that
\begin{align}
P(A_1)  &\leq P\left( \frac{1}{n} \max_{1\leq j \leq q_n} | \eta_{1,j}| \geq \frac{b_n}{2} \right) \nonumber \\
&\leq P\left( \frac{1}{n} \max_{1\leq j \leq q_n} | \eta^{0,\infty}_{1,j}| \geq \frac{b_n}{4} \right) 
+ P\left( \frac{1}{n} \max_{1\leq j \leq q_n} | \eta_{1,j} - \eta^0_{1,j}| \geq \frac{b_n}{8} \right) 
+ P\left( \frac{1}{n} \max_{1\leq j \leq q_n} | \eta^0_{1,j} - \eta^{0,\infty}_{1,j}| \geq \frac{b_n}{8} \right) 
\label{eq_proof_A1}
\end{align}
where $\eta^0_{1,j} = e_{n,j}' (\wtilde{\bsGamma}_n^0(1))^{-1} \bsX_n(1)' (\bsW_n^0)^2 \bseps^0_n$
and $\eta^{0,\infty}_{1,j} = e_{n,j}' (\wtilde{\bsGamma}_n^0(1))^{-1} \bsX_n(1)' (\bsW_n^0)^2 \bseps^0_{n,\infty}$.

 For estimating the first term in \eqref{eq_proof_A1} we observe that
\begin{align*}
 \left\| \frac{1}{\sqrt{n}} e_{n,j}' (\wtilde{\bsGamma}_n^0(1))^{-1} \bsX_n(1)' 
(\bsW_n^0)^2 \right\|_2
 &\leq  \left \| ( \wtilde{\bsGamma}^0_n(1) )^{-1} \right \|_2 \left \| \frac{1}{\sqrt{n}} \bsX_n(1)' \right \|_2 \| \bsW_n^0 \|^2_2  \\
 &\leq \left \| ( \wtilde{\bsGamma}^0_n(1) )^{-1} \right \|_2 \left \|  \bsGamma^0_n(1)   \right \|_2^{\frac{1}{2}}  \left \| \bsW_n^0 \right \|^2_2 
  \leq   \left \| ( \wtilde{\bsGamma}^0_n(1) )^{-1} \right \|_2  \left \|  \bsGamma^0_n(1)   \right \|_2^{\frac{1}{2}}   \sigma^2_{\min}
\end{align*}
for sufficiently large $n$ with $\|\bsW_n^0 \|_2 \leq \sigma_{\min}$ by \ref{asump_variance_bounds}.
Furthermore by assumption \ref{asump_eigenval} we know that 
\begin{equation}
\| \bsGamma_n^0(1)\|^{\frac{1}{2}}_2 = \OO_P(1) \ \text{ and } \
\|  (\wtilde{\bsGamma}_n^{0}(1))^{-1} \|_2 = \OO_P(1) 
\label{eq_proof_matest} .
\end{equation}
Thus we %receive that With assumption \ref{asump_eigenval} we 
get that
$$ P\left( \left\| e_{n,j}' (\wtilde{\bsGamma}_n^0(1))^{-1} \bsX_n(1)' 
(\bsW_n^0)^2 \right\|_2 \leq 
 \lambda_{1,\min}^{-1}  \sqrt{ \lambda_{0,\max} } \sigma_{\min}  \right) \to 1 $$
 for $n\to \infty$.
With Lemma 1 (i) of \cite{huang2006adaptive} and tail assumption \ref{asump_tail} we can deduce that 
\begin{align}
  \left \| \frac{1}{\sqrt{n}} \eta_{1,j}^{0,\infty} \right \|_{\psi_d} \leq
\left \| \frac{1}{\sqrt{n}} e_{n,j}' (\wtilde{\bsGamma}_n^0(1))^{-1} \bsX_n(1)' 
(\bsW_n^0)^2 \bseps_{\infty,n}^0 \right \|_{\psi_d} \leq c_1 \log(n)^{\bsone\{d=1\}}
\label{eq_proof_for_A1}
\end{align}
for sufficiently large $n$, as $\|X\|_2 \leq c \|X\|_{\psi_d}$ for some $c>0$.

Thus, we can conclude with Markov inequality, Lemma 2.2.2 of \cite{van1996weak} and \eqref{eq_proof_for_A1} that
\begin{align}
 P\left( \frac{1}{n} \max_{1\leq j \leq q_n} | \eta^{0,\infty}_{1,j}| \geq \frac{b_n}{4} \right) 
 &\leq P\left( \psi_d \left( \frac{\max_{1\leq j \leq q_n} | \eta^{0,\infty}_{1,j}|}{\|\max_{1\leq j \leq q_n} | \eta^{0,\infty}_{1,j}|\|_{\psi_d}}  \right)
 \geq \psi_d\left( \frac{b_n n}{4 \| \max_{1\leq j \leq q_n} | \eta^{0,\infty}_{1,j}| \|_{\psi_d}} \right) \right) \nonumber \\
&\leq \psi_d\left( \frac{b_n n}{4 \| \max_{1\leq j \leq q_n} | \eta^{0,\infty}_{1,j}| \|_{\psi_d}}    \right)^{-1} \nonumber \\
&\leq \psi_d\left( \frac{b_n n}{4 c_{2} {\psi_d}^{-1}(q_n)  \max_{1\leq j \leq q_n} \|  \eta^{0,\infty}_{1,j} \|_{\psi_d}}    \right)^{-1} \nonumber \\
&\leq \psi_d\left( \frac{b_n \sqrt{n}}{4 c_{2} \log(1+q_n)^{\frac{1}{d}} c_2 \log(n)^{\bsone\{d=1\}}  }    \right)^{-1} 
\label{eq_proof_A1_final}
% &\leq \left(\exp\left(  \frac{b^d_n \sqrt{n}^d}{4^d c^d_{2} \log(1+q_n) c^d_0 \log(n)^{\bsone\{d=1\}}  }  \right) - 1 \right)^{-1}
\end{align}
as $\psi_d^{-1}(x) = \log(1+x)^{\frac{1}{d}}$. 
Hence by assumption \ref{asump_convergence} we have
$P\left( \frac{1}{n} \max_{1\leq j \leq q_n} | \eta^{0,\infty}_{1,j}| \geq \frac{b_n}{4} \right) \to 0$. 

If assumption \ref{asump_tail} is not satisfied we can not use 
equation \eqref{eq_proof_for_A1} to derive that
$P\left( \frac{1}{n} \max_{1\leq j \leq q_n} | \eta^{0,\infty}_{1,j}| \geq \frac{b_n}{4} \right) \to 0$. 
But we can conclude with Chebyshev's inequality and \eqref{eq_proof_B1_conv_to_N01} shown below that
\begin{align*}
P\left( \frac{1}{n} \max_{1\leq j \leq q_n} | \eta^{0,\infty}_{1,j}| \geq \frac{b_n}{4} \right) 
&\leq \frac{16}{n^2 b_n^2} \E( \max_{1\leq j \leq q_n} | \eta^{0,\infty}_{1,j}|^2  ) = \OO_P\left(\frac{1}{n b_n^2}\right).
% &\leq \frac{16}{n^2 b_n^2} \E( \max_{1\leq j \leq q_n} | e_{n,j}' (\wtilde{\bsGamma}_n^0(1))^{-1} \bsX_n(1)' (\bsW_n^0)^2 \bseps_{\infty,n}^0 |^2  ) 
\end{align*}
Thus even without assumption \ref{asump_tail} it holds with assumption \ref{asump_convergence} that
$P\left( \frac{1}{n} \max_{1\leq j \leq q_n} | \eta^{0,\infty}_{1,j}| \geq \frac{b_n}{4} \right) \to 0$. 
However, note that either \ref{asump_tail} or \ref{asump_tailalt} is required for estimating the probability of $A_3$ in a similar situation.

% Additionally we have  
% \begin{equation}
% \| \bseps^0_n - \bseps_{n, \infty}^0 \|_2 = \left \|\sum_{j=p_n+1}^\infty \beta_k X_{i,k} \right \|_2 
% % \leq
% % \left \|\sum_{j=p_n+1}^\infty \beta_k X_{i,k} \right \|_1 
% \to 0
% \label{eq_proof_A1_eps0diff}
% \end{equation}
% as $\sum_{j=p_n+1}^\infty |\beta_k| < \infty $ and $n\to \infty$.

For the second term in \eqref{eq_proof_A1} we proceed as in \cite{wagener2013adaptive}.
% we can show that 
% $P\left( \frac{1}{n} \max_{1\leq j \leq q_n} | \eta^0_{1,j}| \geq \frac{b_n}{4} \right) \to 0 $.
We get
\begin{align}
| \eta_{1,j} - \eta^0_{1,j}| \leq& 
\left| e_{j,n}' \left(  (\wtilde{\bsGamma}_n^{0}(1))^{-1} \bsX_n(1)'  ( (\bsW_n^0)^2 -  (\bsW_n^{[k-1]})^2 )
+ (\wtilde{\bsGamma}_n^{0}(1) - \wtilde{\bsGamma}_n^{[k]}(1) ) \bsX_n(1)' (\bsW_n^{[k-1]})^2 \right) \bseps_n^0 \right| \nonumber \\ 
\leq& 
 \|    ( (\bsW_n^0)^2 -  (\bsW_n^{[k-1]})^2 ) \bsX_n(1) (\wtilde{\bsGamma}_n^{0}(1))^{-1} e_{j,n} \|_2 \| \bseps_n^0 \|_2
+ \|  (\bsW_n^{[k-1]})^2 \bsX_n(1)  \|_2 \| \bseps_n^0 (\wtilde{\bsGamma}_n^{0}(1) - \wtilde{\bsGamma}_n^{[k]}(1) ) \|_2 \nonumber \\ 
\leq& 
\|(\bsW_n^0)^2 -  (\bsW_n^{[k-1]})^2\|_2 \|n \bsGamma_n^0(1)\|^{\frac{1}{2}}_2  
\|  (\wtilde{\bsGamma}_n^{0}(1))^{-1} \|_2 \|  \bseps_n^0\|_2  \nonumber \\
&+ \| (\bsW_n^0)^2 \|_2 \|n \bsGamma_n^0(1)\|^{\frac{1}{2}}_2 \|  (\wtilde{\bsGamma}_n^{0}(1))^{-1} - (\wtilde{\bsGamma}_n^{[k]}(1))^{-1} \|_2 \| 
 \bseps_n^0\|_2 . \label{eq_eta1}
\end{align}
All these appearing single norms we will estimate now.

For the estimation of $\|(\bsW_n^0)^2 -  (\bsW_n^{[k-1]})^2\|_2$ 
% we recall that 
% $\sigma_t = g(\bsalpha_{\infty,t}^0, \bsL_{\infty,t})$ and denote 
% $\wtilde{\sigma}_{n,t}( \bsbeta ) =  g_n( \what{\bsalpha}_n(\bsbeta; \bsX_n, \bsY_n) , \what{\bsL}_{n,t}(\bsbeta; \bsX_n, \bsY_n) ) $.
% %$\sigma_t = g(\bsalpha_{\infty,t}^0, \bsL_{\infty,t})$ 
% Denote $\wtilde{\bsW}_n^0$ the weight matrix that is given by 
% $\diag( \wtilde{\sigma}_{n,1}( \bsbeta_n^{0}),\ldots,\wtilde{\sigma}_{n,n}( \bsbeta_n^{0})  )$.
% % Then we have
% % $$ \left|\frac{1}{\sigma_t^2} - \frac{1}{ \wtilde{\sigma}_{n,t}( \bsbeta_n^{[k-1]} )^2 }\right|
% % \leq <
% % \left|\frac{1}{\sigma_t^2} - \frac{1}{\wtilde{\sigma}_{n,t}( \bsbeta_n^{0} )^2}\right| 
% % + \left|\frac{1}{\wtilde{\sigma}_{n,t}( \bsbeta_n^{0} )^2} - \frac{1}{\wtilde{\sigma}_{n,t}( \bsbeta_n^{[k-1]} )^2}\right| $$
% Then we have
% $$ \|(\bsW_n^0)^2 -  (\bsW_n^{[k-1]})^2\|_2
% \leq 
% \|(\bsW_n^0)^2 -  (\wtilde{\bsW}_n^{0})^2\|_2 
% + \|(\wtilde{\bsW}_n^0)^2 -  (\bsW_n^{[k-1]})^2\|_2 $$
% by triangle inequality.
% Using
we get directly with assumption  \ref{asump_variance_bounds}, 
% series expansion of $x \mapsto \frac{1}{x^2}$ 
 \ref{asump_variance_mom} and \ref{asump_variance_estimators}
that
\begin{align}
\|(\bsW_n^0)^2 -  (\bsW_n^{[k-1]})^2\|_2 = \OO_P \left( \frac{h_n}{\sqrt{n}} \right).
\label{eq_proof_A1_W0_Wk} 
\end{align}

For estimating $\| \bseps_n^0\|_2$  
% remember that the two vectors %$\bseps_n = \bsY_n - \bsX_n \bsbeta_n$, 
% $\bseps_n^0 = \bsY_n - \bsX_n \bsbeta^0_n$ and $\bseps_{\infty,n}^0 = (\eps_1, \ldots, \eps_n)$ with
%  $\eps_t = Y_t - \bsX_{\infty,t} \bsbeta^0_\infty$ for $t\in \{1, \ldots, n\}$ are in general different from each other.
% The
 the triangle inequality yields
$ \| \bseps_n^0\|_2 \leq \| \bseps_n^0 - \bseps_{\infty,n}^0  \|_2 + \| \bseps_{\infty,n}^0 \|_2$.
We have 
\begin{align}
\|\bseps_n^0 - \bseps_{\infty,n}^0  \|_2 = \bigg\|\sum_{j=p_n+1}^\infty \beta_k X_{i,k} \bigg\|_2 \to 0%= \OO_P(1)
\label{eq_proof_A1_eps0diff}
\end{align}
%  $\sum_{j=1}^\infty |\beta_k| < \infty $ 
 as $\sum_{j=p_n+1}^\infty |\beta_k| < \infty $ and $\| \bseps_{\infty,n}^0 \| = \OO_P(\sqrt{n})$ by law of large numbers. Thus we get
\begin{align}
 \| \bseps_n^0\|_2 = \OO_P(\sqrt{n}). 
\label{eq_proof_A1_eps0}
 \end{align}
Further we have $\|\bsW_n^{[k-1]}\|_2 = \OO_P(1)$ by assumption \ref{asump_variance_bounds}.

Next, we have as in \cite{wagener2013adaptive} that with assumption \ref{asump_eigenval},
and equations   \eqref{eq_proof_matest}  and \eqref{eq_proof_A1_W0_Wk}
that
\begin{align*}
 \|  \wtilde{\bsGamma}_n^{0}(1) - \wtilde{\bsGamma}_n^{[k]}(1) \|_2 \leq 
\|\bsGamma_n(1) \|_2 \|(\bsW_n^0)^2 -  (\bsW_n^{[k-1]})^2\|_2
= \OO_P\left( \frac{ h_n}{\sqrt{n}} \right) .
\end{align*}
This leads to 
\begin{align}
\|  (\wtilde{\bsGamma}_n^{0})^{-1} - (\wtilde{\bsGamma}_n^{[k]})^{-1} \|_2
% \leq 
% \OO(\|(\wtilde{\bsGamma}_n^{0})^{-1} - (\wtilde{\bsGamma}_n^{[k]})^{-1}\|_2) 
% + 
= \OO_P\left( \frac{h_n}{\sqrt{n}} \right) 
\label{eq_proof_A1_G0_Gk} 
\end{align}
by using the the triangle inequality 
$$\|A^{-1}  - (A + B)^{-1} \|_2 \leq 
\|A^{-1}  - (A + B)^{-1} + A^{-1} B A^{-1} \|_2  + \|A^{-1} B A^{-1}\|_2 \leq 
\OO_P( \| B\|_2 ) + \|A\|^2_2 \|B\|_2 = \OO_P\left( \frac{ h_n}{\sqrt{n}} \right)
$$ for two matrices $A = \wtilde{\bsGamma}_n^{0}(1) $ 
with $B=\wtilde{\bsGamma}_n^{0}(1) - \wtilde{\bsGamma}_n^{[k]}(1)$
and the Taylor series expansion of $(A+B)^{-1}$ around $A^{-1}$. % [CHECK]

Using all the estimated norms 
(\eqref{eq_proof_matest}, \eqref{eq_proof_A1_W0_Wk}, \eqref{eq_proof_A1_eps0} and \eqref{eq_proof_A1_G0_Gk})
we receive for \eqref{eq_eta1} that
\begin{align*}
| \eta_{1,j} - \eta^0_{1,j}| \leq& 
\|(\bsW_n^0)^2 -  (\bsW_n^{[k-1]})^2\|_2 \|n \bsGamma_n^0(1)\|^{\frac{1}{2}}_2  
\|  (\wtilde{\bsGamma}_n^{0}(1))^{-1} \|_2 \|  \bseps_n^0\|_2  \nonumber \\
&+ \| (\bsW_n^0)^2 \|_2 \|n \bsGamma_n^0(1)\|^{\frac{1}{2}}_2 
\|  (\wtilde{\bsGamma}_n^{0}(1))^{-1} - (\wtilde{\bsGamma}_n^{[k]}(1))^{-1} \|_2 \|  \bseps_n^0\|_2 \nonumber  \\
 \leq& 
 \OO_P(\frac{h_n}{\sqrt{n}}) \OO_P(\sqrt{n}) \OO_P(1) \OO_P(\sqrt{n})
 + \OO_P(1) \OO_P(\sqrt{n} ) \OO_P(\frac{h_n}{\sqrt{n}}) \OO_P(\sqrt{n})
 = \OO_P(h_n \sqrt{n}) 
 . %\label{eq_eta1_est} 
\end{align*}
Thus we get,
$$ 
\frac{1}{n} 
\max_{1\leq j \leq q_n}| \eta_{1,j} - \eta^0_{1,j}|  = \frac{1}{n}\OO_P(h_n \sqrt{n})  =  \OO_P\left( \frac{h_n}{\sqrt{n}}\right).$$
This yields with assumption \ref{asump_convergence} 
that
$P\left( \frac{1}{n} \max_{1\leq j \leq q_n} | \eta_{1,j} - \eta^0_{1,j}| \geq \frac{b_n}{8} \right)
\to 0$ as $n\to \infty$. 
For the third term in \eqref{eq_proof_A1} we get 
with \ref{asump_variance_bounds}, \eqref{eq_proof_matest} and \eqref{eq_proof_A1_eps0diff} that
\begin{align*}
\frac{1}{\sqrt{n}} | \eta^0_{1,j} - \eta^{0,\infty}_{1,j} |
 &\leq \frac{1}{\sqrt{n}}  |e_{n,j}' (\wtilde{\bsGamma}_n^0(1))^{-1} \bsX_n(1)' (\bsW_n^0)^2 (\bseps_n^0 -  \bseps_{n,\infty}^0 ) | \\
 &\leq \| (\wtilde{\bsGamma}_n^0(1))^{-1} \|_2 
 \left\| \frac{1}{\sqrt{n}}  \bsX_n(1)' \right\|_2 \|\bsW_n^0 \|^2_2 \|\bseps_n^0 -  \bseps_{n,\infty}^0 \|_2 \\
 &\leq \| (\wtilde{\bsGamma}_n^0(1))^{-1} \|_2 
  \left\| \bsGamma_n^0 \right\|_2^{\frac{1}{2}}  \|\bsW_n^0 \|^2_2 \|\bseps_n^0 -  \bseps_{n,\infty}^0 \|_2 \to 0
 \end{align*}
as $n\to \infty$. 
So we have that 
$\frac{1}{n} | \eta^0_{1,j} - \eta^{0,\infty}_{1,j} | \to \infty$ as $n\to 0$.
This implies $P(A_1) \to 0$.

Now we consider $P(A_2) \leq P(\frac{\lambda_n}{n} \max_{1\leq j \leq q_n} | \eta_{2,j}| \geq b_n ) $.
We have $| \eta_{2,j}| \leq \| (\wtilde{\bsGamma}_n^{[k]})^{-1} \|_2 \|\bss^0_n(1)\|_2 $ for each $j \in \{1, \ldots, q_n\}$.
By Weyl's perturbation theorem for the matrices
 $(\wtilde{\bsGamma}_n^{0}(1))^{-1}$ and $(\wtilde{\bsGamma}_n^{[k]}(1))^{-1}$
we have for each ordered pair of eigenvalues that
$$ 
\left|\lambda_j\left( (\wtilde{\bsGamma}_n^{0}(1))^{-1} \right) - \lambda_j \left( (\wtilde{\bsGamma}_n^{[k]}(1))^{-1}  \right) \right|  
 \leq  \| (\wtilde{\bsGamma}_n^{0}(1))^{-1} - (\wtilde{\bsGamma}_n^{[k]}(1))^{-1} \|_2 $$
for each $j \in \{1, \ldots, q_n\}$.
As $\| (\wtilde{\bsGamma}_n^{0}(1))^{-1} - (\wtilde{\bsGamma}_n^{[k]}(1))^{-1} \|_2 \to 0$ 
in probability, we get with \ref{asump_eigenval} that
\begin{align}
\| (\wtilde{\bsGamma}_n^{[k]})^{-1} \|_2 \leq \lambda_{1, \min}^{-1} + c_3
\label{eq_proof_A2_Gammakinv} 
\end{align}
with probability arbitrarily close to 1 for sufficiently large $n$.

Furthermore, with assumption \ref{asump_bn} we have
\begin{align}
\|\bss^0_n(1)\|_2 \leq \sqrt{q_n} \sqrt{\max_{1\leq j \leq q_n} |\beta_{\text{init},j}|^{-\tau} } \leq \frac{ \sqrt{b q_n} }{\sqrt{b_n} } 
\label{eq_proof_A2_s1} .
\end{align}
% Thus,
% with a probability of at least $1-C_3$ for all $C_3, C_4>0$ for a sufficiently large $n$.
Hence we have with assumption \ref{asump_convergence}  that
$\leq P\left(\frac{\lambda_n}{n} \max_{1\leq j \leq q_n}  | \eta_{2,j}| \geq b_n \right) 
\leq P\left(\frac{\lambda_n}{n} c_{4} \frac{ \sqrt{b q_n} }{\sqrt{b_n} }   \geq b_n \right) 
\to 0$ as $n\to \infty$. 

For $A_3$ we receive similarly as for $A_1$ that
\begin{align}
P(A_3) \leq& P\left( \max_{q_n <  j\leq p_n} | \eta_{3,j}^{0,\infty} | \geq \frac{\lambda_n r_n} {8} \right)
+P\left( \max_{q_n <  j\leq p_n} | \eta_{3,j}^0 - \eta_{3,j}^{0, \infty} | \geq \frac{\lambda_n r_n} {16} \right)
\nonumber \\
&+P\left( \max_{q_n <  j\leq p_n} | \eta_{3,j} - \eta_{3,j}^0 | \geq \frac{\lambda_n r_n} {16} \right) +
P\left(\max_{q_n <  j\leq p_n} \bsbeta_{j,\text{init}}^{\tau} > r_n^{-1} \right)
\label{eq_proof_A3} 
\end{align}
where
\begin{align*}
\eta_{3,j}^0 &= X_j(1)' (\bsW_n^{0})^2 
(I_n - n^{-1} \bsX_n(1) (\wtilde{\bsGamma}_n^{0})^{-1} \bsX_n(1)'  (\bsW_n^{0})^2 )  \bseps_n^0 , \\
\eta_{3,j}^{0,\infty} &= X_j(1)' (\bsW_n^{0})^2 
(I_n - n^{-1} \bsX_n(1) (\wtilde{\bsGamma}_n^{0})^{-1} \bsX_n(1)'  (\bsW_n^{0})^2 )  \bseps_{n,\infty}^0 . 
\end{align*}

%is in general not independent from $\bseps_n$
As in \cite{wagener2013adaptive} 
we consider $\eta_{3,j}^{0,\infty} = H_{n,j}^0 \bseps_{n,\infty}^0$ with
$$H_{n,j}^0  =  X_j(1)' (\bsW_n^{0})^2 
(I_n - n^{-1} \bsX_n(1) (\wtilde{\bsGamma}_n^{0})^{-1} \bsX_n(1)'  (\bsW_n^{0})^2 ) . $$
Then we have for sufficiently large $n$ that 
\begin{align}
 \|H_{n,j}^0\|_2  &\leq  \|X_j(1)'\|_2 \| (\bsW_n^{0})^2 \|_2 
(1 + \| n^{-1} \bsX_n(1) (\wtilde{\bsGamma}_n^{0})^{-1} \bsX_n(1)' \|_2 \| (\bsW_n^{0})^2  \|_2 ) \nonumber \\
& = \OO_P(\sqrt{n}) \OO_P(1) (1+ \OO_P(1) \OO_P(1) ) = \OO_P(\sqrt{n}).
\label{eq_proof_A3_Hn0}
\end{align}
% . 
% that $\| H_n^0 \|_F \leq c_5 \sqrt{n}$
% for some $c_5 > 0$. 
% Triangle inequality gives
% $$\| \eta_{3,j}^0 \|_{\psi_d} = \|  H_{n,j}^0 \bseps_n^0 \|_{\psi_d} \leq \|  H_{n,j}^0 \bseps_{\infty,n}^0 \|_{\psi_d} 
% + \|  H_{n,j}^0 ( \bseps_n^0  - \bseps_{\infty,n}^0 )\|_{\psi_d} .$$
%$\bseps_n^0$
Now we receive we receive as in equation \eqref{eq_proof_for_A1} with \cite{huang2006adaptive} Lemma 1 (i) and assumption
\ref{asump_tail} that  
\begin{align}
\left\| \frac{1}{\sqrt{n}} \eta_{3,j}^{0,\infty} \right\|_{\psi_d}  
\leq
\left\| \frac{1}{\sqrt{n}} H_{n,j}^0 \bseps_{\infty,n}^0 \right\|_{\psi_d}  
\leq c_6 \log(n)^{\bsone{\{d=1\}}}. 
\label{eq_proof_A3_Hno_norm} 
\end{align}
Thus we get with Markov inequality, Lemma 2.2.2 of \cite{van1996weak}, \eqref{eq_proof_A3_Hno_norm} and 
assumption \ref{asump_convergence} similarly to \eqref{eq_proof_A1_final} 
that
\begin{align}
P\left( \max_{q_n <  j\leq p_n} | \eta_{3,j}^{0,\infty} | \geq \frac{\lambda_n r_n} {8} \right) 
\leq& 
\psi_d\left(  \frac{\lambda_n r_n} {8 c_{6} \psi_d^{-1}(q_n) \max_{q_n <  j\leq p_n} \| \eta_{3,j}^0 \|_{\psi_d} } \right)^{-1} 
\nonumber \\
\leq& 
\psi_d\left(  \frac{\lambda_n r_n} { c_{7} \sqrt{n} \log(1+p_n-q_n)^{\frac{1}{d}} \log(n)^{\bsone{\{d=1\}}} } \right)^{-1} 
\to 0  
\label{eq_proof_A3_eta30}
\end{align}
as $n\to \infty$.

If instead of \ref{asump_tail} the alternative assumption \ref{asump_tailalt} holds we can not use 
equation \eqref{eq_proof_A3_Hno_norm} to derive that it holds
$P\left( \max_{q_n <  j\leq p_n}  | \eta^{0,\infty}_{3,j}| \geq \frac{\lambda_n r_n}{8} \right) \to 0$. 
But can get with Chebyshev's inequality and \eqref{eq_proof_B1_conv_to_N01} shown below that
\begin{align*}
P\left(  \max_{q_n <  j\leq p_n}  | \eta^{0,\infty}_{3,j}| \geq \frac{\lambda_n r_n}{8} \right) 
&\leq \frac{64}{ \lambda_n r_n^2} \sum_{j=q_n+1}^{p_n} \E(  | \eta^{0,\infty}_{3,j}|^2  ) = \OO_P\left(\frac{n(p_n-q_n)}{ \lambda_n^2 r_n^2}\right)
% &\leq \frac{16}{n^2 \lambda_n r_n^2} \E( \max_{1\leq j \leq q_n} | e_{n,j}' (\wtilde{\bsGamma}_n^0(1))^{-1} \bsX_n(1)' (\bsW_n^0)^2 \bseps_{\infty,n}^0 |^2  ) 
.
\end{align*}
Thus with \ref{asump_tailalt} it holds 
$P\left(\max_{q_n <  j\leq p_n}  | \eta^{0,\infty}_{1,j}| \geq \frac{\lambda_n^2 r_n}{8} \right) \to 0$.

For estimating the second term in \eqref{eq_proof_A3} we 
note with \eqref{eq_proof_A1_eps0diff} and \eqref{eq_proof_A3_Hn0}
that
\begin{align*}
\frac{1}{\sqrt{n}} \left| \eta_{3,j}^0 - \eta_{3,j}^{0, \infty} \right| \leq
\left| \frac{1}{\sqrt{n}} H_{n,j}^0 ( \bseps_n^0  - \bseps_{\infty,n}^0 ) \right|
\leq \frac{c_8}{\sqrt{n}} \left\|  H_{n,j}^0 \|_2  \| \bseps_n^0  - \bseps_{\infty,n}^0  \right\|_{2} 
= \frac{1}{\sqrt{n}} \OO_P(\sqrt{n}) . 
\end{align*}
Hence we have with assumption \ref{asump_convergence} that
\begin{equation}
P\left( \max_{q_n <  j\leq p_n} | \eta_{3,j}^0 - \eta_{3,j}^{0, \infty}| \geq \frac{\lambda_n r_n} {16} \right) 
\leq P\left( \frac{\sqrt{n}}{\lambda_n r_n} \geq  c_{9} \right)   \to 0 
\label{eq_proof_A3_eta3diffinfty} 
\end{equation}
as $n\to \infty$.
Now we estimate the third term in \eqref{eq_proof_A3}.
As in \eqref{eq_eta1} we get the estimate % \max_{q_n <  j\leq p_n}
\begin{align}
  | \eta_{3,j}^0 - \eta_{3,j}|
\leq& 
\left| X_j(1)'\left( (\bsW_n^{0})^2 (I_n - n^{-1} \bsX_n(1) (\wtilde{\bsGamma}_n^{0})^{-1} \bsX_n(1)'  (\bsW_n^{0})^2 ) \right. \right. \nonumber \\
&- \left. \left. (\bsW_n^{[k-1]})^2 (I_n - n^{-1} \bsX_n(1) (\wtilde{\bsGamma}_n^{[k]})^{-1} \bsX_n(1)'  (\bsW_n^{[k-1]})^2
\right)\bseps_n^0 \right|  \nonumber \\
 \leq& 
  \| X_j(1) \|_2 \| (\bsW_n^{0})^2 - (\bsW_n^{[k-1]})^2  \|_2 \| \bseps_n^0 \|_2 \label{eq_proof_A3_eta3} \\
  &+  \| X_j(1) \|_2 \left\|  n^{-1} \left( \bsX_n(1) (\wtilde{\bsGamma}_n^{0})^{-1} \bsX_n(1)'  (\bsW_n^{0})^2 - 
  \bsX_n(1) (\wtilde{\bsGamma}_n^{[k]})^{-1} \bsX_n(1)'  (\bsW_n^{[k-1]})^2 \right) \right\|_2 \| \bseps_n^0 \|_2 \nonumber
% \| X_j(1) \|_2 \| (I_n - n^{-1} \bsX_n(1) (\wtilde{\bsGamma}_n^{0})^{-1} \bsX_n(1)'  (\bsW_n^{0})^2 \|_2  \bseps_n^0 -
%  \\
%  &\leq 
%  \| X_j(1) \|_2  \| (\bsW_n^{[k-1]})^2 - (\bsW_n^{0})^2 \|_2 \| \bseps_n^0 \|_2
% \| X_j(1) \|_2 \| (I_n - n^{-1} \bsX_n(1) (\wtilde{\bsGamma}_n^{0})^{-1} \bsX_n(1)'  (\bsW_n^{0})^2 \|_2  \bseps_n^0 -
%  \\
% = \OO\left( \frac{ \sqrt{q_n}}{\sqrt{n}}\right)
\end{align}
using estimates derived for $A_1$. 
For the lengthy norm in \eqref{eq_proof_A3_eta3} we get 
\begin{align*}
 &\left\|  n^{-1} \left( \bsX_n(1) (\wtilde{\bsGamma}_n^{0})^{-1} \bsX_n(1)'  (\bsW_n^{0})^2 - 
  \bsX_n(1) (\wtilde{\bsGamma}_n^{[k]})^{-1} \bsX_n(1)'  (\bsW_n^{[k-1]})^2 \right) \right\|_2 \\
  \leq& 
  \| (\bsW_n^{0})^2 - (\bsW_n^{[k-1]})^2  \|_2 \| \bsGamma_n(1)\|_2 \| (\wtilde{\bsGamma}_n^{[k]})^{-1}  \|_2  \|(\bsW_n^{[k-1]})^2\|_2 \\
&+ \|(\wtilde{\bsGamma}_n^{0})^{-1} - (\wtilde{\bsGamma}_n^{[k-1]})^{-1} \|_2  \| \bsGamma_n(1)\|_2  \|_2  \|(\bsW_n^{[0]})^2\|_2 \|_2  \|(\bsW_n^{[k-1]})^2\|_2 \\
&+ \| (\bsW_n^{0})^2 - (\bsW_n^{[k-1]})^2  \|_2  
\| \bsGamma_n(1)\|_2 \| (\wtilde{\bsGamma}_n^{[0]})^{-1}  \|_2  \|(\bsW_n^{[0]})^2\|_2 \\
 =& \OO_P\left(\frac{h_n}{\sqrt{n}}\right) \OO_P(1) \OO_P(1) \OO_P(1) =  \OO_P\left(\frac{h_n}{\sqrt{n}}\right) 
% +
% \OO_P(\frac{h_n}{\sqrt{n}) \OO_P(1) \OO_P(1) \OO_P(1) + 
% \OO_P(\frac{h_n}{\sqrt{n}) \OO_P(1) \OO_P(1) \OO_P(1) + 
\end{align*}  
by equation \eqref{eq_proof_A1_W0_Wk}, \eqref{eq_proof_A2_Gammakinv}, \eqref{eq_proof_matest}, \eqref{eq_proof_A1_G0_Gk},
$\|W_n^0\|_2 \leq \sigma_{\min}$ and $\|\bsW_n^{[k-1]}\|_2 = \OO_P(1)$ by assumption \ref{asump_variance_bounds}. Thus
we receive for \eqref{eq_proof_A3_eta3} 
with \eqref{eq_proof_A1_eps0} that 
\begin{align}
 | \eta_{3,j}^0 - \eta_{3,j}| = \OO_P(\sqrt{n}) \OO_P\left(\frac{h_n}{\sqrt{n}}\right) \OO_P(\sqrt{n})
 + \OO_P(\sqrt{n}) \OO_P\left(\frac{h_n}{\sqrt{n}}\right) \OO_P(\sqrt{n})  = \OO_P(h_n\sqrt{n})
\label{eq_proof_A3_fin}
 \end{align}
Hence we have with assumption \ref{asump_convergence} and \eqref{eq_proof_A3_fin} that
\begin{equation}
P\left( \max_{q_n <  j\leq p_n} | \eta_{3,j}^0 - \eta_{3,j}| \geq \frac{\lambda_n r_n} {16} \right) 
\leq P\left( \frac{h_n\sqrt{n}}{\lambda_n r_n} \geq  c_{10} \right)   \to 0 
\label{eq_proof_A3_eta3diff} 
\end{equation}
as $n\to \infty$. % by assumption \ref{asump_convergence}. 
Thus we get for \eqref{eq_proof_A3}  with the estimates \eqref{eq_proof_A3_eta30}, \eqref{eq_proof_A3_eta3diffinfty}, \eqref{eq_proof_A3_eta3diff} 
and assumption \ref{asump_rn} that
 $P(A_3) \to 0$.

For missing event $A_4$ the situation is similar. We have %as in \cite{wagener2013adaptive} 
that
\begin{align*}
 P(A_4) \leq 
P\left( \max_{q_n <  j\leq p_n} | \eta_{4,j} | \geq \frac{\lambda_n r_n} {4} \right) 
+ P\left(\max_{q_n <  j\leq p_n} \bsbeta_{j,\text{init}}^{\tau} > r_n^{-1} \right).
\end{align*}
As it holds with \eqref{eq_proof_A2_s1} that
\begin{align}
| \eta_{4,j} | 
&\leq \frac{\lambda_n}{2n} \left\| X_j(1)' (\bsW_n^{[k-1]})^2 
\bsX_n(1) (\wtilde{\bsGamma}_n^{[k]}(1))^{-1} \right\|_2 \| \bss^0_n(1) \|_2 \nonumber \\
&\leq \frac{\lambda_n}{2} 
\left\|\frac{1}{\sqrt{n}} \bsX_n(1)'\right\|_2 
\left\| (\bsW_n^{[k-1]})^2 \right\|_2 
\left\|(\wtilde{\bsGamma}_n^{[k]}(1))^{-1} \right\|_2
\left\|\bss^0_n(1)\right\|_2 \nonumber \\
&= \lambda_n \OO_P(1) \OO_P(1) \OO_P(1) \OO_P\left(\frac{\sqrt{q_n}}{\sqrt{b_n}}\right) 
= \OO_P\left(\frac{\lambda_n \sqrt{q_n} }{\sqrt{b_n}}\right)   \nonumber
\end{align}
we get with assumption \ref{asump_rn} and \ref{asump_convergence} that 
$P(A_4) \to 0$ as $n\to \infty$. Hence, $\bsbeta_n^{[k]}$ is sign consistent.

For the asymptotic normality we use similar concepts as in \cite{wagener2013adaptive}. So given sign consistency of $\bsbeta_n^{[k]}$
we have from equation \eqref{eq_proof_beta_exp} that
\begin{equation}
\bsbeta^{[k]}_n(1) = \bsbeta_n^0(1) + \frac{1}{n} (\wtilde{\bsGamma}_n^{[k]})^{-1} \bsX_n(1)' (\bsW_n^{[k-1]})^2 %\bsSigma_n^0 
\bseps_n^0 - 
\frac{\lambda_n}{2n} (\wtilde{\bsGamma}_n^{[k]})^{-1} \bss^0_n(1) \label{eq_an_beta_decomp} .
\end{equation}
If we subtract $\bsbeta_n^0(1)$ and multiply the result by $\frac{\sqrt{n}}{ s_n(k)}\xi_n'$ we directly get
$$ \frac{\sqrt{n}}{ s_n(k)}\xi_n'( \bsbeta^{[k]}_n(1) - \bsbeta_n^0(1) ) = 
\frac{1}{ \sqrt{n} s_n(k)} \xi_n'(\wtilde{\bsGamma}_n^{[k]})^{-1} \bsX_n(1)' (\bsW_n^{[k-1]})^2 %\bsSigma_n^0 
\bseps_n^0 
-  \frac{\lambda_n}{ 2\sqrt{n} s_n(k)} \xi_n' (\wtilde{\bsGamma}_n^{[k]})^{-1} \bss^0_n(1).$$

For the second term we get with \eqref{eq_proof_A2_Gammakinv}, \eqref{eq_proof_matest} 
 and $\|\xi_n\|_2=1$ that  
\begin{align*}
 \left| \frac{\lambda_n}{2 \sqrt{n} s_n(k)} \xi_n'  (\wtilde{\bsGamma}_n^{[k]})^{-1} \bss^0_n(1) \right| 
  & \leq  \frac{\lambda_n}{2 \sqrt{n} s_n(k)} \|\xi_n\|_2 \| (\wtilde{\bsGamma}_n^{[k]})^{-1} \|_2\|  \bss^0_n(1) \|_2  \\
 &\leq \frac{\lambda_n \sqrt{q_nb}}{2 s_n(k)\sqrt{n b_n} } (\lambda_{1, \min}^{-1} + c_{4}) 
 = \OO_P\left(\frac{\lambda_n \sqrt{q_n}}{\sqrt{n b_n} }\right) .
\end{align*}
With assumption \ref{asump_convergence} this converges to zero.

 For estimating the first term we use the decomposition
 \begin{align*}
(\wtilde{\bsGamma}_n^{[k]})^{-1} \bsX_n(1)' (\bsW_n^{[k-1]})^2 = B_1 + B_2  + B_{3},  \ \ \ &\text{ where } \ \ \
% \label{eq_proof_AN_B123_decomp}  
%  \end{align*}
% \begin{align}
B_1 = (\wtilde{\bsGamma}_n^{0})^{-1} \bsX_n(1)' (\bsW_n^{0})^2,  \\
B_{2} = ( (\wtilde{\bsGamma}_n^{[k-1]})^{-1} - (\wtilde{\bsGamma}_n^{0})^{-1} ) \bsX_n(1)' (\bsW_n^{0})^2, 
\ \ \ &\text{ and } \ \ \
B_{3} = (\wtilde{\bsGamma}_n^{[k]})^{-1} \bsX_n(1)'( (\bsW_n^{[k-1]})^2- \bsW_n^{0})^2 ). 
\end{align*}
Now we decompose 
$\frac{1}{ \sqrt{n} s_n(k)} \xi_n' B_1 \bseps_n^0 = 
\frac{1}{ \sqrt{n} s_n(k)} \xi_n' B_1 \bseps_{\infty,n}^0 + \frac{1}{ \sqrt{n} s_n(k)} \xi_n' B_1 ( \bseps_n^0 - \bseps_{\infty,n}^0)$.
For the first term we have
$\frac{1}{ \sqrt{n} s_n(k)} \xi_n' B_1 \bseps_{\infty,n}^0 =  \sum_{t=1}^n a_t Z_t$
with $a_t = \frac{1}{ \sqrt{n} s_n(k) \sigma_t} \xi_n' ( \wtilde{\bsGamma}_n^{0})^{-1}   \bsX_{n,t}(1) $.
So we can calculate $\E \sum_{t=1}^n a_t Z_t = 0$ and $\E (\sum_{t=1}^n a_t Z_t)^2 = \sum_{t=1}^n \E(a_t)^2 \E(Z_t)^2 =1$.
% With \ref{asump_correlation} we can apply the central limit theorem for $\rho$-mixing
% stationary processes (see e.g. \cite{peligrad1987central}) which gives
% $\frac{1}{ \sqrt{n} s_n(k)} \xi_n' B_1 \bseps_{\infty,n}^0 \to N(0,1)$ in distribution.
It holds
with assumption \ref{asump_convergence} 
that
\begin{align*}
 \max_{1\leq t \leq n} |a_t|
 \leq \frac{1}{ \sqrt{n} s_n(k)}  \| \xi_n \|_2 \|( \wtilde{\bsGamma}_n^{0})^{-1} \|_2 
 \max_{1\leq t \leq n} \| \sigma^{-1}_t  \bsX_{n,t}(1) \|_2 
 \leq \frac{c_{11} }{ \sqrt{n}} \max_{1\leq t \leq n} \| \bsX_{n,t}(1) \|_2 = \OO(\frac{\sqrt{q_n} \vartheta_n }{\sqrt{n}}) \to 0
\end{align*}
for $n\to \infty$. So the Lindeberg condition is satisfied and 
we get with the central limit theorem that
\begin{align}
 \frac{1}{ \sqrt{n} s_n(k)} \xi_n' B_1 \bseps_{\infty,n}^0  \to N(0,1)
\label{eq_proof_B1_conv_to_N01}
\end{align}
in distribution
as $n\to \infty$.
Moreover we obtain %with the central limit theorem
\begin{align}
\left| \frac{1}{ \sqrt{n} s_n(k)} \xi_n' B_1 ( \bseps_n^0 - \bseps_{\infty,n}^0)\right | 
&\leq 
\frac{1}{ \sqrt{n} s_n(k)} \| (\wtilde{\bsGamma}_n^{0})^{-1}  \|_2
\| \bsX_n(1)'  \|_2 \| (\bsW_n^{0})^2 \|_2 \|  \bseps_n^0 - \bseps_{\infty,n}^0 \|_2  \nonumber \\ %O(1)O(sqrt(n))O(1)
&\leq \frac{c_{12}}{ \sqrt{n} s_n(k)}   \sqrt{n}  \|  \bseps_n^0 - \bseps_{\infty,n}^0 \|_2 \to 0
\label{eq_an_B1fin} 
\end{align}
as $\|  \bseps_n^0 - \bseps_{\infty,n}^0 \|_2 \to 0$ as $n\to \infty$.

Regarding $B_2$ we similarly to \cite{wagener2013adaptive} that
\begin{align}
 \left| \frac{1}{ \sqrt{n} s_n(k)} \xi_n' B_2 \bseps_n^0  \right|  
&\leq \frac{1}{ \sqrt{n} s_n(k)} \|\xi_n\|_2 
\|( (\wtilde{\bsGamma}_n^{[k-1]})^{-1} - (\wtilde{\bsGamma}_n^{0})^{-1} ) \bsX_n(1)' (\bsW_n^{0})^2 \bseps_n^0  \|_2 
 \nonumber \\
&\leq 
 \frac{ \sqrt{\lambda_{0, \max}} }{\sigma_{\min} \sqrt{n}} 
 \|(\wtilde{\bsGamma}_n^{[k-1]})^{-1} - (\wtilde{\bsGamma}_n^{0})^{-1} \|_2 \| \bsX_n(1)' (\bsW_n^{0})^2 \bseps_n^0 \|_2 
\label{eq_proof_AN_B2_decomp} 
\end{align}
Using triangle inequality we get
$$\| \bsX_n(1)' (\bsW_n^{0})^2 \bseps_n^0 \|_2
\leq \| \bsX_n(1)' (\bsW_n^{0})^2 (\bseps_n^0 - \bseps_{\infty, n}^0 ) \|_2 + \| \bsX_n(1)' (\bsW_n^{0})^2 \bseps_{\infty, n}^0 \|_2.$$
For the first term we have as above 
$$\| \bsX_n(1)' (\bsW_n^{0})^2 (\bseps_n^0 - \bseps_{\infty, n}^0 ) \|_2 
\leq \| n \bsGamma_n^0 \|^{\frac{1}{2}}_2 \| \bsW_n^{0} \|_2 \|\bseps_n^0 - \bseps_{\infty, n}^0 \|_2
= \OO_P(\sqrt{n})\OO_P(1) \OO_P(1) = \OO_P(\sqrt{n}).$$
For the second term we get with Markov's inequality
$$P\left( \frac{1}{q_n n} \|\bsX_n(1)' (\bsW_n^{0})^2 \bseps_{\infty, n}^0\|^2_2 > c\right)
\leq \frac{1}{c q_n n} \sum_{i=1}^{q_n} \E \left( \sum_{t=1}^n X_{t,i} \frac{Z_t}{\sigma_t} \right)^2 \leq \frac{1}{c \sigma_{\min}^2}$$
for $c>0$. This gives $\| \bsX_n(1)' (\bsW_n^{0})^2 \bseps_{\infty, n}^0 \|_2 = \OO(\sqrt{q_n n})$.
With $ \|(\wtilde{\bsGamma}_n^{[k-1]})^{-1} - (\wtilde{\bsGamma}_n^{0})^{-1} \|_2 = \OO(\frac{h_n}{\sqrt{n}})$ and
the previous estimates it follows for \eqref{eq_proof_AN_B2_decomp}  that
\begin{align}
\left| \frac{1}{ \sqrt{n} s_n(k)} \xi_n' B_2 \bseps_n^0  \right| 
= \OO\left(\frac{1}{\sqrt{n}}\right) 
\OO\left(\frac{h_n}{\sqrt{n}}\right) \OO\left(\sqrt{q_n n} \right) = \OO\left( \frac{h_n \sqrt{q_n}}{ \sqrt{n}}\right).
\label{eq_an_B2fin} 
\end{align}
which converges to 0 with assumption \ref{asump_convergence}. % [[ $\frac{h_n \sqrt{q_n}}{ \sqrt{n}} \to 0$ ]] this.

For the last term that corresponds to $B_3$ we have with \ref{asump_convergence} that
$$ \left| \frac{1}{ \sqrt{n} s_n(k)} \xi_n' B_3 \bseps_n^0  \right|  
\leq 
 \frac{ \sqrt{\lambda_{0, \max}} }{\sigma_{\min} \sqrt{n}} (\lambda_{1, \min}^{-1} + c_4) \| \bsX_n(1)' 
 ( (\bsW_n^{0})^2 - (\bsW_n^{[k-1]})^2 ) \bseps_n^0 \|_2 $$
Again, the second norm can be estimated by
\begin{align*}
&\| \bsX_n(1)' 
 ( (\bsW_n^{0})^2 - (\bsW_n^{[k-1]})^2 ) \bseps_n^0 \|_2 \nonumber \\
 &\leq \| \bsX_n(1)' 
 ( (\bsW_n^{0})^2 - (\bsW_n^{[k-1]})^2 ) (\bseps_n^0 - \bseps_{\infty,n}^0) \|_2 + 
 \| \bsX_n(1)' 
 ( (\bsW_n^{0})^2 - (\bsW_n^{[k-1]})^2 ) \bseps_{\infty,n}^0 \|_2 
\end{align*}
 using the triangle inequality. 
  The first term can be estimated by
\begin{align*}
 \| \bsX_n(1)' 
 ( (\bsW_n^{0})^2 - (\bsW_n^{[k-1]})^2 ) (\bseps_n^0 - \bseps_{\infty,n}^0) \|_2 
 &\leq \| \bsX_n(1)' \|_2 \|  (\bsW_n^{0})^2 - (\bsW_n^{[k-1]})^2  \|_2 \| (\bseps_n^0 - \bseps_{\infty,n}^0)\|_2 \\
 &\leq \OO_P(\sqrt{n}) \OO_P\left( \frac{h_n}{\sqrt{n}}\right) \OO_P(1) = \OO_P(h_n) 
 \end{align*}
 For the second term
 we have again with assumptions \ref{asump_variance_mom}, \ref{asump_variance_estimators} and
and Markov's inequality 
% $$P\left( \frac{1}{q_n n} \|\bsX_n(1)' (\bsW_n^{0})^2 \bseps_{\infty, n}^0\|^2_2 > c\right)
% \leq \frac{1}{c q_n n} \sum_{i=1}^{q_n} \E \left( \sum_{t=1}^n X_{t,i} \frac{Z_t}{\sigma_t} \right)^2 \leq \frac{1}{c \sigma_{\min}^2}$$
\begin{align*}
 P\left(  \|\bsX_n(1)'  ( (\bsW_n^{0})^2 - (\bsW_n^{[k-1]})^2 ) \bseps_{\infty, n}^0\|^2_2 > c\right)
 &\leq  
 \sum_{i=1}^{q_n} \E \left( \sum_{t=1}^n X_{t,i} 
 \frac{1}{\sigma^2_t- (\what{\sigma}^{[k-1]}_t)^2 } \eps_t \right)^2  \\
 &\leq  
 c_{13} \frac{h_n^2}{ n} \sum_{i=1}^{q_n} \E \left( \sum_{t=1}^n X_{t,i} \eps_t \right)^2  = \OO( q_n h_n^2 )
\end{align*}
where $\what{\sigma}^{[k-1]}_t$ for $1\leq t \leq n$ are the diagonal elements of $(\bsW_n^{[k-1]})^{-1}$ and $c>0$.
% \begin{align}
%  \| \bsX_n(1)' 
%  ( (\bsW_n^{0})^2 - (\bsW_n^{[k-1]})^2 ) (\bseps_n^0 - \bseps_{\infty,n}^0) \|_2 
%  &\leq \| \bsX_n(1)' \|_2 \|  (\bsW_n^{0})^2 - (\bsW_n^{[k-1]})^2  \|_2 \| (\bseps_n^0 - \bseps_{\infty,n}^0)\|_2 \\
%  &\leq \OO_P(\sqrt{n}) \OO_P\left( \frac{h_n}{\sqrt{n}}\right) \OO_P(1) = \OO_P(h_n) 
%  \end{align}
%  The second term can be estimated by
%  $$ \| \bsX_n(1)' 
%  ( (\bsW_n^{0})^2 - (\bsW_n^{[k-1]})^2 ) \bseps_{\infty,n}^0 \|_2
%  \leq = \OO(\frac{q_n^{3/2}}{\sqrt{n}}).$$
Hence with assumption \ref{asump_convergence}%[[ cond $h_n \sqrt{q_n}/\sqrt{n}$]] 
we receive 
\begin{align}
| \frac{1}{ \sqrt{n} s_n(k)} \xi_n' B_3 \bseps_n^0  | \leq c_8 \frac{\sqrt{q_n}h_n}{\sqrt{n}} \to 0
\label{eq_an_B3fin} 
\end{align}
as $n\to \infty$.
With the three estimates involving $B_1$, $B_2$ and $B_3$ we receive for 
equation \eqref{eq_an_beta_decomp} together with 
equations \eqref{eq_proof_B1_conv_to_N01}, \eqref{eq_an_B1fin}, \eqref{eq_an_B2fin}, \eqref{eq_an_B3fin} and Slutky's 
theorem that $\frac{\sqrt{n}}{ s_n(k)}\xi_n'( \bsbeta^{[k]}_n(1) - \bsbeta_n^0(1) ) \to N(0,1)$.

At the beginning 
that the theorem is satisfied for $k>1$. So the proof of the inital step with $k=1$ is missing. 
% we assumed that 
% $\| \bsbeta_n^{[k-1]} - \bsbeta_n^0 \|_2 = \OO(\frac{\sqrt{q_n}}{\sqrt{n}})$
% holds for $k>1$. 
% Indeed for $k>1$ we even have $\| \bsbeta_n^{[k-1]} - \bsbeta_n^0 \|_2 = \OO(\frac{1}{\sqrt{n}})$.
% In the case $k=1$ we do not have this property, so the basis case of the induction is missing. 
However, the proof is similar to the sign consistency and asymptotic normality proof with $k>1$,
as \cite{wagener2013adaptive} explained it for the unconstrained weighted adaptive lasso. 
Note that the proof itself is less complex than the case $k>1$, but
involves the eigenvalue assumptions to the unscaled Gramian $\bsGamma^0_n$ 
(i.e. $ \lambda_{0, \min}< \lambda_{\min}(\bsGamma^0_n)$) that were not used in the previous part, 
instead of the assumption to the scaled version $\wtilde{\bsGamma}^0_n$.
% and 
% $g_n(\what{\bsalpha}_n(\bsbeta_n; \bsX_n, \bsY_n) , \what{\bsL}_{n,t}(\bsbeta_n; \bsX_n, \bsY_n)) < \sigma_{\max}$ 
% for all $\bsbeta_n$ in a neighbourhood of $\bsbeta_n^0$

\end{proof}

\small

\section{References}

% \clearpage
   \bibliographystyle{apalike}
 \bibliography{ararch_lasso_final}

\begin{thebibliography}{}

\bibitem[Aknouche and Al-Eid, 2012]{aknouche2012asymptotic}
Aknouche, A. and Al-Eid, E. (2012).
\newblock Asymptotic inference of unstable periodic arch processes.
\newblock {\em Statistical inference for stochastic processes}, 15(1):61--79.

\bibitem[Bardet et~al., 2009]{bardet2009asymptotic}
Bardet, J.-M., Wintenberger, O., et~al. (2009).
\newblock Asymptotic normality of the quasi-maximum likelihood estimator for
  multidimensional causal processes.
\newblock {\em The Annals of Statistics}, 37(5B):2730--2759.

\bibitem[Bien et~al., 2013]{bien2013lasso}
Bien, J., Taylor, J., Tibshirani, R., et~al. (2013).
\newblock A lasso for hierarchical interactions.
\newblock {\em The Annals of Statistics}, 41(3):1111--1141.

\bibitem[Chan et~al., 2013]{chan2013group}
Chan, N.~H., Yau, C.~Y., and Zhang, R.-M. (2013).
\newblock Group lasso for structural break time series.
\newblock {\em Journal of the American Statistical Association},
  (just-accepted).

\bibitem[Chen and Chan, 2011]{chen2011subset}
Chen, K. and Chan, K.-S. (2011).
\newblock Subset arma selection via the adaptive lasso.
\newblock {\em Statistics and its Interface}, 4(2):197--205.

\bibitem[Choi et~al., 2010]{choi2010variable}
Choi, N.~H., Li, W., and Zhu, J. (2010).
\newblock Variable selection with the strong heredity constraint and its oracle
  property.
\newblock {\em Journal of the American Statistical Association},
  105(489):354--364.

\bibitem[Efron et~al., 2004]{efron2004least}
Efron, B., Hastie, T., Johnstone, I., Tibshirani, R., et~al. (2004).
\newblock Least angle regression.
\newblock {\em The Annals of statistics}, 32(2):407--499.

\bibitem[Francq and Zako{\"\i}an, 2013]{francq2013optimal}
Francq, C. and Zako{\"\i}an, J.-M. (2013).
\newblock Optimal predictions of powers of conditionally heteroscedastic
  processes.
\newblock {\em Journal of the Royal Statistical Society: Series B (Statistical
  Methodology)}, 75(2):345--367.

\bibitem[Friedman et~al., 2007]{friedman2007pathwise}
Friedman, J., Hastie, T., H{\"o}fling, H., Tibshirani, R., et~al. (2007).
\newblock Pathwise coordinate optimization.
\newblock {\em The Annals of Applied Statistics}, 1(2):302--332.

\bibitem[Gefang, 2014]{gefang2014bayesian}
Gefang, D. (2014).
\newblock Bayesian doubly adaptive elastic-net lasso for var shrinkage.
\newblock {\em International Journal of Forecasting}, 30(1):1--11.

\bibitem[Harchaoui and L{\'e}vy-Leduc, 2010]{harchaoui2010multiple}
Harchaoui, Z. and L{\'e}vy-Leduc, C. (2010).
\newblock Multiple change-point estimation with a total variation penalty.
\newblock {\em Journal of the American Statistical Association}, 105(492).

\bibitem[Hsu et~al., 2008]{hsu2008subset}
Hsu, N.-J., Hung, H.-L., and Chang, Y.-M. (2008).
\newblock Subset selection for vector autoregressive processes using lasso.
\newblock {\em Computational Statistics \& Data Analysis}, 52(7):3645--3657.

\bibitem[Huang et~al., 2006]{huang2006adaptive}
Huang, J., Ma, S., and Zhang, C.-H. (2006).
\newblock Adaptive lasso for sparse high-dimensional regression models.
\newblock Technical report, The University of Iowa, Department of Statistics
  and Actuarial Science.
\newblock Technical Report No. 374.

\bibitem[Huang et~al., 2008]{huang2008adaptive}
Huang, J., Ma, S., and Zhang, C.-H. (2008).
\newblock Adaptive lasso for sparse high-dimensional regression models.
\newblock {\em Statistica Sinica}, 18(4):1603.

\bibitem[Kim et~al., 2012]{kim2012consistent}
Kim, Y., Kwon, S., and Choi, H. (2012).
\newblock Consistent model selection criteria on high dimensions.
\newblock {\em The Journal of Machine Learning Research}, 98888(1):1037--1057.

\bibitem[Lawson and Hanson, 1995]{lawson1974solving}
Lawson, C.~L. and Hanson, R.~J. (1995).
\newblock {\em {Solving Least Squares Problems}}.
\newblock SIAM.

\bibitem[Levy-leduc and Harchaoui, 2008]{levy2008catching}
Levy-leduc, C. and Harchaoui, Z. (2008).
\newblock Catching change-points with lasso.
\newblock In {\em Advances in Neural Information Processing Systems}, pages
  617--624.

\bibitem[Ling, 2007]{ling2007self}
Ling, S. (2007).
\newblock Self-weighted and local quasi-maximum likelihood estimators for
  arma-garch/igarch models.
\newblock {\em Journal of Econometrics}, 140(2):849--873.

\bibitem[Mak et~al., 1997]{mak1997estimation}
Mak, T., Wong, H., and Li, W. (1997).
\newblock Estimation of nonlinear time series with conditional heteroscedastic
  variances by iteratively weighted least squares.
\newblock {\em Computational statistics \& data analysis}, 24(2):169--178.

\bibitem[Medeiros and Mendes, 2012]{medeiros2012estimating}
Medeiros, M.~C. and Mendes, E. (2012).
\newblock Estimating high-dimensional time series models.
\newblock {\em CREATES Research Paper}, 37.

\bibitem[Meinshausen et~al., 2013]{meinshausen2013sign}
Meinshausen, N. et~al. (2013).
\newblock Sign-constrained least squares estimation for high-dimensional
  regression.
\newblock {\em Electronic Journal of Statistics}, 7:1607--1631.

\bibitem[Nardi and Rinaldo, 2011]{nardi2011autoregressive}
Nardi, Y. and Rinaldo, A. (2011).
\newblock Autoregressive process modeling via the lasso procedure.
\newblock {\em Journal of Multivariate Analysis}, 102(3):528--549.

\bibitem[Rabemananjara and Zakoian, 1993]{rabemananjara1993threshold}
Rabemananjara, R. and Zakoian, J.-M. (1993).
\newblock Threshold arch models and asymmetries in volatility.
\newblock {\em Journal of Applied Econometrics}, 8(1):31--49.

\bibitem[Ren et~al., 2013]{ren2013two}
Ren, Y., Xiao, Z., and Zhang, X. (2013).
\newblock Two-step adaptive model selection for vector autoregressive
  processes.
\newblock {\em Journal of Multivariate Analysis}, 116:349--364.

\bibitem[Ren and Zhang, 2010]{ren2010subset}
Ren, Y. and Zhang, X. (2010).
\newblock Subset selection for vector autoregressive processes via adaptive
  lasso.
\newblock {\em Statistics \& probability letters}, 80(23):1705--1712.

\bibitem[Slawski et~al., 2013]{slawski2013non}
Slawski, M., Hein, M., et~al. (2013).
\newblock {Non-negative least squares for high-dimensional linear models:
  Consistency and sparse recovery without regularization}.
\newblock {\em Electronic Journal of Statistics}, 7:3004--3056.

\bibitem[Tibshirani, 1996]{tibshirani1996regression}
Tibshirani, R. (1996).
\newblock Regression shrinkage and selection via the lasso.
\newblock {\em Journal of the Royal Statistical Society. Series B
  (Methodological)}, pages 267--288.

\bibitem[Tibshirani et~al., 2005]{tibshirani2005sparsity}
Tibshirani, R., Saunders, M., Rosset, S., Zhu, J., and Knight, K. (2005).
\newblock Sparsity and smoothness via the fused lasso.
\newblock {\em Journal of the Royal Statistical Society: Series B (Statistical
  Methodology)}, 67(1):91--108.

\bibitem[Van~der Vaart and Wellner, 1996]{van1996weak}
Van~der Vaart, A.~W. and Wellner, J.~A. (1996).
\newblock {\em Weak Convergence}.
\newblock Springer.

\bibitem[Wagener and Dette, 2012]{wagener2012bridge}
Wagener, J. and Dette, H. (2012).
\newblock Bridge estimators and the adaptive lasso under heteroscedasticity.
\newblock {\em Mathematical Methods of Statistics}, 21(2):109--126.

\bibitem[Wagener and Dette, 2013]{wagener2013adaptive}
Wagener, J. and Dette, H. (2013).
\newblock The adaptive lasso in high-dimensional sparse heteroscedastic models.
\newblock {\em Mathematical Methods of Statistics}, 22(2):137--154.

\bibitem[Wang et~al., 2007a]{wang2007robust}
Wang, H., Li, G., and Jiang, G. (2007a).
\newblock Robust regression shrinkage and consistent variable selection through
  the lad-lasso.
\newblock {\em Journal of Business \& Economic Statistics}, 25(3):347--355.

\bibitem[Wang et~al., 2007b]{wang2007regression}
Wang, H., Li, G., and Tsai, C.-L. (2007b).
\newblock Regression coefficient and autoregressive order shrinkage and
  selection via the lasso.
\newblock {\em Journal of the Royal Statistical Society: Series B (Statistical
  Methodology)}, 69(1):63--78.

\bibitem[Yoon et~al., 2013]{yoon2013penalized}
Yoon, Y.~J., Park, C., and Lee, T. (2013).
\newblock Penalized regression models with autoregressive error terms.
\newblock {\em Journal of Statistical Computation and Simulation},
  83(9):1756--1772.

\bibitem[Zhang et~al., 2010]{zhang2010regularization}
Zhang, Y., Li, R., and Tsai, C.-L. (2010).
\newblock Regularization parameter selections via generalized information
  criterion.
\newblock {\em Journal of the American Statistical Association},
  105(489):312--323.

\bibitem[Zhao and Yu, 2006]{zhao2006model}
Zhao, P. and Yu, B. (2006).
\newblock On model selection consistency of lasso.
\newblock {\em The Journal of Machine Learning Research}, 7:2541--2563.

\bibitem[Ziel, 2015]{ziel2015quasi}
Ziel, F. (2015).
\newblock Quasi-maximum likelihood estimation of periodic autoregressive,
  conditionally heteroscedastic time series.
\newblock In {\em Stochastic Models, Statistics and Their Applications}, pages
  207--214. Springer.

\bibitem[Ziel et~al., 2015]{ziel2015efficient}
Ziel, F., Steinert, R., and Husmann, S. (2015).
\newblock Efficient modeling and forecasting of electricity spot prices.
\newblock {\em Energy Economics}, 47:98--111.

\bibitem[Zou, 2006]{zou2006adaptive}
Zou, H. (2006).
\newblock The adaptive lasso and its oracle properties.
\newblock {\em Journal of the American statistical association},
  101(476):1418--1429.

\bibitem[Zou and Hastie, 2005]{zou2005regularization}
Zou, H. and Hastie, T. (2005).
\newblock Regularization and variable selection via the elastic net.
\newblock {\em Journal of the Royal Statistical Society: Series B (Statistical
  Methodology)}, 67(2):301--320.

\bibitem[Zou et~al., 2007]{zou2007degrees}
Zou, H., Hastie, T., Tibshirani, R., et~al. (2007).
\newblock On the “degrees of freedom” of the lasso.
\newblock {\em The Annals of Statistics}, 35(5):2173--2192.

\end{thebibliography}
% \clearpage
\end{document}